\documentclass[letterpaper,11pt]{article}
\pdfoutput=1
\usepackage{jheppub}
\usepackage{multirow}
\usepackage{shuffle}
\usepackage{tipa}
\usepackage{upgreek}
\usepackage{breqn}
\usepackage{caption}
\usepackage{subcaption}
\usepackage{xspace}
\usepackage[countmax]{subfloat}
\usepackage{enumitem}
\usepackage{amssymb}
\usepackage{amsmath}
\usepackage{cancel}
\usepackage{color}
\usepackage{braket}
\usepackage{graphicx}
\usepackage{multirow}
\usepackage{verbatim}
\usepackage{amsthm}
\usepackage{slashed}
\usepackage{wasysym}
\usepackage{simplewick}
\usepackage{mathtools}
\usepackage{soul}
\usepackage{xspace}
\usepackage{hyperref}
\usepackage{graphicx}
\usepackage{amsmath}
\usepackage{mathrsfs}
\usepackage{xspace}
\usepackage{amsfonts}
\usepackage{color}
\usepackage{booktabs,siunitx}
\usepackage{mathrsfs}

\usepackage{tikz}
\usepackage{tikzit}
\usetikzlibrary{snakes}
\usetikzlibrary{decorations.markings}
\tikzset{snake it/.style={decorate, decoration=snake}}

\usepackage{subcaption}
\usetikzlibrary{decorations.pathmorphing}
\usetikzlibrary{positioning,decorations.pathreplacing}
\tikzset{snake it/.style={decorate, decoration=snake}}

\definecolor{lPurple}{RGB}{135,0,255}
\definecolor{forest}{RGB}{34,139,34}
\newcommand{\eq}[1]{Eq.~\eqref{eq:#1}}
\newcommand{\eqs}[2]{Eqs.~\eqref{eq:#1} and \eqref{eq:#2}}
\newcommand{\fig}[1]{Fig.~\ref{fig:#1}}
\newcommand{\figs}[2]{Figs.~\ref{fig:#1} and \ref{fig:#2}}

\newcommand{\gco}{\Phi}

\def\cE{\mathcal{E}}

\newcommand{\nn}{\nonumber}

\allowdisplaybreaks[2]

\usepackage{xcolor}
\usepackage{marginnote}
\usepackage[normalem]{ulem}
\usepackage{dsfont,mathrsfs}

\definecolor{darkgreen}{rgb}{0.13,0.55,0.13}
\definecolor{pur}{rgb}{0.6,0.,1.}

\DeclareRobustCommand{\Sec}[1]{sec.~\ref{#1}}

\preprint{MIT-CTP-5819}

\title{Generalized Detectors at Colliders}
\author[a]{Mark Gonzalez,} \author[a,b,c]{Kyle Lee,}
\author[a]{Ian Moult} 
\affiliation[a]{Department of Physics, Yale University, New Haven, CT 06511}
\affiliation[b]{High Energy Physics Division, Argonne National Laboratory, Lemont, IL 60439, USA}
\affiliation[c]{Center for Theoretical Physics -- a Leinweber Institute, Massachusetts Institute of Technology, Cambridge, MA 02139}
\emailAdd{mark.gonzalez@yale.edu, kyle@anl.gov, ian.moult@yale.edu}

\abstract{Recent progress in collider physics has reformulated phenomenological questions in terms of detector correlation functions and advanced their theoretical understanding. These advances have primarily focused on correlators of the average null energy operator, $\mathcal{E}(\vec{n})$, known as energy correlators. However, colliders have access to a much broader class of detector operators, $\mathcal{E}^n_R(\vec{n})$, which measure powers of the energy on a subset of hadrons $R$, such as charged hadrons. These generalized detectors are generically not infrared and collinear safe, and their description requires nonperturbative matching between the hadronic detectors measured in the infrared and the partonic detectors used in ultraviolet calculations. We develop a framework for computing their multi-point correlation functions. We introduce universal nonperturbative matching coefficients, termed ``detector functions'', that implement this infrared-ultraviolet matching. For the $\mathcal{E}_R^n$ operators studied here, these coefficients are represented by energy-weighted moments of single- and multi-hadron fragmentation functions. We present their renormalization group structure, derive QCD factorization theorems for the projected correlators, and compute jet functions through next-to-leading order. In the fixed-coupling pure Yang-Mills limit, we derive the light-ray OPE of hadronic detectors and connect it to the QCD factorization framework. We also identify universal nonperturbative power corrections generated by soft radiation, which are enhanced in the collinear limit and can substantially modify the perturbative angular scaling. A parton shower study finds qualitative agreement with the predicted perturbative and nonperturbative scaling behaviors. Our work significantly broadens the space of detector operators under theoretical control, with potential phenomenological applications.}
\begin{document} 
\maketitle
\section{Introduction}
At their core, collider experiments measure correlations in the asymptotic fluxes of energy and quantum numbers produced in high-energy collisions. In quantum field theory, such flux measurements are precisely formulated in terms of ``detector operators", and the outcomes of collider measurements are expressed as correlation functions of these detector operators.

Among these, the flux of energy plays a distinguished role and its associated detector operator is the average null energy (ANE) operator~\cite{Sterman:1975xv,Sveshnikov:1995vi,Tkachov:1995kk,Korchemsky:1997sy}
\begin{align}
\label{eq:energyflow}
\mathcal{E}(\vec{n})=\lim _{r \rightarrow \infty} \int_0^\infty dt\, r^2\, n^i\, T_{0 i}(t, r \vec{n})\,.
\end{align}
Its study has a long history, and is one of the primary means by which we have understood the microscopic structure of the Standard Model. Its correlation functions, $\langle \mathcal{E}(\vec{n}_1) \mathcal{E}(\vec{n}_2)\cdots \mathcal{E}(\vec{n}_K) \rangle$, are the celebrated ``energy correlators" \cite{Basham:1978zq,Basham:1979gh,Basham:1977iq,Basham:1978bw} introduced in the late 1970s to study correlations in energy flux in QCD.

Such correlations of energy flux have recently received newfound attention, leading to tremendous developments in theoretical tools for understanding correlation functions of detector operators, including the light-ray OPE, celestial block decompositions, and explicit perturbative calculations~\cite{Hofman:2008ar, DelDuca:2016csb, Kravchuk:2018htv, Dixon:2018qgp, Kologlu:2019mfz, Dixon:2019uzg, Caron-Huot:2022eqs, Chen:2022jhb, Chang:2022ryc, Duhr:2022yyp, Chen:2023zzh, Dempsey:2025yiv,Li:2025knf,Mecaj:2025ecl,Lee:2026zyl}.
At the same time, the study of energy correlators has been extended to multiple phenomenological settings, including their use as jet substructure observables at the LHC~\cite{Lee:2022uwt,Komiske:2022enw}, where it has become possible to experimentally study the detailed flow of energy within jets. The application of these newly developed theoretical tools to phenomenological studies has enabled significant progress across both high energy and nuclear collider physics~\cite{Craft:2022kdo, Ricci:2022htc, Yang:2022tgm, Liu:2022wop, Andres:2022ovj, Devereaux:2023vjz, Jaarsma:2023ell, Kang:2023big, Barata:2023bhh, Gao:2023ivm, Liu:2024kqt, Chen:2024nfl, Lee:2024esz, Andres:2024ksi, Holguin:2024tkz, Budhraja:2024tev, Lee:2024jnt, Dai:2024wff, Gao:2024wcg, Mantysaari:2025mht, Bhattacharya:2025bqa, Barata:2025uxp, Cao:2025icu, Ke:2025ibt, Lee:2026hub, Gonzalez:2026dzy, Gao:2026xuq, Kang:2026hig}. See~\cite{Moult:2025nhu} for a recent review.

The ANE detector is the simplest detector one can imagine from a phenomenological perspective. Due to the conservation of the stress tensor, it is not renormalized, and multi-point correlators of the ANE are infrared and collinear (IRC) safe observables, allowing them to be computed entirely in perturbation theory. It is for this reason that it was singled out in early studies, and its correlators provide a useful way of organizing IRC safe observables. However, in collider experiments it is possible to measure correlation functions of a broader range of detector operators. Exploring this broader space of detector operators and their correlators is interesting both phenomenologically, as they can lead to new ways to study the Standard Model, and theoretically. 
\begin{figure}[t]
\begin{center}
	\includegraphics[width =2 in]{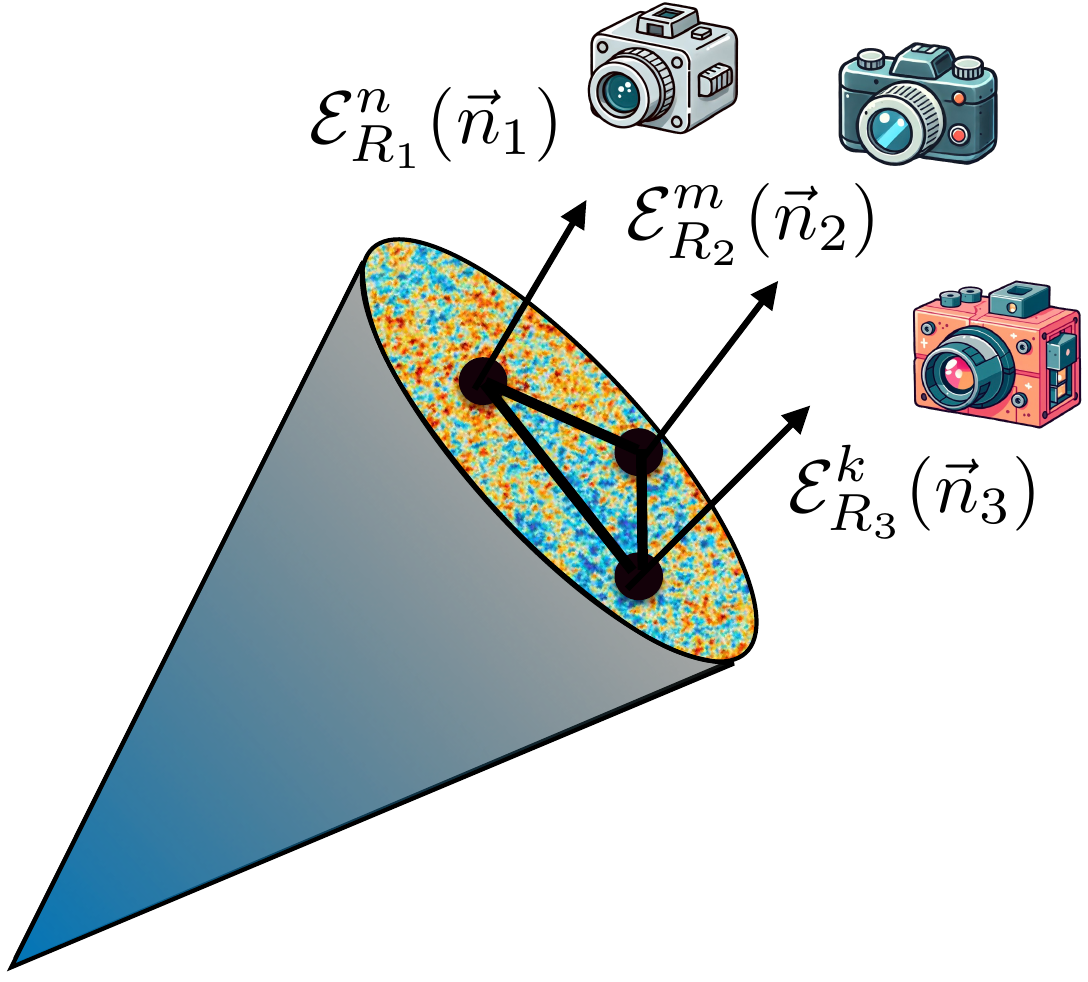}
\end{center}
\caption{A multi-point correlator of generalized detector operators which measure the energy flux on hadrons with property $R_i$ weighted to generic powers $n,m,k$. As compared to the standard ANE detector, the calculation of more general detector operators requires the introduction of ``detector functions", which provide the non-trivial matching between the perturbative and hadronic detectors.}
\label{fig:det}
\end{figure}

While there is a broad space of general detector operators in quantum field theory, here we will focus on a subclass that is particularly natural to measure at current collider experiments. Colliders can identify species of hadrons using their particle identification capabilities and measure their energies.\footnote{Beyond energy flux, correlators of other conserved fluxes such as electric charge and flavor have also been considered recently~\cite{Lee:2026FlavorFlux,Riembau:2024tom,Zhang:2026esd,Zhang:2026vdo,Zhang:2026emt}.} Using this information, we can generalize the ANE in two different ways. First, we can restrict to detectors that measure energy flux on a subset of hadrons, e.g. (positively/negatively) charged hadrons. We will denote such a detector by $\mathcal{E}_R$, where $R$ denotes the restricted set of hadrons.  Such detectors have important practical applications since the restriction to charged hadrons enables improved angular resolution. Recently, track-based measurements of the two-point energy correlator have been performed at the $Z$-pole using archival LEP data~\cite{Electron-PositronAlliance:2025fhk,Electron-PositronAlliance:2025wzh,Zhang:2025nlf}. 

Second, one can study not just the energy flux, but powers of the energy flux, which we will denote $\mathcal{E}^n$. Such generalized detectors are of clear interest phenomenologically, as their correlations enable one to image not only how energy is distributed as a function of the angular scale, but also how hard and soft particles are distributed as a function of angle. The one-point function of $\mathcal{E}^n$ is simply a moment of the fragmentation function and was studied recently from a formal perspective in \cite{Caron-Huot:2022eqs} and in QCD~\cite{Chang:2025zib}. Multi-point correlators with different energy weights were introduced already in the 1980s \cite{Strharsky:1984fy}, where they were studied in the back-to-back limit. In the modern jet substructure context, they have been proposed as phenomenological observables in \cite{Devereaux:2023vjz,Bossi:2024qho}, and measured in \cite{CMS:2025ydi}.

Combining these two generalizations of the ANE detector, we are led to explore the generalized detector operators $\mathcal{E}_R^n$, and their multipoint correlators, $\langle \mathcal{E}_{R_1}^{n_1} \mathcal{E}_{R_2}^{n_2}\cdots \mathcal{E}_{R_N}^{n_N} \rangle$. Such a correlation function of generalized detector operators is shown schematically in \fig{det}. Such correlators naturally generalize the energy correlators, which organize IRC safe observables, to correlations that organize a broader class of observables that are not IRC safe. In particular, they provide clean probes of hadronization, as they directly access correlations of hadronic operators with different energy weights and quantum numbers.

Understanding these multi-point correlators of generalized detector operators is significantly more complicated than their ANE counterparts. First, they are not infrared and collinear safe. This means that one must perform a nontrivial matching between the perturbative partonic detectors in the UV and the nonperturbative hadronic detectors in the IR. Such hadronic IR detectors are a distinctive feature of confining, gapped theories such as QCD. We use ``detector matching functions", or simply ``detector functions", as the general name for the universal nonperturbative coefficients that match these hadronic detectors onto partonic UV detectors. For the $\mathcal{E}_R^n$ operators studied here, these coefficients are represented by energy-weighted moments of single- and multi-hadron fragmentation functions~\cite{Pitonyak:2025lin,Jaffe:1997hf,Pitonyak:2023gjx,Collins:1981uw,Rogers:2024nhb,vonKuk:2025kbv,Rogers:2026jca,Majumder:2004wh,Bianconi:1999cd,deFlorian:2003cg}. Furthermore, building on the track-function formalism~\cite{Chang:2013rca,Chang:2013iba,Li:2021zcf,Jaarsma:2022kdd,Chen:2022pdu,Lee:2023xzv,Lee:2023tkr,Barata:2024nqo,Barata:2026kkv}, we introduce generalized track functions that compactly organize the combinations of single- and multi-hadron fragmentation-function moments entering this detector matching. We derive their renormalization group evolution through NLO and show how they enter QCD factorization theorems for generalized detector correlators.

Second, the leading nonperturbative power corrections to the correlation functions of generalized detectors are modified relative to the ANE case. For projected correlators of the ANE, which measure only the largest angle $x_L = (1-{\rm min}\{\vec{n}_i\cdot\vec{n}_j\})/2$ amongst the detectors, the leading nonperturbative corrections have received significant attention \cite{Korchemsky:1999kt,Lee:2024esz,Schindler:2023cww,Chen:2023wah} and are well understood: while suppressed by the ratio of $\Lambda_{\text{QCD}}$ to the hard scale, they exhibit an enhanced power-law scaling in the collinear limit, and therefore eventually dominate the correlators at small angles. By studying different quantum numbers and energy weights of the generalized detector operators, we will see that the leading nonperturbative corrections to their projected correlators have in general different scaling behaviors than those of the ANE case.

The outline of this paper is as follows. In \Sec{sec:d_functions}, we introduce detector functions as the general matching coefficients between hadronic and partonic detectors. For the $\mathcal{E}_R^n$ detectors, we identify their representation in terms of moments of multi-hadron fragmentation functions, study their renormalization group structure, and analyze the light-ray OPE, restricted to the pure YM case. In \Sec{sec:GenTrack}, we introduce one- and multi-dimensional generalized track functions, show how their moments organize and encode the linear combinations of multi-hadron fragmentation functions relevant for detector matching, for general energy weights and hadron selections, and discuss their renormalization group evolution. In \Sec{sec:fact}, we present collinear factorization theorems in full QCD for general $x_L$-projected correlators, compute the associated jet functions through NLO, and compare with the light-ray OPE framework of \Sec{sec:d_functions}. In \Sec{sec:NP}, we discuss the leading nonperturbative power corrections. In \Sec{sec:MC}, we compare the scaling behaviors observed in parton shower simulations with the expectations of our framework, arising from both the purely perturbative contributions and the nonperturbative power corrections, and find consistent results. We conclude in \Sec{sec:conc}.

\section{UV and IR Detectors, Matching, and the Light-Ray OPE}\label{sec:d_functions}
As compared to the case of CFTs, in a generic confining QFT, the space of detector operators will be different in the UV and the IR. This is particularly true in QCD, where the IR detectors should be expressed in terms of hadrons due to the mass gap set by the scale of order $\mathcal{O}(\Lambda_{\rm QCD})$, while the UV detectors are expressed in terms of quarks and gluons. The fact that the space of detectors differs in the UV and the IR has practical consequences, namely, that experimental measurements are performed on hadrons. On the other hand, all practical calculations involving energetic scattering of some perturbative scale $Q$ are performed in terms of the UV partonic detectors. Therefore, to be able to understand detector operators in generic QFTs, we must understand precisely the mapping between UV detectors (\emph{what is calculable}) and IR detectors (\emph{what is measurable}).

In this section, we set up the UV and IR detectors and introduce detector matching functions (or simply detector functions) that carry out the matching between partonic UV detectors and the hadronic IR detectors that emerge after confinement. Throughout, ``detector function" denotes the role of an object as a matching coefficient. For correlations of $\mathcal{E}_R^m$ detectors considered here, detector functions are given by moments of single-hadron and multi-hadron fragmentation functions. We further discuss how this matching enables the light-ray OPE of multi-point correlators of IR detectors.

\subsection{Matching between UV and IR Detectors}\label{sec:matching}
\begin{figure}
\centering
\includegraphics[width=0.45\textwidth]{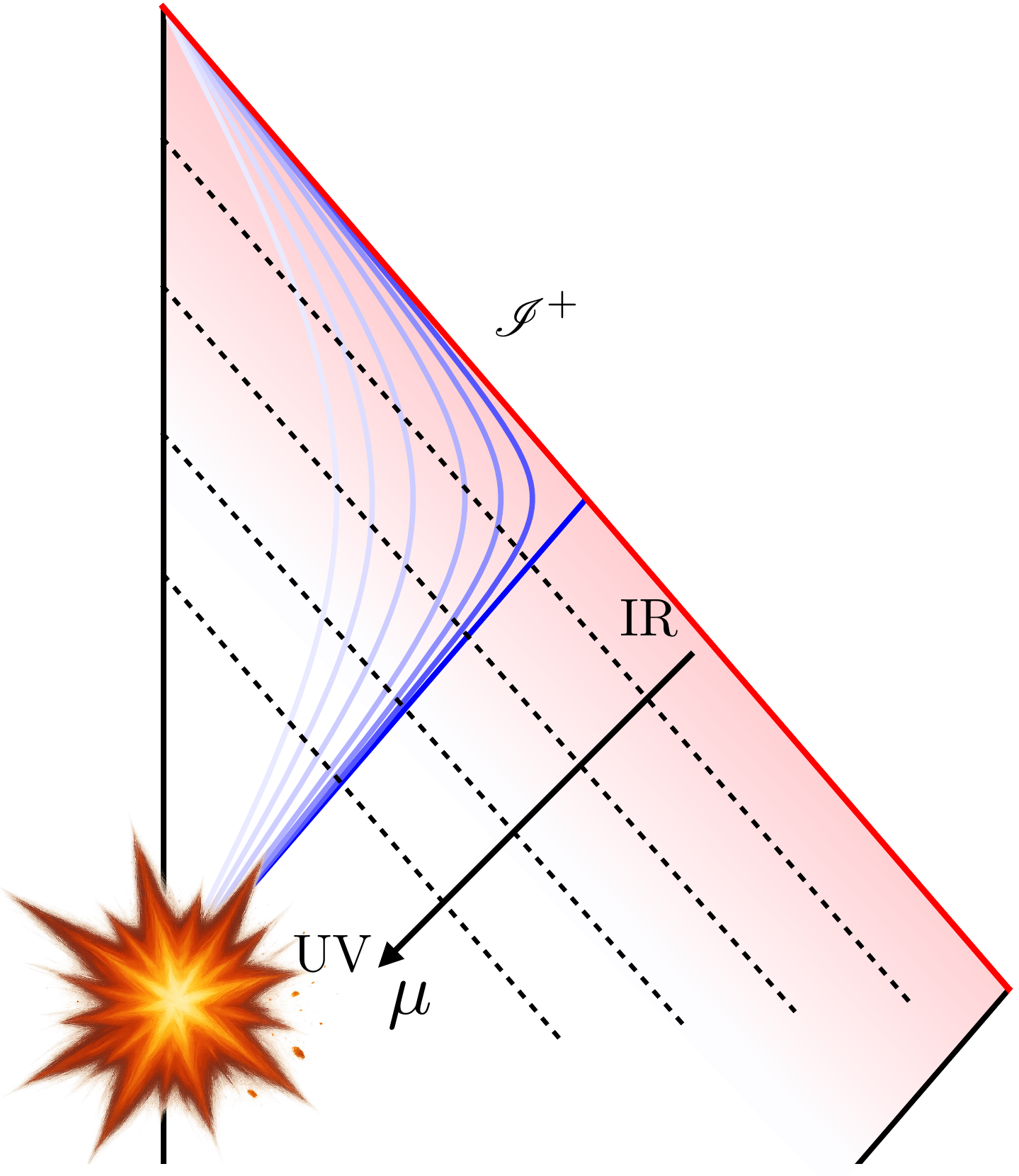}
\caption{Schematic Penrose diagram illustrating the detector‐based renormalization group. The red line at future null infinity, $\mathscr{I}^+$, denotes UV detectors sensitive only to massless, null trajectories. As the null sheet is moved inward, more timelike massive trajectories can be treated as approximately null. \eq{matching} then formalizes how an IR detector on a finite null sheet is matched onto the UV detectors at $\mathscr{I}^+$ by integrating out (``moving inward'') long‐distance dynamics.%
}
\label{fig:penrose}
\end{figure}

\begin{figure}
\centering
\includegraphics[width=0.5\textwidth]{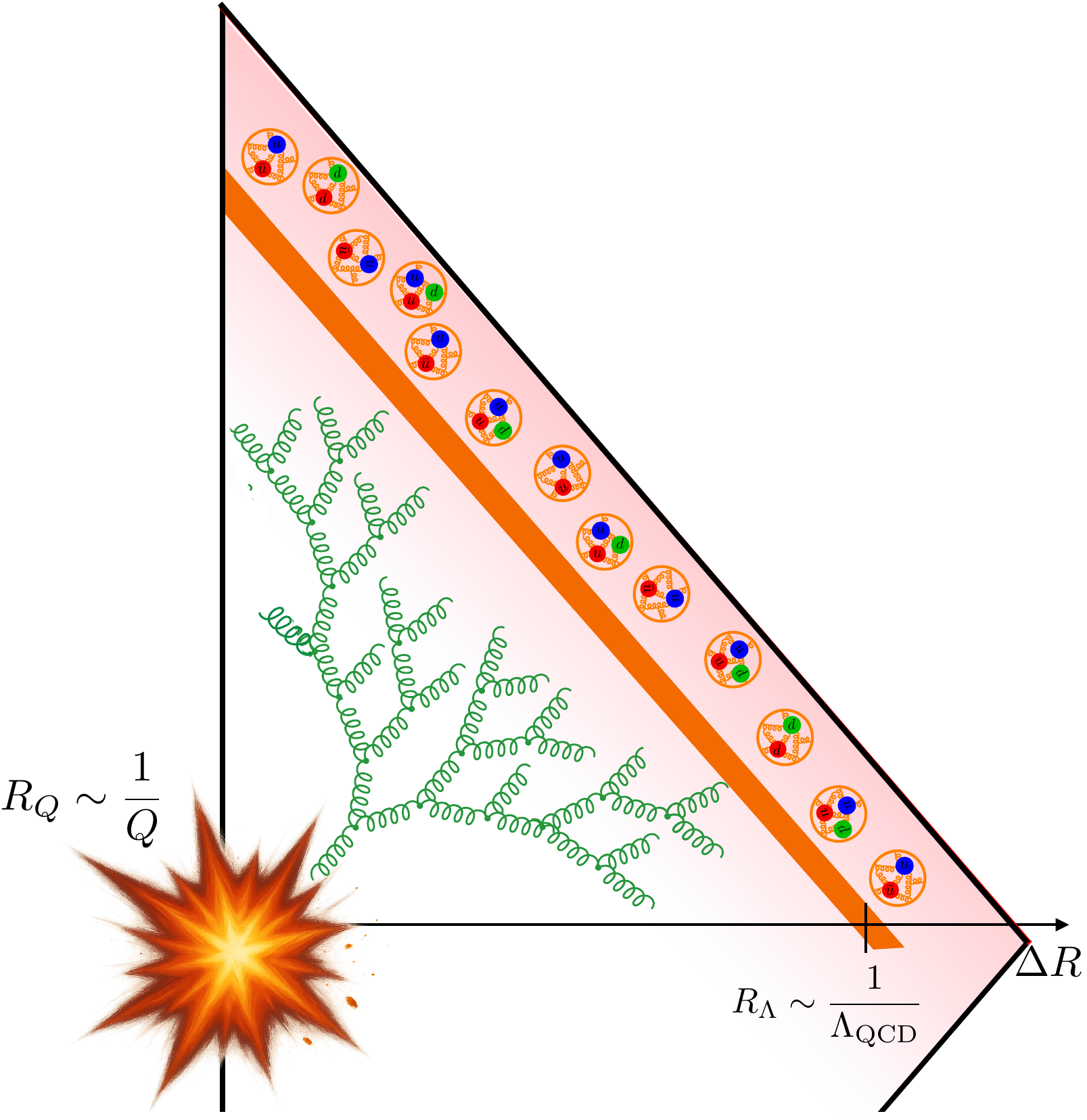}
\caption{Schematic Penrose diagram illustrating the detector‐matching procedure for QCD. A hard scattering at the short‐distance scale $1/Q$ produces energetic partons that propagate along nearly null trajectories until they hadronize at the long distance associated with confinement scale $1/\Lambda_{\rm QCD}$. The resulting hadrons follow timelike trajectories and are recorded by detectors at finite radius. The orange shading denotes the nonperturbative detector functions, which encode the mapping from hadronic IR detectors onto partonic UV detectors.
}
\label{fig:penroseQCD}
\end{figure}

In general, an IR detector in a general gapped theory can be expanded over UV detectors in the theory as
\begin{align}
\label{eq:matching}
\mathcal{D}_{\text{IR}}=  \sum_i \mathcal{D}^{(i)}_{\text{UV}} (\mu) \cdot F^{(i)}_{\mathcal{D}}(\mu)\,, 
\end{align}
where $i$ labels the distinct UV detectors and $\mathcal{D}_{\rm IR}$ is a single IR detector.\footnote{The matching of a product of several IR detectors generalizes~\eq{matching} and develops additional contact terms, as we discuss in detail in Sec.~\ref{sec:matchmultiple}.} This equation introduces the concept of ``detector functions" $F^{(i)}_{\mathcal{D}}(\mu)$, which are matching coefficients between the IR and the UV detectors. As is familiar in any effective field theory, or matching calculation, the split of IR detector into a UV detector and a detector function introduces a renormalization group scale $\mu$. This, however, is in contrast to the usual Wilsonian picture, where UV dynamics are integrated out to match to IR operators. Instead, \eq{matching} is achieving the reverse: matching the IR detector to the UV detectors by integrating out the long distance dynamics. This reflects the detector-oriented perspective on the renormalization group (RG) that we advocate, illustrated schematically in \fig{penrose}. The UV detectors are positioned at null infinity, shown in red, where only massless particles---traveling along null trajectories---can arrive after being produced in short-distance processes. In contrast, massive particles, following timelike paths, never reach the null infinity. Yet as we move the null sheet inward from infinity, an increasing number of trajectories can be approximated as null and are thus captured by the UV detectors. The matching relation in \eq{matching} precisely formalizes this transition, expressing IR detectors in terms of UV detectors through the matching coefficients, i.e. detector functions.

If the flow between the UV and IR is under perturbative control, these detector functions can be computed (or are completely trivial, as for the ANE, which is protected by energy conservation) using perturbation theory. In the case of QCD, where the UV and IR detectors are separated by the confinement transition at~$\mathcal{O}(\Lambda_{\rm QCD})$, the detector functions are in general nonperturbative matching coefficients. The schematic matching procedure for the QCD case is sketched in \fig{penroseQCD}. A hard scattering at the short-distance scale $\sim 1/Q$ produces energetic partons, which then propagate outward along nearly null trajectories until eventually hadronizing at the long distance $\sim 1/\Lambda_{\rm QCD}$. The resulting hadrons then follow timelike trajectories and are recorded by physical detectors at finite distance. This is analogous to moving the null sheet (celestial sphere) inward. The nonperturbative detector functions (shown in orange) precisely encode how to match these hadronic IR detectors at finite distance onto the UV detector description.

The general mapping~\eq{matching} between the UV and IR detectors is far from fully understood. In this paper we work it out for the hadronic IR detectors of QCD, $\mathcal{E}_R^{m}(\vec{n})$, which measure the energy flow of hadrons with quantum number $R$ in the direction $\vec{n}$ weighted to a complex power $m\in \mathbb{C}$. The matching of a single such detector was examined in~\cite{Chang:2025zib}. Going beyond this, we also treat products of several detectors, $\mathcal{E}_{R_1}^{m_1}(\vec{n}_1)\cdots\mathcal{E}_{R_N}^{m_N}(\vec{n}_N)$, whose matching generalizes~\eq{matching}: alongside the individual matching of each detector it develops contact terms, localized when the detectors are too close to be resolved by the UV (Sec.~\ref{sec:matchmultiple}). This multi-detector matching is a central new ingredient of our analysis, and it is what makes the light-ray OPE of the correlators below possible. In this paper, we will mainly focus on the case where the leading UV detectors onto which we match the hadronic detectors are twist-$2$ DGLAP partonic detectors.\footnote{Our general philosophy should apply for general $m\in \mathbb{C}$, but we need to consider mixing~\cite{Chang:2025zib} between DGLAP and BFKL branch near $m=4-d$ and mixing with shadow operators near $m=3-d$.}

Once the matching is complete, we can use the light-ray operator product expansion (OPE) and study the general multi-point correlators of IR detectors
\begin{align} 
\label{eq:generalcorr}
\langle \Psi| \mathcal{E}_{R_1}^{m_1}(\vec{n}_1) \mathcal{E}_{R_2}^{m_2}(\vec{n}_2)\cdots \mathcal{E}_{R_N}^{m_N}(\vec{n}_N) |\Psi\rangle\,,
\end{align}
where $|\Psi\rangle$ denotes some energetic state characterized by $Q$ in which correlations are measured.
In \Sec{sec:fact}, we will present the explicit all-order QCD factorization required to study such correlators with the one-dimensional projection onto the largest pairwise angle. In this section, we focus primarily on the light-ray OPE for the simpler case of pure Yang-Mills (YM). We will also ignore the coupling evolution from the beta functions, again just in the context of the light-ray OPE discussion, for simplicity.

\subsection{IR Detectors of Hadrons}
In order to match IR detectors with UV detectors, we first need to understand IR detectors of hadrons. For simplicity, let us assume the IR theory of hadrons is given by free scalar field theory. If a scalar field $\phi(x)$ describes creation of a hadron, we can represent our primary IR detector $\mathcal{E}^m$ as\footnote{It is trivial to extend to $\mathcal{E}_R^m$ by including a hadron flavor index $i$ on the fields $\phi_i$ and adding the contributions of different hadron flavors with the same quantum number $R$. For simplicity of notation, we restrict to the single-flavor case and drop the label $R$.}
\begin{align}
\label{eq:LT}
\hspace{-0.5cm}\mathcal{E}^m(\vec{n})  = c_m\, \mathbf{L}[\mathcal{O}_{J=m+1}^{\tau=2}](\infty,z) = c_m \lim_{x \rightarrow \infty}\, (x^2)^{1-J} \int_{-\infty}^{\infty} d \alpha\,(-\alpha)^{-\Delta-J}\, \mathcal{O}_{J=m+1}^{\tau=2}\left(x-\frac{z}{\alpha}, z\right)\,,
\end{align}
where $z=(1,\vec n)$ is the future-pointing null vector pointing along $\vec n\in S^{d-2}$ and $c_m$ is a normalization. The right-hand side is the \emph{light-transform}~\cite{Kravchuk:2018htv} of the spin-$J$ twist-$2$ local primary $\mathcal{O}_J^{\tau=2}$, defined as the null integral of $\mathcal{O}_J^{\tau=2}$. For~\eq{LT} to involve a genuine local operator we take $J=m+1$ to be a non-negative even integer, equivalently $m\in\{-1,1,3,\dots\}$. For general complex $m$ the light transform is defined by analytic continuation and yields the continuous-spin light-ray operator discussed below. The light-transform is a conformally covariant operation that sends a local primary of dimension $\Delta$ and spin $J$ to a primary operator on the celestial sphere with corresponding quantum numbers
\begin{align}
\Delta_L = 1-J\,,\qquad J_L = 1-\Delta \,.
\end{align}
For our twist-2 operators, $J_L = 1-\Delta = 3-d-J = 2-d-m$. 

The twist-2 spin-$J$ local operator is given by~\cite{Caron-Huot:2022eqs}
\begin{align}
\mathcal{O}_J^{\tau=2}(x, z)=\frac{2 \sqrt{2} \pi^{3 / 2} 4^J \Gamma\left(J+\frac{1}{2}\right)}{\sqrt{(2 J)!} \Gamma(J+1)}\left[: \phi(x)(z \cdot \partial)^J \phi(x):+(z \cdot \partial)(\cdots)\right ]
\end{align}
where the term $(z\!\cdot\!\partial)(\cdots)$ collects the total-derivative pieces required for primariness; these drop out under the null integral in~\eqref{eq:LT}. Substituting the mode expansion of $\phi$
\begin{align}
\phi(x)=\int \frac{\mathrm{d}^{d-1} p}{(2 \pi)^{d-1} 2 E_p}\left[a(p) e^{-i p \cdot x}+a^{\dagger}(p) e^{i p \cdot x}\right]
\end{align}
into~\eqref{eq:LT} produces
\begin{align}
\mathcal{E}^m(\vec{n})= \frac{1}{2(2 \pi)^{d-1}} \int_0^{\infty} dE\, E^{m+d-3} a^{\dagger}(p) a(p)\bigg|_{p=Ez}\,,
\end{align}
which counts each outgoing massless particle in the direction $\vec n$ with weight $E^m$. 

More generally, one can construct primary detectors directly at null infinity. Define the scalar field at null infinity by
\begin{align}
\phi(\alpha,z) = \lim_{L\to\infty} L^{\Delta_\phi} \phi(x+Lz), \qquad \Delta_\phi=\frac{d-2}{2}, \qquad \alpha=-2x\cdot z\,.
\end{align}
Then generic primary detectors with Lorentz spin $J_L$ with $n$ scalar fields at null infinity take the form~\cite{Caron-Huot:2022eqs}
\begin{align}
\mathcal{D}^{J_L}_{\psi,n}(\vec{n})=\int d \alpha_1 \ldots d \alpha_n \,\psi(\alpha_1, \ldots, \alpha_n): \phi\left(\alpha_1, z\right) \cdots \phi\left(\alpha_n, z\right):\,.
\end{align}
The primariness condition for such detector placed on $\mathscr{I}^+$
\begin{align}\label{eq:primary-condition}
[P^{\mu},\,\mathcal{D}_{\psi,n}^{J_L}(z)]\;=\;0
\end{align}
together with the requirement that they transform irreducibly under the Lorentz group impose conditions on the wavefunction $\psi$: 
\begin{itemize}
\item[(i)] \emph{Translation invariance.}
\begin{align}
\psi(\alpha_1 + \lambda, \ldots, \alpha_n+ \lambda)  = \psi(\alpha_1, \ldots, \alpha_n)\,,
\end{align}
\item[(ii)] \emph{Homogeneity.}
\begin{align}
\psi( \lambda\alpha_1, \ldots, \lambda\alpha_n)= \lambda^{{\rm deg }_\alpha\psi} \psi(\alpha_1, \ldots, \alpha_n)\,.
\end{align}
\item[(iii)] \emph{Permutation invariance.} Bose-symmetry of the scalar field also implies
\begin{align}
\psi\left(\alpha_{\sigma(1)}, \ldots, \alpha_{\sigma(n)}\right)=\psi\left(\alpha_1, \ldots, \alpha_n\right), \quad \sigma \in S_n\,.
\end{align}
\end{itemize}
The Lorentz spin $J_L$ and scaling dimension $\Delta_L$ of the primary detector $\mathcal{D}^{J_L}_\psi(z)$ are then given by
\begin{align}
\label{eq:JLDL}
J_L &=  n\left(1-\Delta_\phi\right)+\operatorname{deg}_\alpha \psi\,,\\
\Delta_L &=  n+\operatorname{deg}_\alpha \psi\,.\nn
\end{align}

For $n=2$, the constraints $(i)-(iii)$ uniquely fix $\psi(\alpha_1,\alpha_2)$ up to normalization as~\cite{Caron-Huot:2022eqs}
\begin{align}
\label{eq:psi_n2}
\psi_{J_L}\left(\alpha_1, \alpha_2\right)=\frac{1}{C_{J_L}}\left|\alpha_1-\alpha_2\right|^{d-4+J_L}, \quad J_L \in \mathbb{C},
\end{align}
where we added subscript label $J_L$ on $\psi_{J_L}$ to make the Lorentz spin explicit. The convenient normalization choice is given by 
\begin{align}
C_{J_L} = 2^{J_L+d-1} \pi \sin \left(\pi \frac{J_L+2 \Delta_\phi}{2}\right) \Gamma\left(2 \Delta_\phi+J_L-1\right)\,.
\end{align}
When the energy weight $m = 2-d -J_L  \in \{-1,1,3,5,\cdots\}$, the wavefunction collapses to a derivative of a delta function. Continuing in $J_L$, the simple pole of $|\alpha_1-\alpha_2|^{d-4+J_L}=|\alpha_1-\alpha_2|^{-2-m}$ at these integer points is exactly cancelled by the simple zero of $1/C_{J_L}$, that is by the pole of $C_{J_L}$ coming from $\Gamma(2\Delta_\phi+J_L-1)$, leaving the finite result
\begin{align}
\label{eq:psidelta}
\psi_{J_L}\left(\alpha_1, \alpha_2\right) = K_m\,\delta^{(m+1)}(\alpha_1-\alpha_2)\,, \qquad K_m = \frac{(-1)^{(m+1)/2}\,2^{m}}{\pi}\,, \qquad m \in \{-1,1,3,5,\cdots\}\,,
\end{align}
where $\delta^{(m+1)}$ denotes the $(m+1)$-th derivative of the delta function. For even integer $J=m+1$, substituting $\psi_{J_L}=K_m\delta^{(m+1)}$ into the $n=2$ primary detector and expanding the fields in modes gives
\begin{align}
\label{eq:EmIR}
\mathcal{D}^{J_L}_{\psi_{J_L},2}(\vec{n}) &= \int d \alpha_1  d \alpha_2 \,\psi_{J_L}(\alpha_1,  \alpha_2): \phi\left(\alpha_1, z\right) \phi\left(\alpha_2, z\right): \nn\\
&= K_m\int d \alpha \, : \big(\partial_\alpha^{m+1} \phi\big)\left(\alpha, z\right)  \phi\left(\alpha, z\right): \,\propto\, \int_0^{\infty} d E\, E^{\,m+d-3}\, a^{\dagger}(E z) a(E z) \,.
\end{align}
This is the energy-weighted number operator that counts each outgoing particle along $\vec{n}$ with weight $E^m$, with the total-derivative terms dropping under the $\alpha$ integral. Crucially, the primary detector in the first line is built from the general wavefunction~\eq{psi_n2} and is well-defined for \emph{any} complex $m$ (or equivalently complex $J_L$), and we take this construction to define $\mathcal{E}^m(\vec{n})$ for all $m$. At even integer $J=m+1$ it coincides with the light transform of a twist-2 local operator, as just shown, while for general complex $m$ it remains a well-defined continuous-spin light-ray operator with no local-operator counterpart. For $n\geq3$, properties $(i)-(iii)$ no longer fix the wavefunction, but in this paper we only consider IR detectors of the form $\mathcal{E}^m(\vec{n})$ and thus only need detectors built from two scalar fields at null infinity.

\subsection{Matching A Single Hadronic Detector  $\mathcal{E}_R^{m}(\vec{n})$ with UV Detectors}
\label{sec:matchingsingle}

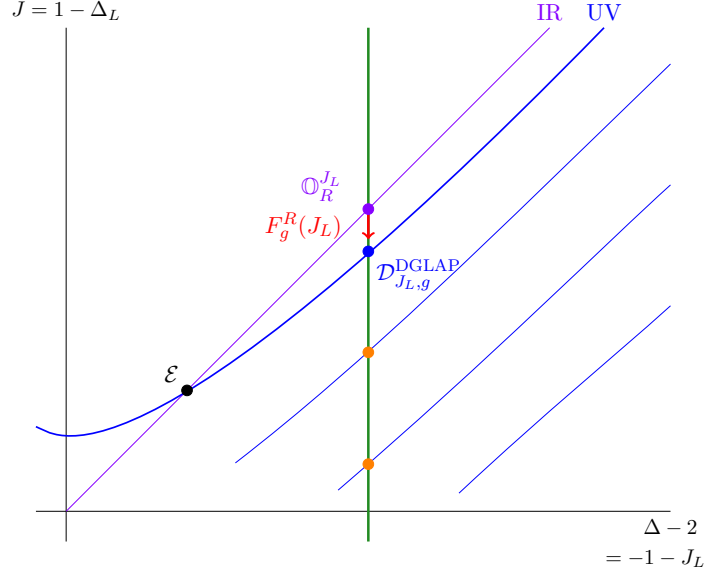
\begin{figure}
\centering
\scalebox{0.8}{
\begin{tikzpicture}
	\draw [] (-0.5,0) -- (10,0);
	\draw [] (0,-0.5) -- (0,8);
	\draw [lPurple] (0,0) -- (8,8);
	\draw [blue, thick] (-0.5,1.4) .. controls (-0.2,1.3) and (1,-0.1) .. (8.9,8);
	\draw [blue] (2.8,0.8) .. controls (5,2.5) and (7,4.5) .. (10,7.4);
	\draw [blue] (4.5,0.35) .. controls (7,2.5) and (9,4.5) .. (10,5.4);
	\draw [blue] (6.5,0.3) .. controls (8,1.7) and (9,2.5) .. (10,3.4);
	\draw [forest, very thick] (5,8) -- (5,-0.5);
	\filldraw[] (2,2) circle (2.5pt);
	\draw[very thick, ->, red] (5,5) -- (5,4.5);
	\filldraw[blue] (5,4.3) circle (2.5pt);
	\filldraw[lPurple] (5,5) circle (2.5pt);
	\filldraw[orange] (5,2.63) circle (2.5pt);
	\filldraw[orange] (5,0.78) circle (2.5pt);
	\node [below] at (10,0) {\small $\Delta - 2 $};
	\node [below] at (9.75,-0.5) {\small $= -1-J_L$};
	\node [above] at (0,8) {\small $J = 1 - \Delta_L$};
	\node [above] at (8,8) {\color{lPurple}IR};
	\node [above] at (8.9,8) {\color{blue}UV};
	\node [above left] at (2,2) {$\mathcal{E}$};
	\node [lPurple,above left] at (4.7,5) {$\mathbb{O}^{J_L}_R$};
	\node [blue,below right] at (5,4.3) {$\mathcal{D}_{J_L,g}^{\rm DGLAP}$};
	\node [red,left] at (4.7,4.7) {$F_{g}^R(J_L)$};
\end{tikzpicture}
}
\caption{Illustration of the matching procedure at fixed $J_L$, or equivalently at fixed $\Delta$, for $\mathbb{O}^{J_L}_R$. The IR detector $\mathbb{O}_R^{J_{L}}$ can be matched to UV detectors that lie on the same $J_L$ value. The dominant contribution comes from the twist-$2$ UV trajectory through matching via the detector function $F_{g}^R(J_L)$, whereas the higher-twist matchings, shown by the orange points, give $\Lambda_{\rm QCD}$ power suppressed contributions. Note also that the matching between IR and UV detector for the ANEC operator $\mathcal{E}$ is trivial.
}
\label{fig:matching}
\end{figure}

We begin by considering the matching of the single hadronic IR detector $\mathcal{E}_R^{m}(\vec{n})$, which was also examined in~\cite{Chang:2025zib}. The expectation value of the IR detector in a high-energy perturbative state $|\Psi\rangle$ with associated perturbative scale $Q$ is best characterized by measurements of UV detectors. Since UV detectors must preserve Lorentz boost symmetry, they must have the same Lorentz spin $J_L$ as the IR detector $\mathcal{E}_R^{m}$. The Lorentz weight $J_L$ of the IR detector $\mathcal{E}_R^{m}$ is given by
\begin{align}
\label{eq:JLmrel}
J_L[\mathcal{E}_R^{m}] = 2-d-m\,.
\end{align}
To make this clear, we use another notation to denote the IR detector $\mathcal{E}_R^{m}$ when we want to emphasize its Lorentz spin $J_L$ 
\begin{align}
\mathbb{O}_R^{J_L} \equiv \mathcal{E}_R^{m}\,,
\end{align}
where $J_L$ is understood to be $2-d-m$ as given in~\eq{JLmrel}.

The matching between the IR detector $\mathbb{O}_R^{J_L}$ and UV detectors is then given by  
\begin{align}
\label{eq:singleHad}
\mathcal{E}_R^m(\vec{n})= \mathbb{O}_R^{J_L}(\vec{n})=  \sum_i F_{i}^{R}(J_L,\mu)\, \mathcal{D}_{i}^{J_L,\Delta_L} (\vec{n},\mu)\approx F_{g}^{R}(J_L,\mu)\,\mathcal{D}^{\rm DGLAP}_{J_L,g}(\vec{n},\mu)\,,
\end{align}
where the sum runs over the distinct UV detectors $\mathcal{D}_{i}^{J_L,\Delta_L}$ that share the boost weight $J_L$ of the IR detector, and $F_{i}^{R}(J_L,\mu)$ is the detector matching function. A UV detector carries two labels, the boost weight $J_L$ and the scaling dimension $\Delta_L$. The matching fixes $J_L$ to be common to both sides, as it preserves the Lorentz boost, while $\Delta_L$ distinguishes the different trajectories, or twists, that sit at that $J_L$. The superscript ``DGLAP'' in the second equality selects the leading-twist (twist-2) trajectory, on which $\Delta_L$ is fixed in terms of $J_L$,
\begin{align}
\label{eq:DGLAPDeltaL}
\Delta_L = J_L + (d-2) + 2\gamma_T(3-d-J_L)\,,
\end{align}
with $\gamma_T$ the timelike anomalous dimension, known to three-loop accuracy in QCD~\cite{Mitov:2006ic,Mitov:2006wy,Moch:2007tx,Chen:2020uvt}. Equivalently, the matching is at fixed $\Delta = 1-J_L$.

The matching procedure at fixed $J_L$, or equivalently at the corresponding fixed $\Delta$, is illustrated in~\fig{matching}. As can be seen in the figure, at fixed $J_L$ the IR detector can actually be matched to multiple UV trajectories with different twists. Taking the detector matching functions for twist $\tau$ to be naturally of $\Lambda_{\rm QCD}$ scale, higher twist contributions, given by orange points in the figure, will be suppressed by classical scaling $\Lambda_{\rm QCD}^{\tau-2}$ relative to the leading twist contribution given by the purple point. In the second equality ($\approx$) of~\eq{singleHad}, we focused on matching to the leading twist UV detector $\mathcal{D}^{\rm DGLAP}_{J_L,g}(\vec{n},\mu)$, whose scaling dimension is fixed by~\eq{DGLAPDeltaL}. Again, we focus on the gluon detector for the pure YM case considered here, but in QCD there will also be a quark detector on a leading twist trajectory. 

When we wrote the dominant contribution in the second equality of~\eq{singleHad}, we implicitly assumed that the matching is away from the BFKL/DGLAP intersection point~\cite{Chang:2025zib} $J_L=-2,-1$, in order to focus on the region where the DGLAP detector gives the leading contribution. The bare definition of twist-2 (even-spin branch) UV partonic (DGLAP) detectors with Lorentz spin $J_L$ are given by (focusing only on gluon for pure YM case considered here)
\begin{align}
& \mathcal{D}_{J_L,g}^{\rm DGLAP}(\vec{n})=\left.\sum_{\lambda, c} \int_0^{\infty} \frac{E^{-J_L} d E}{(2 \pi)^{d-1} 2 E}\left[a_{\lambda, c}^{\dagger}(p) a_{\lambda, c}(p)\right]\right|_{p=E n}\,, 
\end{align}
where $n = (1,\vec{n})$. After renormalization, their renormalization group evolutions are given by
\begin{align}
\label{eq:Devol}
\frac{d}{d\ln\mu^2}\mathcal{D}_{J_L,g}^{\rm DGLAP}(\vec{n},\mu) = \gamma_T(3-d-J_L)\, \mathcal{D}_{J_L,g}^{\rm DGLAP}(\vec{n},\mu)\,.
\end{align}
For twist-2 operators, the boost weight $J_L$ and the spin $J$ are related by
\begin{align}
\label{eq:JLandJ}
J + 2\gamma_S(J) = 3-d-J_L\,,
\end{align}
where $2\gamma_S(J)$ is the \emph{spacelike} anomalous dimension of the twist-2 local operator of spin $J$. From the local operator viewpoint,~\eq{JLandJ} is simply the standard twist relation $\Delta-J = d-2+2\gamma_S(J)$ rewritten through the light transform dictionary $J_L=1-\Delta$. One can picture the trajectory by starting from local operators of integer spin $J$, light transforming them into light-ray operators, and analytically continuing in $J$. From the detector viewpoint, it is instead more natural to fix the boost weight $J_L$ and read off which operator sits there. Combining~\eq{JLandJ} with the reciprocity relation $\gamma_S(J)=\gamma_T(J+2\gamma_S(J))$, the spin of the leading-twist UV detector matched to $\mathcal{E}_R^{m}$ is
\begin{align}
\label{eq:JdglapJL}
J = 3-d-J_L - 2\gamma_T(3-d-J_L) = m+1 - 2\gamma_T(m+1)\,,
\end{align}
where we used $3-d-J_L = m+1$ at $J_L = J_L[\mathcal{E}_R^m]$. Away from the ANEC point $m=1$, where $\gamma_T(2)=0$ by energy conservation, giving $J=2$, $J$ is generally non-integer. The UV detector we match to is then a genuine continuous-spin light-ray operator with no local-operator counterpart.

The detector functions $F_{g}^R(J_L,\mu)$ in~\eq{singleHad} that match between the UV and IR detectors entirely carry the dependence on the hadronic quantum number $R$, since the UV dynamics are insensitive to the IR selection of the hadronic species on which the energy flow is measured (unlike Lorentz symmetry, which must be preserved between the UV and IR detectors). Physically, the detector function describes how weighted energy of a parton, a gluon here, converts into weighted energy of a hadron to the power $m=2-d-J_L$. It therefore measures the overlap between the gluon state $|g\rangle$ and the inclusive hadronic state $|hX\rangle$, and how energy of a hadron from the gluon fragmentation is distributed. This is described by the well-known non-perturbative functions in QCD called the single-hadron fragmentation functions, defined through hadronic matrix elements of partonic fields. For a gluon, the operator definition of the single-hadron fragmentation function is
\begin{align}
  \label{eq:singlefrag}
D_{g\to h} (z,\mu) =&\,  \frac{-1}{(d-2)(N_c^2-1)\, k^-}
\int dy^+\, d^{d-2}y_\perp\;e^{\,i k^-  y^+/2} \nn\\
& \times \sum_X \langle 0 | G^a_{- \lambda}(y^+, 0, y_\perp ) | h X
  \rangle \langle h X | G_-^{\lambda,a}(0) |0 \rangle\,\delta\left(z-\frac{p_h^-}{k^-}\right) \,.
\end{align}
These single-hadron fragmentation functions $D_{i\to h}(z,\mu)$ describe the number density in the momentum fraction $z= p_h^-/k^-$ of a hadron $h$ produced from a parton $i$ of momentum $k^-$. The single-hadron detector function $F_{g}^R(J_L,\mu)$ is then simply its energy-weighted moment,
\begin{align}
\label{eq:F1}
F_{g}^R(J_L,\mu)\equiv \sum_{h\in R} \int_0^1 dz\, z^{2-d-J_L} \,  D_{g\to h}(z,\mu) \equiv D_{g}^R(m=2-d-J_L,\mu)\,,
\end{align}
which gives the hadronic energy of hadrons with quantum number $R$, converted from the fragmenting parton, raised to the power $2-d-J_L$. As for IR detectors, we denote these matching coefficients in two different ways. The notation $D_g^R(m,\mu)$ emphasizes the actual energy power $m$ that IR detector was raised to. On the other hand, $F_{g}^R(J_L,\mu)$ emphasizes the role of the Lorentz spin in the matching equation~\eq{singleHad}. In Tab.~\ref{tab:map}, we summarize the map between Lorentz spin $J_L$ and energy weight $m$ for both IR detectors and detector functions, for the single-hadron case discussed in this subsection. The multi-hadron case will be discussed in the next subsection. 

\begin{table}[t]
\centering
\renewcommand{\arraystretch}{2.2} 

\begin{tabular}{c || c || c @{\quad}c@{\quad} c} 
\textbf{Category} & \textbf{Type} & \textbf{Lorentz Spin  ($J_L$)} & & \textbf{Energy Weight ($m$)} \\ 
\hline\hline 
\multirow{2}{*}{IR Detectors} & Single-Hadron & $\mathbb{O}^{J_L}_R$ & $=$ & $\mathcal{E}_R^m$ \\ 
                              & Multi-Hadron  & $:\mathbb{O}^{J_{L_1}}_{R_1}\cdots \mathbb{O}^{J_{L_N}}_{R_N}:$ & $=$ & $:\mathcal{E}_{R_1}^{m_1}\cdots \mathcal{E}_{R_N}^{m_N}:$\\ 
\hline
\multirow{2}{*}{\begin{tabular}[c]{@{}c@{}}Detector Functions\\ (Matching Coeffs.)\end{tabular}} & Single-Hadron & $F_{i}^R(J_L)$ & $=$ & $D_{i}^R(m)$\\ 
                              & Multi-Hadron  & $F_{i}^{R_1\cdots R_N}(J_{L_1},\cdots,J_{L_N})$ & $=$ & $\tilde{D}_{i}^{R_1\cdots R_N}(m_1,\dots,m_N)$
\end{tabular}
\caption{Explicit mapping of IR detectors and detector functions between Lorentz spin $J_L$ and energy weight $m$ representations. We will only use Lorentz spin notation for UV detectors.}
\label{tab:map}
\end{table}

The RG evolution of $F_{g}^R(J_L,\mu)$ is derived from the well-known RG evolution of the single hadron fragmentation functions and is given as
\begin{align}
\label{eq:Tevol}
\frac{d}{d\ln\mu^2}F_{g}^R(J_L,\mu) =- \gamma_T(3-d-J_L)\,F_{g}^R(J_L,\mu)\,,
\end{align}
which is opposite to that of the UV detector in~\eq{Devol}, so that the RG invariance of the IR detector $\mathbb{O}_R^{J_L}(\vec{n})$ is ensured by the pair of RG equations~\eqs{Devol}{Tevol}.

\subsection{Matching Multiple Hadronic Detectors  $\mathcal{E}_{R_1}^{m_1}(\vec{n}_1)\cdot\cdot\cdot\mathcal{E}_{R_N}^{m_N}(\vec{n}_N)$ with UV detectors}
\label{sec:matchmultiple}

To probe more interesting dynamics, we now consider correlations of multiple hadronic detectors. This requires matching multiple IR detectors, which introduces the angular separation between IR detectors $x_{ij} = (n_i\cdot n_j)/2$. While the matching itself is an operator statement, we ultimately study correlations in a high-energy state $|\Psi\rangle$ with hard scale $Q$, so that the relevant transverse scale set by a pair of detectors is $Q\sqrt{x_{ij}}$. The matching depends critically on whether we resolve the scale near or far above the confinement scale, $Q\sqrt{x_{ij}} \lesssim \Lambda_{\rm QCD}$ or $Q\sqrt{x_{ij}} \gg \Lambda_{\rm QCD}$. If we want to resolve the region $x_{ij}\lesssim \Lambda_{\rm QCD}^2/Q^2$, the detector matching functions themselves depend on the separation $x_{ij}$~\cite{Chang:2025kgq,Lee:2025okn}. In this paper we restrict to resolving angles well above the confinement scale, $x_{ij} \gg \Lambda_{\rm QCD}^2/Q^2$. In this regime the matching of two hadronic detectors takes the form
\begin{align}
\label{eq:twoHad}
\mathcal{E}_{R_1}^{m_1}(\vec{n}_1) \mathcal{E}_{R_2}^{m_2}(\vec{n}_2) =&\ F_g^{R_1}(J_{L_1},\mu)F_g^{R_2}(J_{L_2},\mu)\, \mathcal{D}_{J_{L_1},g}^{\rm DGLAP}(\vec{n}_1,\mu)\,\mathcal{D}_{J_{L_2},g}^{\rm DGLAP}(\vec{n}_2,\mu) \\
&\hspace{-1.7cm}+ \left[F_g^{R_1 R_2}( J_{L_1},J_{L_2},\mu)+ \delta_{R_1 R_2}\, F_g^{R_1}(\tilde{J}_{L_{12}},\mu)\right]
\mathcal{D}_{\tilde{J}_{L_{12}},g}^{\rm DGLAP}(\vec{n}_1,\mu)\, \delta^{(d-2)}(\vec{n}_1- \vec{n}_2)\,,\nn
\end{align}
where we introduced the shorthand for the boost weight of the two merged detectors,
\begin{align}
\label{eq:Jtilde12}
\tilde{J}_{L_{12}} \equiv J_{L_1}+J_{L_2}+d-2 = 2-d-(m_1+m_2)\,.
\end{align}
The matching separates into two types of term. In the first, the two IR detectors are well separated and each is matched individually as in~\eq{singleHad}. The second is a contact term, localized at $\vec{n}_1=\vec{n}_2$ by $\delta^{(d-2)}(\vec{n}_1-\vec{n}_2)$, arising from configurations in which the two IR detectors are so close that the UV detectors cannot resolve the angle between them. Since we assume we will only resolve an angle away from the confinement scale, i.e. $Q\sqrt{x_{ij}} \gg \Lambda_{\rm QCD}$, this contact term may seem unimportant. This is indeed the case if we only consider two detectors. However, if we want to consider three or more detectors and resolve the largest angle, $Q\sqrt{x_L} \gg \Lambda_{\rm QCD}$, then such a contact term becomes important for giving a finite contribution.

When two IR detectors are matched individually, they follow the single IR detector matching discussed in Sec.~\ref{sec:matchingsingle}. The single-hadron detector function $F_g^{R_1}(J_{L_1},\mu)$ in the first term is exactly the detector function given in~\eq{F1}. The contact term involves two detector matching functions, $F_g^{R_1 R_2}(J_{L_1},J_{L_2},\mu)$ and $\delta_{R_1 R_2}F_g^{R_1}(\tilde{J}_{L_{12}},\mu)$, both of which match to the same UV detector $\mathcal{D}_{\tilde{J}_{L_{12}},g}^{\rm DGLAP}$ and delta function. Since this UV detector carries boost weight $\tilde{J}_{L_{12}}$, RG consistency forces both detector functions to evolve as
\begin{align}
\label{eq:Tevolcontact}
\frac{d}{d\ln\mu^2}F_g^{R_1}(\tilde{J}_{L_{12}},\mu) &=- \gamma_T(3-d-\tilde{J}_{L_{12}})\,F_g^{R_1}(\tilde{J}_{L_{12}},\mu)\,,\nn\\
\frac{d}{d\ln\mu^2}F_g^{R_1 R_2}(J_{L_1},J_{L_2},\mu) &=- \gamma_T(3-d-\tilde{J}_{L_{12}})\,F_g^{R_1 R_2}(J_{L_1},J_{L_2},\mu)\,,
\end{align}
compensating the RG evolution equations of $\mathcal{D}_{\tilde{J}_{L_{12}},g}^{\rm DGLAP}$ in~\eq{Devol}.

Although both of these detector functions for the contact term evolve identically, they match to qualitatively different IR detectors. The single-hadron coefficient $F_g^{R_1}(\tilde{J}_{L_{12}},\mu)$, as in~\eq{F1}, is built from single-hadron fragmentation functions. When the two detectors carry the same quantum number $R_1=R_2$, it captures the configuration in which both detectors act on the \emph{same} hadron, raising it to the combined weight, $\mathcal{E}_{R_1}^{m_1+m_2} = \mathbb{O}_{R_1}^{\tilde{J}_{L_{12}}}$. The detector matching function $F_g^{R_1 R_2}(J_{L_1},J_{L_2},\mu)$ is instead built from so-called di-hadron fragmentation functions, discussed more in detail in Sec.~\ref{sec:matchmultifrag}, and matches to the normal-ordered product of two IR detectors $:\mathcal{E}_{R_1}^{m_1}\mathcal{E}_{R_2}^{m_2}: = :\mathbb{O}_{R_1}^{J_{L_1}}\mathbb{O}_{R_2}^{J_{L_2}}:$, which acts on \emph{distinct} hadrons and gives a nonzero result only starting from two-hadron states. In terms of creation and annihilation operators, they are given as
\begin{align}
:\mathcal{E}_{R_1}^{m_1}\mathcal{E}_{R_2}^{m_2}: =  \frac{1}{4(2 \pi)^{2(d-1)}} \int_0^{\infty} dE_1 dE_2\, E_1^{m_1+d-3} E_2^{m_2+d-3} a_1^{\dagger}(p_1) a_2^{\dagger}(p_2) a_2(p_2) a_1(p_1)\bigg|_{p_1=E_1z,p_2=E_2z}\,,
\end{align}
and their action on asymptotic hadronic states gives a non-vanishing contribution starting from two-hadron states
\begin{align}
:\mathcal{E}_{R_1}^{m_1}\mathcal{E}_{R_2}^{m_2}: | h \rangle &= 0 |h\rangle\,,\nn\\
:\mathcal{E}_{R_1}^{m_1}\mathcal{E}_{R_2}^{m_2}: | h_a,h_b \rangle &=\left( E_a^{m_1}E_b^{m_2} +E_a^{m_2}E_b^{m_1} \right)|h_a,h_b\rangle\,.
\end{align}

The matching generalizes straightforwardly to three or more detectors. It is convenient to extend the shorthand of~\eq{Jtilde12} to the merged boost weights of any pair or triple of coincident detectors,
\begin{align}
\label{eq:Jtildeshort}
\tilde{J}_{L_{ij}} \equiv J_{L_i}+J_{L_j}+d-2\,,\qquad
\tilde{J}_{L_{123}} \equiv J_{L_1}+J_{L_2}+J_{L_3}+2(d-2)\,.
\end{align}
The three-detector matching then reads
\begin{align}
\label{eq:threeHad}
\hspace{-1.2cm}\mathcal{E}_{R_1}^{m_1}(\vec{n}_1) \mathcal{E}_{R_2}^{m_2}(\vec{n}_2) \mathcal{E}_{R_3}^{m_3}(\vec{n}_3) =&\ F_g^{R_1}(J_{L_1})F_g^{R_2}(J_{L_2})F_g^{R_3}(J_{L_3})\, \mathcal{D}_{J_{L_1},g}^{\rm DGLAP}(\vec{n}_1)\mathcal{D}_{J_{L_2},g}^{\rm DGLAP}(\vec{n}_2)\mathcal{D}_{J_{L_3},g}^{\rm DGLAP}(\vec{n}_3) \nn\\
&\hspace{-3.4cm}+ \Big[F_g^{R_1 R_2}(J_{L_1},J_{L_2})+ \delta_{R_1 R_2}\, F_g^{R_1}(\tilde{J}_{L_{12}})\Big] F_g^{R_3}(J_{L_3})\,\mathcal{D}_{\tilde{J}_{L_{12}},g}^{\rm DGLAP}(\vec{n}_1)\mathcal{D}_{J_{L_3},g}^{\rm DGLAP}(\vec{n}_3)\, \delta^{(d-2)}(\vec{n}_1- \vec{n}_2)\nn\\
&\hspace{-3.4cm}+ \Big[F_g^{R_1 R_3}(J_{L_1},J_{L_3})+ \delta_{R_1 R_3}\, F_g^{R_1}(\tilde{J}_{L_{13}})\Big] F_g^{R_2}(J_{L_2})\,\mathcal{D}_{\tilde{J}_{L_{13}},g}^{\rm DGLAP}(\vec{n}_1)\mathcal{D}_{J_{L_2},g}^{\rm DGLAP}(\vec{n}_2)\, \delta^{(d-2)}(\vec{n}_1- \vec{n}_3)\nn\\
&\hspace{-3.4cm}+ \Big[F_g^{R_2 R_3}(J_{L_2},J_{L_3})+ \delta_{R_2 R_3}\, F_g^{R_2}(\tilde{J}_{L_{23}})\Big] F_g^{R_1}(J_{L_1})\,\mathcal{D}_{\tilde{J}_{L_{23}},g}^{\rm DGLAP}(\vec{n}_2)\mathcal{D}_{J_{L_1},g}^{\rm DGLAP}(\vec{n}_1)\, \delta^{(d-2)}(\vec{n}_2- \vec{n}_3)\nn\\
&\hspace{-3.4cm}+ \Big[F_g^{R_1 R_2 R_3}(J_{L_1},J_{L_2},J_{L_3})+ \delta_{R_1 R_2}\, F_g^{R_1 R_3}(\tilde{J}_{L_{12}},J_{L_3})+ \delta_{R_1 R_3}\, F_g^{R_1 R_2}(\tilde{J}_{L_{13}},J_{L_2})\\
&\hspace{-3.4cm}+ \delta_{R_2 R_3}\, F_g^{R_1 R_2}(J_{L_1},\tilde{J}_{L_{23}})+ \delta_{R_1 R_2} \delta_{R_2 R_3}\, F_g^{R_1}(\tilde{J}_{L_{123}})\Big]\, \mathcal{D}_{\tilde{J}_{L_{123}},g}^{\rm DGLAP}(\vec{n}_1)\, \delta^{(d-2)}(\vec{n}_1- \vec{n}_2)\delta^{(d-2)}(\vec{n}_2- \vec{n}_3)\,,\nn
\end{align}
where we have suppressed the matching scale $\mu$ for brevity. The physical content of each term is transparent. The first term, proportional to $F_g^{R_1} F_g^{R_2} F_g^{R_3}$, is the configuration in which all three detectors are well separated and each is matched individually. The next three terms, of the schematic form $(F_g^{R_i R_j} + \delta_{R_i R_j} F_g^{R_i})F_g^{R_k}$, are the configurations in which one pair of detectors is unresolved and matched together, exactly as in the two-detector case, while the third detector stays separate. Although these are contact terms between two of the three detectors, when we project onto the largest pairwise angle $x_L$ with $Q\sqrt{x_L}\gg\Lambda_{\rm QCD}$, the resolved angle is the one between the unresolved pair and the far-away third detector. Therefore, these two-detector contact terms feed into finite-$x_L$ observables for the three-detector case, and thus contact terms matter in general, as discussed above. The final bracket collects the configuration in which all three detectors are unresolved, localized by the two delta functions $\delta^{(d-2)}(\vec{n}_1- \vec{n}_2)\delta^{(d-2)}(\vec{n}_2- \vec{n}_3)$. Again, this becomes important once a fourth detector is added and the largest angle separates it from the three unresolved ones. Within this bracket, $F_g^{R_1 R_2 R_3}$ matches the normal-ordered three-hadron detector $:\mathcal{E}_{R_1}^{m_1}\mathcal{E}_{R_2}^{m_2}\mathcal{E}_{R_3}^{m_3}:= :\mathbb{O}_{R_1}^{J_{L_1}}\mathbb{O}_{R_2}^{J_{L_2}}\mathbb{O}_{R_3}^{J_{L_3}}:$, which is nonzero only on states of three or more hadrons; its matching function is represented by the so-called tri-hadron fragmentation-functions discussed more in detail in \Sec{sec:matchmultifrag}. The three terms with a single Kronecker delta match to normal-ordered two-hadron detectors, such as $:\mathcal{E}_{R_1}^{m_1+m_2}\mathcal{E}_{R_3}^{m_3}:=:\mathbb{O}_{R_1}^{\tilde{J}_{L_{12}}}\mathbb{O}_{R_1}^{J_{L_3}}:$ when $R_1=R_2$, while the doubly-restricted term $\delta_{R_1 R_2}\delta_{R_2 R_3}F_g^{R_1}$ matches the single-hadron detector $\mathcal{E}_{R_1}^{m_1+m_2+m_3}=\mathbb{O}_{R_1}^{\tilde{J}_{L_{123}}}$, which survives only when $R_1=R_2=R_3$.

The tri-hadron detector function $F_g^{R_1 R_2 R_3}(J_{L_1},J_{L_2},J_{L_3})$ evolves as
\begin{align}
\label{eq:trihadRG}
\frac{d}{d\ln\mu^2} F_g^{R_1 R_2 R_3}(J_{L_1},J_{L_2},J_{L_3},\mu) = -\gamma_T(3-d-\tilde{J}_{L_{123}})\, F_g^{R_1 R_2 R_3}(J_{L_1},J_{L_2},J_{L_3},\mu)\,.
\end{align}
Matching four or more IR detectors proceeds in a similar way, involving higher-multiplicity hadron detector functions $F_g^{R_1 \cdots R_N}(J_{L_1},\cdots,J_{L_N},\mu)$, where their evolution is given by
\begin{align}
\label{eq:multihadRG}
\frac{d}{d\ln\mu^2} F_g^{R_1 \cdots R_N}(J_{L_1},\cdots,J_{L_N},\mu) = -\gamma_T(3-d-\tilde{J}_{L_{12\cdots N}})\, F_g^{R_1  \cdots R_N}(J_{L_1},\cdots,J_{L_N},\mu)\,,
\end{align}
where  we introduced the shorthand notation
\begin{align}
\tilde{J}_{L_{12\cdots N}} \equiv J_{L_1}+J_{L_2}+\cdots + J_{L_N}+(N-1)(d-2)\,.
\end{align}

\subsection{Multi-hadron Fragmentation Functions}\label{sec:matchmultifrag}

In Sec.~\ref{sec:matchingsingle}, we saw that a moment of the single-hadron fragmentation function, with the operator definition given in~\eq{singlefrag}, defines the single-hadron detector function given in~\eq{F1}. Such single-hadron detector function provides the matching between the single hadronic detector $\mathbb{O}_R^{J_L}=\mathcal{E}_R^{m}$ in the IR and the twist-2 DGLAP partonic detector in the UV. In Sec.~\ref{sec:matchmultiple}, we then saw that the matching of multiple hadronic detectors requires multi-hadron detector functions $F_g^{R_1 \cdots R_N}(J_{L_1},\cdots,J_{L_N},\mu)$, which we identify below with the appropriate RG-diagonalized moments of multi-hadron fragmentation functions.

The $N$-hadron fragmentation function describes the energy distribution of $N$ identified hadrons produced in the fragmentation of a parton. The number of hadrons in the final state may exceed $N$, in which case one is inclusive over the remainder. Restricting to pure Yang-Mills for concreteness, the gluon $N$-hadron fragmentation function is
\begin{align}
\label{eq:gluonNhadron}
D_{g\to h_1 \cdots h_N}(x_1, \ldots, x_N,\mu) =&\ \frac{-1}{(d-2)\left(N_c^2-1\right) k^{-}}\int \mathrm{d} y^{+} \mathrm{d}^{d-2} y_{\perp}\, e^{i k^{-} y^{+}/2} \\
&\hspace{-1cm}\times \sum_X \langle 0| G_{-\lambda}^a(y^{+}, 0, y_{\perp})|h_1\cdots h_N X\rangle\langle h_1\cdots h_N X| G_{-}^{\lambda, a}(0)|0\rangle\, \prod_{i=1}^{N}\delta\left(x_i-\frac{p_{h_i}^-}{k^-}\right) ,\nn
\end{align}
where $x_i=p_{h_i}^-/k^-$ is the momentum fraction of the identified hadron $h_i$. For $N=1$ this reduces to the single-hadron fragmentation function~\eq{singlefrag}.

The renormalization group evolution describes how these fragmentation functions change with the scale $\mu$ through perturbative splittings. Crucially, the initial parton can undergo several splittings before hadronizing, so the evolution of an $N$-hadron fragmentation function mixes with lower-multiplicity ones. For $N=2$, for example, there are two distinct contributions. The gluon may evolve as a single parton that fragments into both observed hadrons, or its two daughter partons from partonic splitting may each fragment into one of them. The first is the homogeneous, DGLAP-like evolution of the di-hadron fragmentation function itself; the second is an inhomogeneous term built from a product of two single-hadron fragmentation functions,
\begin{align}
\label{eq:dihadronevol}
\frac{d}{d\ln\mu^2} D_{g\to h_1 h_2}(x_1, x_2, \mu) =&\ \int_{x_1+x_2}^1 \frac{dz}{z^2}\, P^T_{gg}(z)\, D_{g\to h_1 h_2}\left(\frac{x_1}{z}, \frac{x_2}{z}, \mu\right) \\
&\hspace{-2cm}+ \int_{x_1}^{1-x_2} \frac{dz}{z(1-z)}\, \hat{P}^T_{g\to gg}(z) \left[ D_{g\to h_1}\left(\frac{x_1}{z}, \mu\right) D_{g\to h_2}\left(\frac{x_2}{1-z}, \mu\right) + (z\leftrightarrow 1-z) \right] \nn,
\end{align}
where $P^T_{gg}$ is the timelike splitting function and $\hat{P}^T_{g\to gg}$ describes $1\to2$ partonic splitting that separately fragments to observed hadrons. In full QCD, there will also be quark channels involving quark fragmentation functions as well. For the $N$-hadron fragmentation functions, the same mechanism generates additional inhomogeneous terms: each way of distributing the $N$ observed hadrons among the daughter partons of a splitting contributes a term built from lower-multiplicity fragmentation functions.

As in the single-hadronic detector case, the matching with two hadronic detectors involves \emph{moments} of these fragmentation functions. For the matching of $:\mathcal{E}_{R_1}^{m_1}\mathcal{E}_{R_2}^{m_2}:$, the relevant di-hadron moment is
\begin{align}
\label{eq:dihadronmoment}
D_{g}^{R_1 R_2}(m_1,m_2,\mu) \equiv \sum_{h_1\in R_1,\, h_2\in R_2} \int_0^1 dx_1\, dx_2\; x_1^{m_1} x_2^{m_2}\, D_{g\to h_1 h_2}(x_1, x_2, \mu)\,.
\end{align}
Taking the corresponding moment of~\eq{dihadronevol}, the homogeneous and inhomogeneous terms become
\begin{align}
\label{eq:dihadronmomevol}
\frac{d}{d\ln\mu^2} D_{g}^{R_1 R_2}(m_1,m_2,\mu) =&\ -\gamma_T(m_1+m_2+1)\, D_{g}^{R_1 R_2}(m_1,m_2,\mu) \nn\\
&+ \hat{\gamma}_g(m_1,m_2)\, D_{g}^{R_1}(m_1,\mu)\,D_{g}^{R_2}(m_2,\mu)\,,
\end{align}
where the homogeneous term carries the timelike anomalous dimension at the combined weight $m_1+m_2$. The timelike anomalous dimension is itself a moment of the splitting function, which in pure Yang-Mills reads
\begin{align}
\label{eq:gammaTfromP}
\gamma_T(k) = -\int_0^1 dz\, z^{k-1}\, P^T_{gg}(z)\,,
\end{align}
so that the homogeneous term of~\eq{dihadronevol}, weighted by $z^{m_1+m_2}$, produces $-\gamma_T(m_1+m_2+1)$. The inhomogeneous mixing kernel is instead the double moment of the real splitting function,
\begin{align}
\label{eq:mixingkernel}
\hat{\gamma}_g(m_1,m_2) \equiv \int_0^1 dz\, \hat{P}^T_{g\to gg}(z)\left[ z^{m_1}(1-z)^{m_2} + z^{m_2}(1-z)^{m_1}\right]\,,
\end{align}
where the two terms are the two daughter assignments of~\eq{dihadronevol} and make $\hat{\gamma}_g$ manifestly symmetric under $m_1\leftrightarrow m_2$. The single-hadron moments evolve linearly as in~\eq{Tevol}, $\tfrac{d}{d\ln\mu^2}D_g^{R_i}(m_i,\mu) = -\gamma_T(m_i+1)D_g^{R_i}(m_i,\mu)$, so their product evolves with $-[\gamma_T(m_1+1)+\gamma_T(m_2+1)]$. The evolution therefore closes on the pair $\{D_g^{R_1 R_2},\,D_g^{R_1}D_g^{R_2}\}$ as the triangular system
\begin{align}
\label{eq:dihadronmatrix}
\frac{d}{d\ln\mu^2} \begin{pmatrix} D_{g}^{R_1 R_2} \\[3pt] D_{g}^{R_1}\,D_{g}^{R_2} \end{pmatrix} = \begin{pmatrix} -\gamma_T(m_1+m_2+1) & \hat{\gamma}_g(m_1,m_2) \\[3pt] 0 & -\gamma_T(m_1+1)-\gamma_T(m_2+1) \end{pmatrix} \begin{pmatrix} D_{g}^{R_1 R_2} \\[3pt] D_{g}^{R_1}\,D_{g}^{R_2} \end{pmatrix}\,,
\end{align}
The mixing into the product $D_g^{R_1}D_g^{R_2}$ has a clear physical origin: it is the independent fragmentation of the two hadrons from separate partons created by a perturbative splitting. This can be removed by passing to the combination that diagonalizes~\eq{dihadronmatrix}. Ignoring the running of the coupling, we can define the diagonalized di-hadron moment as
\begin{align}
\label{eq:dihadrondiag}
\tilde{D}_{g}^{R_1 R_2}(m_1,m_2,\mu)\equiv&\ D_{g}^{R_1 R_2}(m_1,m_2,\mu) \nn\\
&- \frac{\hat{\gamma}_g(m_1,m_2)}{\gamma_T(m_1+m_2+1) - \gamma_T(m_1+1)-\gamma_T(m_2+1)}\, D_{g}^{R_1}(m_1,\mu)\,D_{g}^{R_2}(m_2,\mu)\,,
\end{align}
whose coefficient is fixed precisely so as to cancel the mixing with $D_{g}^{R_1}(m_1)D_{g}^{R_2}(m_2)$. Then $\tilde{D}_{g}^{R_1 R_2}$ evolves multiplicatively as
\begin{align}
\label{eq:dihadrondiagevol}
\frac{d}{d\ln\mu^2} \tilde{D}_{g}^{R_1 R_2}(m_1,m_2,\mu) = -\gamma_T(m_1+m_2+1)\, \tilde{D}_{g}^{R_1 R_2}(m_1,m_2,\mu)\,.
\end{align}
Physically, $\tilde{D}_{g}^{R_1 R_2}$ isolates the genuinely correlated two-hadron fragmentation. Another way to say this is that, if at any scale it vanishes, $\tilde{D}_{g}^{R_1 R_2}(\mu_0)=0$, then we are unable to generate nonzero $\tilde{D}_{g}^{R_1 R_2}(\mu)$ at another scale $\mu$ through perturbative splittings. It is exactly this genuine two-hadron fragmentation piece that enters the simultaneous matching of the two IR detectors $:\mathcal{E}_{R_1}^{m_1}\mathcal{E}_{R_2}^{m_2} = :\mathbb{O}_{R_1}^{J_{L_1}}\mathbb{O}_{R_2}^{J_{L_2}}:$.

Since $\tilde{D}_{g}^{R_1 R_2}$ evolves with $\gamma_T(m_1+m_2+1)$, it matches a UV detector of boost weight $\tilde{J}_{L_{12}}=2-d-(m_1+m_2)=J_{L_1}+J_{L_2}+d-2$. We therefore identify the diagonalized di-hadron moment with the di-hadron detector function of Sec.~\ref{sec:matchmultiple}, written in the Lorentz-spin notation for its arguments,
\begin{align}
\label{eq:FtoDdihadron}
F_g^{R_1 R_2}(J_{L_1},J_{L_2},\mu) \equiv \tilde{D}_{g}^{R_1 R_2}(m_1,m_2,\mu)\,,
\end{align}
with RG evolution given in~\eq{Tevolcontact} expected of a detector function matching to the UV detector with $J_L = \tilde{J}_{L_{12}}$. It is precisely the object appearing in the contact term of~\eq{twoHad}. The same diagonalization can be carried out at higher multiplicity, producing the genuinely correlated multi-hadron fragmentation-function moments $\tilde{D}_g^{R_1\cdots R_N}(m_1,\dots,m_N,\mu) = F_g^{R_1 \cdots R_N}(J_{L_1},\cdots,J_{L_N},\mu)$ that match the contact terms of higher-multiplicity of IR detectors, such as $F_g^{R_1 R_2 R_3}$ in~\eq{threeHad}. Their RG evolution is given by~\eq{multihadRG} and will thus match to the UV DGLAP detector with $\mathcal{D}_{\tilde{J}_{L_{1\cdots N}},g}^{\rm DGLAP}(\vec{n}_1,\mu)$. In Tab.~\ref{tab:map}, we give a full summary of the dictionary of Lorentz spin $J_L$ and energy weight $m$ representations of IR detectors and detector functions. For a single hadron there is no mixing to remove, so the diagonalization is trivial, $\tilde{D}_g^{R}(m,\mu)=D_g^{R}(m,\mu)$, and the single-hadron detector function $F_g^R(J_L,\mu)$ of~\eq{F1} is recovered.

\begin{figure}
\centering
\scalebox{0.8}{
\begin{tikzpicture}
	\draw [] (-0.5,0) -- (10,0);
	\draw [] (0,-0.5) -- (0,8);
	\draw [lPurple] (0,0) -- (8,8);
	\draw [blue, thick] (-0.5,1.4) .. controls (-0.2,1.3) and (1,-0.1) .. (8.9,8);
	\draw [blue] (2.8,0.8) .. controls (5,2.5) and (7,4.5) .. (10,7.4);
	\draw [blue] (4.5,0.35) .. controls (7,2.5) and (9,4.5) .. (10,5.4);
	\draw [blue] (6.5,0.3) .. controls (8,1.7) and (9,2.5) .. (10,3.4);
	\draw [forest, very thick] (-1,4.3) -- (10,4.3);
	\filldraw[] (2,2) circle (2.5pt);
	\filldraw[blue] (5,4.3) circle (2.5pt);
	\filldraw[orange] (6.8,4.3) circle (2.5pt);
	\filldraw[orange] (8.85,4.3) circle (2.5pt);
	\node [below] at (10,0) {\small $\Delta - 2 $};
	\node [below] at (9.75,-0.5) {\small $= -1-J_L$};
	\node [above] at (0,8) {\small $J = 1 - \Delta_L$};
	\node [forest,below] at (1.6,4.3) {\small $J = J_1 + J_2 -1$};
	\node [forest,above] at (1.7,4.3) {\small $\Delta_L = \Delta_{L_1} + \Delta_{L_2}$};
	\node [above] at (8,8) {\color{lPurple}IR};
	\node [above] at (8.9,8) {\color{blue}UV};
	\node [above left] at (2,2) {$\mathcal{E}$};
\end{tikzpicture}
}
\caption{Illustration of the light-ray operator product expansion (OPE) of two detectors with scaling dimension $ \Delta_{L_1} = 1-J_1$ and $\Delta_{L_2}=1-J_2$. The operator appearing from the OPE has a selection rule that constrains its scaling dimension to $\Delta_L = \Delta_{L_1}+\Delta_{L_2}$, or equivalently its spin $J = J_1+J_2-1$. This fixes the vertical position of the operators that can appear in the OPE, but not its horizontal position. The dominant contribution comes from the twist-$2$ UV detector given by the blue point, whereas higher twist contributions from orange points are further suppressed by the classical scaling $x_L^{(-4+\tau)/2}$.
}
\label{fig:OPE}
\end{figure}
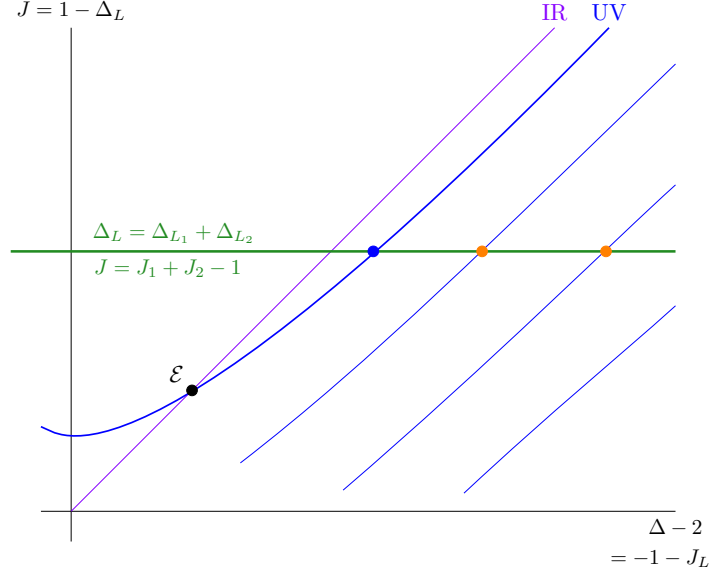

An important consistency check follows in the simultaneous all-hadron, unit-weight limit. Denoting the set of all hadrons by $R=\forall$ and setting $m_1=m_2=1$, we have $J_{L_1}=J_{L_2}=1-d$, $\tilde{J}_{L_{12}}=-d$, and $\delta_{R_1R_2}=1$. Momentum conservation then fixes the separately matched coefficient and the complete two-detector contact coefficient to be
\begin{align*}
F_g^{\forall}(1-d,\mu)&=1\,,&
F_g^{\forall\forall}(1-d,1-d,\mu)+F_g^{\forall}(-d,\mu)&=0\,.
\end{align*}
In Sec.~\ref{sec:trackmatch}, we derive this sum rule explicitly.

\subsection{Light-Ray Operator Product Expansion of Hadronic Detectors}
\label{sec:LROPE}
Let us now consider correlations of generalized detectors carrying independent quantum numbers $R_i$ and energy weights $m_i$,
\begin{align}
\langle \mathcal{E}_{R_1}^{m_1}(\vec{n}_1) \mathcal{E}_{R_2}^{m_2}(\vec{n}_2)\cdots \mathcal{E}_{R_N}^{m_N}(\vec{n}_N) \rangle \equiv \langle \Psi| \mathcal{E}_{R_1}^{m_1}(\vec{n}_1) \mathcal{E}_{R_2}^{m_2}(\vec{n}_2)\cdots \mathcal{E}_{R_N}^{m_N}(\vec{n}_N) |\Psi\rangle\,,
\end{align}
where $|\Psi\rangle$ is a high-energy state with associated perturbative scale $Q$. We study the one-dimensional projection onto the largest pairwise angle $x_L={\rm max}[z_{ij}]$, where $z_{ij}=(n_i\cdot n_j)/2$,
\begin{align}
\label{eq:xLmeas}
\langle \mathcal{E}_{R_1}^{m_1}(&\vec{n}_1) \mathcal{E}_{R_2}^{m_2}(\vec{n}_2)\cdots \mathcal{E}_{R_N}^{m_N}(\vec{n}_N) \rangle_{x_L}\\
&\equiv \frac{1}{Q^{\,m_1+\cdots+m_N}\langle\Psi|\Psi\rangle}\int d^2\Omega_{\vec{n}_1}\cdots d^2\Omega_{\vec{n}_N}\, \langle \mathcal{E}_{R_1}^{m_1}(\vec{n}_1)\cdots \mathcal{E}_{R_N}^{m_N}(\vec{n}_N) \rangle\, \delta(x_L- {\rm{max}}[(n_i\cdot n_j)/2])\,.\nn
\end{align}
As in the rest of this section, we work in pure Yang-Mills at fixed coupling.

We begin with the two-point case, $N=2$, at angular separation $x_L=(n_1\cdot n_2)/2\gg \Lambda_{\rm QCD}^2/Q^2$. Our starting point is the operator matching of~\eq{twoHad},
\begin{align}
\label{eq:twopointmatch}
\mathcal{E}_{R_1}^{m_1}(\vec{n}_1) \mathcal{E}_{R_2}^{m_2}(\vec{n}_2) =&\ F_g^{R_1}(J_{L_1},\mu)F_g^{R_2}(J_{L_2},\mu)\, \mathcal{D}_{J_{L_1},g}^{\rm DGLAP}(\vec{n}_1,\mu)\,\mathcal{D}_{J_{L_2},g}^{\rm DGLAP}(\vec{n}_2,\mu) \\
&\hspace{-1.7cm}+ \left[F_g^{R_1 R_2}( J_{L_1},J_{L_2},\mu)+ \delta_{R_1 R_2}\, F_g^{R_1}(\tilde{J}_{L_{12}},\mu)\right]
\mathcal{D}_{\tilde{J}_{L_{12}},g}^{\rm DGLAP}(\vec{n}_1,\mu)\, \delta^{(d-2)}(\vec{n}_1- \vec{n}_2)\,.\nn
\end{align}
As discussed in Sec.~\ref{sec:matchmultiple}, the contact term is irrelevant for the two-point correlator at finite $x_L$, but becomes important for the three-point correlator, where a finite-$x_L$ contribution arises from its angular separation with a third detector. We now carry out the light-ray OPE of the well-separated term. The OPE imposes a selection rule on the scaling dimension $\Delta_L$, or equivalently the spin $J=1-\Delta_L$. For two operators of dimensions $\Delta_{L_1}=1-J_1$ and $\Delta_{L_2}=1-J_2$, the operator produced has
\begin{align}
\label{eq:DeltaLSelection}
\Delta_L = \Delta_{L_1} + \Delta_{L_2}\,,
\end{align}
or equivalently, in terms of its spin,\footnote{In QCD, scale invariance is broken, so the selection rule is only approximate. The resulting scaling violations can be incorporated through an appropriate ansatz~\cite{Chen:2023zzh}. We neglect these effects here as is consistent with the fixed-coupling pure Yang-Mills setup.}
\begin{align}
\label{eq:Jselection}
J = J_1 + J_2 - 1\,.
\end{align}
This is illustrated in~\fig{OPE}. The rule fixes $\Delta_L$, the vertical position in~\fig{OPE}, but not the twist, so operators of different twists can appear in the OPE. Therefore, for OPE of two twist-2 UV detectors with spins $J_i=m_i+1-2\gamma_T(m_i+1)$, the OPE selects operators with spin $J=J_{12}$ given by
\begin{align}
\label{eq:J12}
J_{12}\equiv J_1+J_2-1 = m_1+m_2+1-2\gamma_T(m_1+1)-2\gamma_T(m_2+1)\,.
\end{align}
We note that the convergence of the light-ray OPE requires the selected spin to lie above the Regge intercept $J_0$ of the theory, $J_{12} > J_0$~\cite{Kologlu:2019mfz}. The leading contribution comes from the twist-2 operator, the blue point in~\fig{OPE}. Focusing on the twist-2 contribution, the OPE therefore produces a UV detector of spin $J_{L_{12}}$ fixed by~\eq{JLandJ}, $3-d-J_{L_{12}} = J_{12}+2\gamma_S(J_{12})$.\footnote{$J_{L_{12}}$ is the boost weight of the operator produced in the well-separated OPE, not to be confused with the contact weight $\tilde J_{L_{12}}$ of~\eq{Jtilde12} in matching. The two coincide when anomalous dimensions are dropped but differ in general.} Therefore, performing the OPE of~\eq{twopointmatch} gives
\begin{align}
\label{eq:twopointOPE}
\mathcal{E}_{R_1}^{m_1}(\vec{n}_1) \mathcal{E}_{R_2}^{m_2}(\vec{n}_2) \approx&\ F_g^{R_1}(J_{L_1},\mu)\,F_g^{R_2}(J_{L_2},\mu)\, C_{12}(x_{12},\mu)\, \mathcal{D}^{\rm DGLAP}_{J_{L_{12}},g}(\vec{n}_1,\mu) \\
&+ \big[F_g^{R_1 R_2}(J_{L_1},J_{L_2},\mu)+\delta_{R_1 R_2}\,F_g^{R_1}(\tilde{J}_{L_{12}},\mu)\big]\, \mathcal{D}^{\rm DGLAP}_{\tilde{J}_{L_{12}},g}(\vec{n}_1,\mu)\, \delta^{(d-2)}(\vec{n}_1-\vec{n}_2)\,.\nn
\end{align}
where $C_{12}(x_{12},\mu)$ is the OPE coefficient describing the dependence on the angular separation $x_{12}=(n_1\cdot n_2)/2$ of the two detectors. Let us consider both the RG invariance and Lorentz symmetry constraints on the OPE.

The contact term matches to the UV detector with the Lorentz spin $\tilde{J}_{L_{12}}$ and, by~\eq{Devol}, evolves as
\begin{align}
\frac{d}{d\ln\mu^2}\mathcal{D}^{\rm DGLAP}_{\tilde{J}_{L_{12}},g}(\vec{n},\mu) = \gamma_T(3-d-\tilde{J}_{L_{12}})\, \mathcal{D}^{\rm DGLAP}_{\tilde{J}_{L_{12}},g}(\vec{n},\mu)\,.
\end{align}
This is exactly opposite to the evolution of the contact coefficient: both $F_g^{R_1 R_2}(J_{L_1},J_{L_2})$ and $F_g^{R_1}(\tilde{J}_{L_{12}})$ evolve with $-\gamma_T(3-d-\tilde{J}_{L_{12}})$, by~\eq{Tevolcontact}, so the contact term is RG invariant. In terms of Lorentz transformation, the LHS transforms manifestly by $J_{L_1}+J_{L_2}$ under boosts. On the other hand, the UV detector $\mathcal{D}^{\rm DGLAP}_{\tilde{J}_{L_{12}},g}(\vec{n},\mu)$ transforms by $\tilde{J}_{L_{12}}= J_{L_1}+J_{L_2} + (d-2)$ and the delta function $\delta^{(d-2)}(\vec{n}_1-\vec{n}_2)$ by $-(d-2)$, so the contact term is also boost invariant. 

The $x_{12}$ dependence of the OPE coefficient is fixed by Lorentz symmetry and RG invariance. Lorentz symmetry sets its classical scaling, $C_{12}(x_L,\mu)\sim 1/x_L$, while RG invariance of~\eq{twopointOPE} fixes its evolution,
\begin{align}
\frac{d}{d\ln\mu^2} C_{12}(x_{12},\mu) = \big[\gamma_T(3-d-J_{L_1})+\gamma_T(3-d-J_{L_2})-\gamma_T(3-d-J_{L_{12}})\big]\, C_{12}(x_{12},\mu)\,,
\end{align}
which compensates the evolution of $F_g^{R_1}(J_{L_1})F_g^{R_2}(J_{L_2})$ and of $\mathcal{D}^{\rm DGLAP}_{J_{L_{12}},g}$.

To read off the physical scaling of the projected correlator~\eq{xLmeas}, we note that the UV-detector expectation value in $|\Psi\rangle$ is naturally taken at $\mu\sim Q$, while the OPE coefficient lives at the angular scale $\mu\sim Q\sqrt{x_L}$ and the detector functions at $\mu \sim \Lambda_{\rm QCD}$. If we evaluate both the OPE coefficient and the detector functions at $\mu=Q\sqrt{x_L}$,\footnote{Detector functions evaluated at $\mu=Q\sqrt{x_L}$ of course change their values with $x_L$. Here, we display the scaling obtained when the result is expressed in terms of detector functions at $\mu=Q\sqrt{x_L}$. To make the additional scaling from the RG evolution of the detector functions explicit, one can instead fix them at $\mu\sim\Lambda_{\rm QCD}$ and evolve them up to $Q\sqrt{x_L}$.} the entire $x_L$ scaling beyond the classical $1/x_L$ comes from running the UV detector from $Q$ to $Q\sqrt{x_L}$, and is controlled by $\gamma_T(3-d-J_{L_{12}})$. By reciprocity, this is just the spacelike anomalous dimension at the OPE spin,
\begin{align}
\gamma_T(3-d-J_{L_{12}})= \gamma_T(J_{12}+2\gamma_S(J_{12})) = \gamma_S(J_{12})\,,
\end{align}
where the first equality uses~\eq{JLandJ} and the second the reciprocity relation
\begin{align}
\label{eq:reciprocity}
\gamma_S(J) = \gamma_T(J+2\gamma_S(J))\,.
\end{align}
For a clear derivation of reciprocity, see~\cite{Dixon:2019uzg,Lee:2024icn}. The anomalous scaling of the two-point $x_L$ distribution is therefore
\begin{align}
\label{eq:gammaSJ'}
\gamma_S(J_{12}) = \gamma_S\big(m_1+m_2+1-2\gamma_T(m_1+1)-2\gamma_T(m_2+1)\big)\,,
\end{align}
which for equal weights $m_1=m_2=m$ reduces to $\gamma_S(2m+1-4\gamma_T(m+1))$.

For the ordinary two-point energy correlator, $m_1=m_2=1$ and $R_1=R_2=\forall$. Momentum conservation gives $\gamma_T(2)=0$ and $F_g^\forall(1-d,\mu)=1$, while the all-hadron sum rule of Sec.~\ref{sec:matchmultifrag}, derived explicitly in Sec.~\ref{sec:trackmatch}, sets the contact coefficient in~\eq{twopointOPE} to zero. The surviving well-separated channel therefore has $J_{12}=3$, and its leading-twist anomalous scaling is $\gamma_S(3)$, as required for the standard two-point energy correlator.

Finally, while we dropped the higher-twist operators in~\eq{twopointOPE}, their OPE coefficients scale classically as $x_L^{(\tau-4)/2}$, and so are power suppressed in $x_L$ relative to the leading twist-2 contribution. In addition to these higher-twist corrections, the light-ray OPE also receives nonperturbative power corrections in $\Lambda_{\rm QCD}$, which supplement~\eq{twopointOPE} with additional nonperturbative channels; these will be discussed in Sec.~\ref{sec:NP}.

We now turn to the three-point OPE. With three detectors we form three pairwise angles, ordered $x_S\le x_M\le x_L$, and our starting point is the operator matching of~\eq{threeHad}. Projecting onto the largest angle $x_L$, three configurations appear: all three detectors well separated, a single pair $(ij)$ unresolved with the third separate, and all three unresolved, with the last not contributing to the finite-$x_L$ measurement for three detectors. Carrying out the OPE of the UV detectors in each case, and writing $\sum_{(ij)k}$ for the sum over the three merged pairs with $\{i,j,k\}=\{1,2,3\}$ and $i<j$,
\begin{align}
\label{eq:threepointOPE}
\hspace{-1.5cm}\mathcal{E}_{R_1}^{m_1}(\vec{n}_1) \mathcal{E}_{R_2}^{m_2}(\vec{n}_2)\,\mathcal{E}_{R_3}^{m_3}(\vec{n}_3) \approx&\ \bigg[\prod_{i=1}^3 F_g^{R_i}(J_{L_i})\bigg]\, C_{123}\Big(x_L,\tfrac{x_S}{x_L},\tfrac{x_M}{x_L}\Big)\,\mathcal{D}^{\rm DGLAP}_{J_{L_{123}},g}(\vec{n}_1) \\
&\hspace{-4.5cm}+ \sum_{(ij)k}\big[F_g^{R_i R_j}(J_{L_i},J_{L_j})+\delta_{R_i R_j}\,F_g^{R_i}(\tilde{J}_{L_{ij}})\big]\, F_g^{R_k}(J_{L_k})\, C_{(ij)k}(x_{ik})\,\mathcal{D}^{\rm DGLAP}_{J_{L_{(ij)k}},g}(\vec{n}_i)\, \delta^{(d-2)}(\vec{n}_i-\vec{n}_j) \nn\\
&\hspace{-4.5cm}+ \bigg[F_g^{R_1 R_2 R_3}(J_{L_1},J_{L_2},J_{L_3})+\sum_{(ij)k}\delta_{R_i R_j}\,F_g^{R_i R_k}(\tilde{J}_{L_{ij}},J_{L_k})+\delta_{R_1 R_2}\delta_{R_2 R_3}\,F_g^{R_1}(\tilde{J}_{L_{123}})\bigg]\nn\\
&\hspace{-4.5cm}\quad\times\,\mathcal{D}^{\rm DGLAP}_{\tilde{J}_{L_{123}},g}(\vec{n}_1)\, \delta^{(d-2)}(\vec{n}_1-\vec{n}_2)\delta^{(d-2)}(\vec{n}_2-\vec{n}_3)\,.\nn
\end{align}
where the matching and OPE scale $\mu$ is suppressed. This is just the matching~\eq{threeHad} with  UV-detector products replaced by their OPEs, so the detector functions are unchanged. The OPE coefficient $C_{123}$ carries the dependence on the three resolved angles, while $C_{(ij)k}$ depends on the single surviving angle $x_{ik}=(n_i\cdot n_k)/2$ between the IR detector pair that is matched together and the third detector. The last line is a pure contact term, proportional to $\delta^{(d-2)}(\vec{n}_1-\vec{n}_2)\delta^{(d-2)}(\vec{n}_2-\vec{n}_3)$. Since all three IR detectors are matched together to a single UV detector for this contact term, there is no additional OPE to be carried out.

The fully separated term involves the simultaneous OPE of three leading-twist detectors with spins $J_i=m_i+1-2\gamma_T(m_i+1)$ and selects detectors of spin $J=J_{123}$ given by
\begin{align}
\label{eq:J123}
J_{123} = J_1+J_2+J_3-2 = m_1+m_2+m_3+1-2\sum_{i=1}^3\gamma_T(m_i+1)\,.
\end{align}
This follows from generalizing~\eq{DeltaLSelection} to the three detector case $\Delta_L = \Delta_{L_1}+\Delta_{L_2}+\Delta_{L_3}$. This OPE should be thought of as considering the three detectors as a single cluster and performing the OPE of the three together with respect to $x_L$ while keeping the shape fixed (i.e. fixed $x_S/x_L$ and $x_M/x_L$). The second term in~\eq{threepointOPE} arises from performing the OPE between a UV detector with spin $J=m_i+m_j+1 - 2\gamma_T(m_i+m_j+1)$ from a pair of IR detectors $\mathcal{E}_{R_i}^{m_i}$ and $\mathcal{E}_{R_j}^{m_j}$ matched together and a third separate detector matched to a separate UV detector with spin $J = m_k + 1 - 2\gamma_T(m_k+1)$. Using the selection rule~\eq{Jselection}, the OPE selects detectors of spin $J=J_{(ij)k}$ given by
\begin{align}
\label{eq:Jijkselection}
J_{(ij)k} = m_1+m_2+m_3+1-2\gamma_T(m_i+m_j+1)-2\gamma_T(m_k+1)\,.
\end{align}
Focusing on the leading twist contribution, the associated boost weights $J_{L_{123}}$ and $J_{L_{(ij)k}}$ follow from~\eq{JLandJ}.

RG invariance fixes the evolution of the OPE coefficients exactly as for $C_{12}$ in the two-point case. In the boost-weight form,
\begin{align}
\frac{d}{d\ln\mu^2} C_{123} &= \bigg[\sum_{i=1}^3\gamma_T(3-d-J_{L_i})-\gamma_T(3-d-J_{L_{123}})\bigg]\, C_{123}\,,\nn\\
\frac{d}{d\ln\mu^2} C_{(ij)k} &= \big[\gamma_T(3-d-\tilde{J}_{L_{ij}})+\gamma_T(3-d-J_{L_k})-\gamma_T(3-d-J_{L_{(ij)k}})\big]\, C_{(ij)k}\,.
\end{align}
We now project onto $x_L$ via~\eq{xLmeas}. Ordering the angles as $z_{12}\le z_{13}\le z_{23}$ so that $x_L=z_{23}$, the fully separated coefficient is integrated over the two cross-ratios at fixed $x_L$,
\begin{align}
\tilde{C}_{123}(x_L,\mu) \equiv \int d^2\Omega_{\vec{n}_1} d^2\Omega_{\vec{n}_2} d^2\Omega_{\vec{n}_3}\, \delta(x_L-z_{23})\, \theta(x_L-z_{12})\, \theta(x_L-z_{13})\,C_{123}\Big(x_L,\tfrac{z_{12}}{x_L},\tfrac{z_{13}}{x_L},\mu\Big)\,,
\end{align}
while the single angle $x_{ik}$ dependence in the OPE coefficient $C_{(ij)k}(x_{ik})$ becomes $x_L$ by the projection. This contribution is precisely the contact term of the two IR detector case given in~\eq{twopointOPE} being OPE-ed with the third detector. Evaluating the OPE coefficients and detector functions at $\mu=Q\sqrt{x_L}$, the anomalous $x_L$ scaling comes from running the produced UV detector from $Q$ to $Q\sqrt{x_L}$, controlled by $\gamma_T(3-d-J_L)$, which by reciprocity is the spacelike anomalous dimension at the produced spin,
\begin{align}
\label{eq:gammaSJThree}
\gamma_S(J_{123}) &= \gamma_S\Big(m_1+m_2+m_3+1-2\sum_{i=1}^3\gamma_T(m_i+1)\Big)\,,\\
\gamma_S(J_{(ij)k}) &= \gamma_S\big(m_1+m_2+m_3+1-2\gamma_T(m_i+m_j+1)-2\gamma_T(m_k+1)\big)\,,\nn
\end{align}
and thus have two distinctive anomalous scalings, one from the OPE of three well-separated detectors and one from performing the pairwise OPE between a pair of detectors matched together and a third. For equal weights $m_1=m_2=m_3=m$ these reduce to $\gamma_S(3m+1-6\gamma_T(m+1))$ and $\gamma_S(3m+1-2\gamma_T(2m+1)-2\gamma_T(m+1))$. The generalization to higher points is straightforward, requiring the integration of more cross-ratios in the $x_L$ projection and accumulating contributions from the iterated OPEs of the new contact terms generated at each level.

The inclusive unit energy-weight limit eliminates the apparent contact-induced scaling channels. For $m_i=1$ and $R_i=\forall$, every pair-contact coefficient in~\eq{threepointOPE} contains the vanishing two-detector combination stated in Sec.~\ref{sec:matchmultifrag}, while the three-detector combination likewise vanish as shown in Sec.~\ref{sec:trackmatch}. The channels governed by $\gamma_S(J_{(ij)k})$ therefore carry zero matching coefficient and there is no pair-contact scaling. Since $\gamma_T(2)=0$, the surviving fully separated channel has $J_{123}=4$ and anomalous scaling $\gamma_S(4)$, as expected for the standard three-point energy correlator. More generally, products of $N$ inclusive unit-weight detectors receive no additional contact corrections from the IR-to-UV matching, and the fully separated leading-twist channel gives $J_{1\cdots N}=N+1$ and anomalous scaling $\gamma_S(N+1)$ at finite $x_L$.

\section{Generalized Track Function Formalism}\label{sec:GenTrack}
In Sec.~\ref{sec:matchmultifrag}, we introduced multi-hadron fragmentation functions whose moments provide the detector matching functions between hadronic IR detectors and partonic UV detectors. Here we reorganize the same nonperturbative information in terms of generalized track functions, whose moments directly supply the combinations of fragmentation-function moments that enter the matching. This representation also allows us to obtain the NLO evolution of the required combinations from the recently derived NLO track-function kernels~\cite{Li:2021zcf,Jaarsma:2022kdd,Chen:2022muj,Chen:2022pdu}. After reviewing standard track functions in Sec.~\ref{sec:trackreview}, we introduce one-dimensional generalized track functions for products of identical detectors with arbitrary energy weight in Sec.~\ref{sec:gentrack1d}, followed by multi-dimensional generalized track functions for correlators with arbitrary energy weights and hadron selections in Sec.~\ref{sec:gentrackmulti}.

\subsection{Review of the Track Function Formalism}
\label{sec:trackreview}

In this subsection we briefly review the track function formalism. Track functions were first introduced in~\cite{Chang:2013rca,Chang:2013iba}, and have since been developed significantly, including the calculation of their renormalization group evolution to NLO~\cite{Li:2021zcf,Jaarsma:2022kdd,Chen:2022muj,Chen:2022pdu}, their application to energy correlators~\cite{Chen:2020vvp,Jaarsma:2023ell,Lee:2026hub}, and their extraction from data~\cite{Lee:2023xzv,ATLAS:2024jrp}. In Sec.~\ref{sec:trackdef} we define the track functions and relate their moments to the moments of the multi-hadron fragmentation functions of Sec.~\ref{sec:matchmultifrag}. In Sec.~\ref{sec:trackRGsec} we discuss their renormalization group evolution and its shift symmetry. Finally, in Sec.~\ref{sec:trackmatch} we show how track function moments reproduce the detector functions appearing in the matching of multiple IR detectors with unit energy weights.

\subsubsection{Definition and Relation to Multi-Hadron Fragmentation Functions}
\label{sec:trackdef}

Physically, the track function $T_{iR}(z,\mu)$ describes the distribution of the total momentum fraction $z=\sum_{h\in R} p^-_{h}/k^-$ carried by all hadrons with quantum number $R$ produced in the fragmentation of a parton $i$ with large light-cone momentum $k^-$. This should be contrasted with the single-hadron fragmentation function of~\eq{singlefrag}, which measures the momentum fraction of a single identified hadron $h$ at a time. The gluon track function is obtained from~\eq{singlefrag} by making the sum over final states fully inclusive and replacing the measurement $\delta(z-p_h^-/k^-)$ with $\delta(z-\sum_{h\in R} p_h^-/k^-)$, which adds up the momentum fractions of all hadrons carrying the quantum number $R$,
\begin{align}
\label{eq:trackdefg}
T_{gR}(z,\mu)=&\ \frac{-1}{(d-2)\left(N_c^2-1\right) k^{-}}\int \mathrm{d} y^{+} \mathrm{d}^{d-2} y_{\perp}\, e^{i k^{-} y^{+} / 2} \nn\\
&\times \sum_X \langle 0| G_{-\lambda}^a\left(y^{+}, 0, y_{\perp}\right)|X\rangle\langle X| G_{-}^{\lambda, a}(0)|0\rangle\,  \delta\left(z-\frac{\sum_{h\in R} p_h^-}{k^{-}}\right) \,.
\end{align}
To consider the full QCD case, we also need quark track functions, which are given by
\begin{align}
\label{eq:trackdefq}
T_{qR}(z,\mu)=&\ \int \mathrm{d} y^{+} \mathrm{d}^{d-2} y_{\perp}\, e^{i k^{-} y^{+} / 2} \nn\\
&\times\sum_X
\frac{1}{2 N_c} \operatorname{tr}\left[\frac{\gamma^{-}}{2}\langle 0| \psi\left(y^{+}, 0, y_{\perp}\right)|X\rangle\langle X| \bar{\psi}(0)|0\rangle\right]  \delta\left(z-\frac{\sum_{h\in R} p_h^-}{k^{-}}\right) \,,
\end{align}
where we work in light-cone gauge to suppress gauge links. An important phenomenological motivation for restricting measurements to charged hadrons is that experimental tracking systems provide far better angular resolution than calorimetry. Recently, this enabled precise measurements of the two-point energy correlator in $e^+e^-$ annihilation at the $Z$-pole, $Q=m_Z$, using archival LEP data~\cite{Electron-PositronAlliance:2025fhk,Electron-PositronAlliance:2025wzh,Zhang:2025nlf}. The name ``track function'' derives from the original case in which $R$ is the set of electrically charged hadrons (tracks), but the formalism applies equally to any quantum number $R$, e.g. strangeness, heavy flavor~\cite{Craft:2022kdo}, baryon number, or isospin. Unlike the fragmentation functions, which are number densities, the track functions are probability distributions, normalized as $\int_0^1 dz\, T_{iR}(z,\mu)=1$. In particular, when $R$ is the set of all hadrons, which we denote by $R=\forall$, momentum conservation fixes the track function to be a delta function at the endpoint, $T_{i\forall}(z,\mu)=\delta(1-z)$, since the full set of hadrons carries all of the momentum of the fragmenting parton. A nontrivial track function is therefore a direct consequence of restricting the measurement to a subset of hadrons, and for any restricted $R$ the track function is a genuinely nonperturbative distribution.

We define the moments of track functions as
\begin{align}
\label{eq:trackmom}
T_{iR}(n,\mu)\equiv \int_0^1 dz\, z^n\, T_{iR}(z,\mu) \,,
\end{align}
which are normalized as $T_{iR}(0,\mu)=1$. Since the track functions and the multi-hadron fragmentation functions of Sec.~\ref{sec:matchmultifrag} are built from the same operators and differ only in the measurement performed on the final state, their moments are directly related. The relation follows from expanding the $n$-th power of the total momentum fraction multinomially, organizing the terms by the number $M$ of distinct hadrons that are weighted. For integer $n\geq 1$,
\begin{align}
\label{eq:multinomexp}
\Big(\sum_{h \in R} z_h\Big)^{n} = \sum_{M=1}^{n}\, \frac{1}{M!}\sum_{\substack{h_1,\ldots, h_M\in R \\ \text{pairwise distinct}}}\ \sum_{\substack{m_j\in\mathbb{N}^+ \\ m_1+\cdots+m_M =n}} \frac{n!}{m_1!\cdots m_M!}\; z_{h_1}^{m_1} \cdots z_{h_M}^{m_M}\,,
\end{align}
where $z_h = p_h^-/k^-$, the middle sum runs over ordered $M$-tuples of distinct hadrons in the final state, and the factor $1/M!$ compensates for the ordering. Since each $m_j\geq 1$, at most $n$ hadrons can be weighted, so the sum over $M$ terminates at $M=n$. Inserting~\eq{multinomexp} into the moments of~\eqs{trackdefg}{trackdefq}, we find that each term in which $M$ hadrons are weighted, summed inclusively over the remaining hadrons, produces the corresponding moment of the $M$-hadron fragmentation function of~\eq{gluonNhadron} (and of its quark analogue). Denoting the moments of the $M$-hadron fragmentation functions in analogy with the di-hadron moments of~\eq{dihadronmoment},
\begin{align}
\label{eq:multihadmom}
D_i^{R_1\cdots R_M}(m_1,\ldots,m_M,\mu) \equiv& \sum_{h_1\in R_1}\cdots\sum_{h_M\in R_M}\int_0^1 dx_1 \cdots dx_M\; x_1^{m_1}\cdots x_M^{m_M}\nn\\
&\times D_{i\to h_1\cdots h_M}(x_1,\ldots,x_M,\mu)\,,
\end{align}
we then find
\begin{align}
\label{eq:trackNhad}
T_{iR}(n,\mu) = \sum_{M=1}^{n}\, \frac{1}{M!}\sum_{\substack{m_j\in\mathbb{N}^+ \\ m_1+\cdots+m_M =n}} \frac{n!}{m_1!\cdots m_M!}\; D_i^{R\cdots R}(m_1,\ldots,m_M,\mu)\,,
\end{align}
where the fragmentation function moment in the summand carries $M$ identical labels $R$. The first few moments are given explicitly by
\begin{align}
\label{eq:tracklowmom}
T_{iR}(1,\mu) &= D_i^{R}(1,\mu)\,,\nn\\
T_{iR}(2,\mu) &= D_i^{R}(2,\mu) + D_i^{RR}(1,1,\mu)\,,\nn\\
T_{iR}(3,\mu) &= D_i^{R}(3,\mu) + 3\,D_i^{RR}(2,1,\mu) + D_i^{RRR}(1,1,1,\mu)\,,
\end{align}
where we used the permutation symmetry of the moments, $D_i^{RR}(1,2,\mu) = D_i^{RR}(2,1,\mu)$. We therefore see that track function moments provide a convenient way to package multi-hadron fragmentation-function moments: the $n$-th track function moment is a sum of fragmentation-function moments involving up to $n$ identified hadrons.

\subsubsection{Renormalization Group Evolution and Shift Symmetry}
\label{sec:trackRGsec}

The RG evolution of the track functions has a non-linear structure, mirroring that of the multi-hadron fragmentation functions. As discussed in Sec.~\ref{sec:matchmultifrag}, the RG evolution of an $N$-hadron fragmentation function involves lower-multiplicity fragmentation functions, with terms containing up to the product of $N$ single-hadron fragmentation functions. Since the track function packages fragmentation functions of all multiplicities, its RG evolution involves products of arbitrarily many track functions, with higher non-linear terms appearing at higher orders in $\alpha_s$. The evolution takes the general form
\begin{align}
\label{eq:trackRGE}
\frac{d}{d \ln \mu^2} T_{iR}(z,\mu)=\sum_{\left\{i_f\right\}} \sum_{\ell=0}^{\infty} a_s^{\ell+1}\, K_{i \rightarrow\left\{i_f\right\}}^{(\ell)} \otimes \prod_{j \in\left\{i_f\right\}}  T_{jR}\,,
\end{align}
where $a_s=\alpha_s/(4\pi)$ and $K_{i\to i_1 i_2 \cdots i_M}$ are kernels describing the splitting of $1\to M$ partons. These evolution kernels were derived to NLO in \cite{Chen:2022muj,Chen:2022pdu}, at which order up to three partons are involved. The virtual term $K_{i\to i}$ has a trivial convolution structure (a simple product), while the convolutions involving two and three track functions are given by
\begin{align}
\label{eq:trackconvol}
 K_{i \rightarrow i_1 i_2} &\otimes T_{i_1R} T_{i_2R}\, (z) \nn\\
 =&\int_0^1 dx_1\, dx_2\ \delta\left(1-x_1-x_2\right) K_{i \rightarrow i_1 i_2}\left(x_1, x_2\right) \nonumber\\
 &\times \int_0^1 dy_1\,dy_2 \,\delta (z-x_1 y_1-x_2 y_2)\,T_{i_1R}(y_{1})\, T_{i_2R}(y_{2})\,, \nonumber\\
 K_{i \rightarrow i_1 i_2 i_3} &\otimes T_{i_1R}T_{i_2R}T_{i_3R}\,(z) \nonumber\\
=&\int_0^1 dx_1\, dx_2\, dx_3\ \delta\left(1-x_1-x_2-x_3\right)K_{i \rightarrow i_1 i_2 i_3}\left(x_1, x_2, x_3\right)\nonumber\\
& \times \int_0^1 dy_1\,dy_2\,dy_3\ \delta (z-x_1 y_{1}-x_2 y_{2}-x_3 y_{3})\, T_{i_1R}(y_{1})\,T_{i_2R}(y_{2})\,T_{i_3R}(y_{3})\,,
\end{align}
where $x_j$ denotes the momentum fraction of the daughter parton $i_j$ produced in the splitting and $y_j$ denotes the momentum fraction of $R$-hadrons produced in the fragmentation of that daughter parton. The delta function then builds up the measured $R$-hadron momentum fraction $z$ of the parent parton
\begin{align}
\label{eq:trackzbuild}
z = \sum^M_{j=1} x_j\, y_{j}\,,\qquad \sum_{j=1}^M x_j = 1\,,
\end{align}
additively from the daughter partons. Higher convolutions appearing in higher-order kernels generalize straightforwardly.

Taking moments of~\eq{trackRGE}, the structure of the convolutions~\eq{trackconvol} implies that the evolution of any given moment involves the moment itself, together with products of lower moments whose orders sum to that of the original moment. This is consistent with the fact that the $n$-th moment of the track function contains up to $n$-hadron fragmentation functions in~\eq{trackNhad}, while the RG evolution of multi-hadron fragmentation functions involves only fragmentation functions of the same or lower multiplicity. Importantly, the linear term in the evolution of the $n$-th moment is governed by the timelike DGLAP anomalous dimensions,
\begin{align}
\label{eq:trackmomlinear}
\frac{d}{d\ln\mu^2} T_{iR}(n,\mu) \supset -\gamma^T_{ji}(n+1)\,  T_{jR}(n,\mu)\,,
\end{align}
where a sum over the parton flavor $j$ is implied and $\gamma^T_{ji}$ is the matrix of timelike anomalous dimensions in flavor space, whose gluon-gluon entry, relevant for the pure YM case, appeared in~\eq{gammaTfromP}. For details, see~\cite{Jaarsma:2022kdd,Li:2021zcf}.

The RG evolution of the track functions is further constrained by a shift symmetry~\cite{Jaarsma:2022kdd,Li:2021zcf}, which follows directly from the additive structure of~\eq{trackzbuild}: since the parton momentum fractions satisfy $\sum_{j} x_j =1$, the relation~\eq{trackzbuild} is invariant under the simultaneous shift
\begin{align}
\label{eq:trackshift}
z &\to z + a\,, \nonumber\\
y_{j} &\to y_{j}+a\,.
\end{align}
In other words, the evolution commutes with a common shift of all track functions. It was shown in~\cite{Jaarsma:2022kdd,Li:2021zcf} that this symmetry uniquely determines the RG evolution of the first three moments of the track functions in terms of the timelike anomalous dimensions, up to terms proportional to the difference of first moments, $T_{qR}(1,\mu)-T_{gR}(1,\mu)$. The shift symmetry also constrains the UV asymptotics of the track functions: it ensures that the central moments evolve homogeneously, and, since the corresponding timelike anomalous dimensions are positive, they decay away under evolution towards the UV. Meanwhile, the difference between the quark and gluon first moments decays, while momentum conservation forces the matrix of timelike anomalous dimensions $\gamma^T_{ji}(2)$ to have a vanishing eigenvalue, whose eigenvector combination of the quark and gluon first moments remains constant and the first moments therefore approach this common conserved value. In the deep UV, the track functions of all parton species therefore collapse onto a common delta function centered at this value.

\subsubsection{Track Function Moments as Detector Matching Functions}
\label{sec:trackmatch}

Let us now see how the track functions efficiently capture the matching between IR and UV detectors. To connect with the discussion in Sec.~\ref{sec:matchmultifrag}, we again restrict to the pure YM case, where only the gluon track function appears. For the matching of two IR detectors, we need track function moments up to the second. From~\eq{tracklowmom}, the second moment contains the di-hadron moment $D_g^{RR}(1,1,\mu)$, whose evolution~\eq{dihadronmomevol} mixes into the product of single-hadron moments $D_g^{R}(1,\mu)^2 = T_{gR}(1,\mu)^2$. The first and second moments of $T_{gR}(z,\mu)$ therefore evolve as the closed triangular system
\begin{align}
\label{eq:trackmatrix2}
\hspace{-1cm}\frac{d}{d \ln \mu^2}\left(\begin{array}{c}
T_{gR}(2) \\
T_{gR}(1)^2
\end{array}\right)=
\left(\begin{array}{cc}
-\gamma_{T}(3) & \hat{\gamma}_g(1,1)\\
0& -2\gamma_{T}(2)
\end{array}\right)
\left(\begin{array}{c}
T_{gR}(2) \\
T_{gR}(1)^2
\end{array}\right)\,,
\end{align}
where we suppress the $\mu$ dependence for brevity. The diagonal entries are given exactly by the timelike anomalous dimensions, in accordance with~\eq{trackmomlinear}, while the off-diagonal mixing is given precisely by the kernel $\hat{\gamma}_g(1,1)$ of~\eq{mixingkernel}, as follows from applying the evolution equations \eqs{Tevol}{dihadronmomevol} to the moment relations~\eq{tracklowmom}. For unit energy weight, the shift symmetry relates this mixing entry to the timelike anomalous dimension~\cite{Jaarsma:2022kdd,Li:2021zcf},
\begin{align}
\label{eq:trackshiftvals}
\gamma_T(2) &= 0\,,\nn\\
\hat{\gamma}_g(1,1) &=\gamma_T(3)\,,
\end{align}
where the vanishing of $\gamma_T(2)$ also follows directly from energy conservation.

The role of the off-diagonal mixing is to single out the particular linear combination of $T_{gR}(2)$ and $T_{gR}(1)^2$ that evolves purely multiplicatively with a single timelike anomalous dimension. Ignoring the $\mu$ dependence of $\alpha_s(\mu)$ for simplicity, this combination is given by
\begin{align}
\label{eq:trackeigen2}
\tilde{T}_{gR}(2) = T_{gR}(2) - \frac{\hat{\gamma}_g(1,1)}{\gamma_T(3)-2\gamma_T(2)}\,T_{gR}(1)^2\,,
\end{align}
in terms of which the system~\eq{trackmatrix2} diagonalizes,
\begin{align}
\label{eq:trackdiag2}
\hspace{-1cm}\frac{d}{d \ln \mu^2}\left(\begin{array}{c}
\tilde{T}_{gR}(2) \vspace{0.05cm}\\
T_{gR}(1)^2
\end{array}\right)=
\left(\begin{array}{cc}
-\gamma_{T}(3) & 0\\
0& -2\gamma_{T}(2)
\end{array}\right)
\left(\begin{array}{c}
\tilde{T}_{gR}(2) \vspace{0.05cm}\\
T_{gR}(1)^2
\end{array}\right)\,,
\end{align}
where we keep the (vanishing) $\gamma_T(2)$ entries explicit for later comparison with the case of general energy weights. Inserting the shift symmetry values~\eq{trackshiftvals} into~\eq{trackeigen2} gives $\tilde{T}_{gR}(2) = T_{gR}(2)-T_{gR}(1)^2$: for track functions, the diagonalized combination is nothing but the second central moment (the variance) of the track function.

We can now make contact with the detector functions of Sec.~\ref{sec:matchmultifrag}. Substituting the relations~\eq{tracklowmom} between track function moments and multi-hadron fragmentation function moments into~\eq{trackeigen2}, we find
\begin{align}
\label{eq:trackeigen2frag}
\tilde{T}_{gR}(2,\mu) &= D_g^{R}(2,\mu) + D_g^{RR}(1,1,\mu) - \frac{\hat{\gamma}_g(1,1)}{\gamma_T(3)-2\gamma_T(2)}\, D_g^{R}(1,\mu)^2\nn\\
&= D_g^{R}(2,\mu) + \tilde{D}_g^{RR}(1,1,\mu) = F_g^{R}(-d,\mu) + F_g^{RR}(1-d,1-d,\mu)\,,
\end{align}
where in the second line we recognized the diagonalized di-hadron moment $\tilde{D}_g^{RR}(1,1,\mu)$ of~\eq{dihadrondiag} evaluated at $m_1=m_2=1$, and then translated to the Lorentz-spin notation using the map of Tab.~\ref{tab:map}. This is exactly the matching coefficient of the contact term in~\eq{twoHad} for the case $m_1=m_2=1$ and $R_1=R_2=R$. For the well-separated term, we similarly have $T_{gR}(1,\mu) = D_g^{R}(1,\mu) = F_g^{R}(1-d,\mu)$. The matching of the two unit-weight IR detectors $\mathcal{E}_{R}(\vec{n}_1)\mathcal{E}_{R}(\vec{n}_2)$, where we abbreviate $\mathcal{E}_{R}\equiv\mathcal{E}_{R}^{m=1}$, a special case of~\eq{twoHad}, can thus be written compactly in terms of track function moments as
\begin{align}
\label{eq:EEmatchtrack}
\mathcal{E}_{R}(\vec{n}_1) \mathcal{E}_{R}(\vec{n}_2) =&\ T_{gR}(1,\mu)\,T_{gR}(1,\mu)\, \mathcal{D}_{1-d,g}^{\rm DGLAP}(\vec{n}_1,\mu)\,\mathcal{D}_{1-d,g}^{\rm DGLAP}(\vec{n}_2,\mu) \nn\\
&+ \tilde{T}_{gR}(2,\mu)\,
\mathcal{D}_{-d,g}^{\rm DGLAP}(\vec{n}_1,\mu)\, \delta^{(d-2)}(\vec{n}_1- \vec{n}_2)\,.
\end{align}
We emphasize that it is the diagonalized combination $\tilde{T}_{gR}(2,\mu)$, rather than the non-diagonalized $T_{gR}(2,\mu)$, that appears in the contact term: RG consistency requires the contact coefficient to evolve homogeneously with $-\gamma_T(3)$, compensating the evolution~\eq{Devol} of the UV detector $\mathcal{D}_{-d,g}^{\rm DGLAP}$.

The extension to higher moments of track functions, relevant for higher-point correlators, proceeds in the same way. We define $\tilde{T}_{gR}(3)$ as the linear combination of $T_{gR}(1)^3$, $T_{gR}(2)\, T_{gR}(1)$ and $T_{gR}(3)$ that diagonalizes the RG evolution of the third moment; the explicit construction, for general energy weight, is given in Sec.~\ref{sec:gentrackmatch}. An analogous computation using the third-moment relation of~\eq{tracklowmom} then yields
\begin{align}
\label{eq:trackeigen3frag}
\tilde{T}_{gR}(3,\mu) &= D_g^{R}(3,\mu) + 3\,\tilde{D}_g^{RR}(2,1,\mu) + \tilde{D}_g^{RRR}(1,1,1,\mu)\nn\\
&= F_g^{R}(-1-d,\mu) + 3\,F_g^{RR}(-d,1-d,\mu) + F_g^{RRR}(1-d,1-d,1-d,\mu)\,,
\end{align}
which is precisely the matching coefficient of the fully unresolved contact term in~\eq{threeHad} for $m_i=1$ and $R_i=R$. The matching of three IR detectors, a special case of~\eq{threeHad}, is then recast in terms of track function moments as
\begin{align}
\label{eq:EEEmatchtrack}
\hspace{-1.2cm}\mathcal{E}_{R}(\vec{n}_1) \mathcal{E}_{R}(\vec{n}_2) \mathcal{E}_{R}(\vec{n}_3) =&\ T_{gR}(1,\mu)T_{gR}(1,\mu)T_{gR}(1,\mu)\, \mathcal{D}_{1-d,g}^{\rm DGLAP}(\vec{n}_1)\mathcal{D}_{1-d,g}^{\rm DGLAP}(\vec{n}_2)\mathcal{D}_{1-d,g}^{\rm DGLAP}(\vec{n}_3) \nn\\
&+ \tilde{T}_{gR}(2,\mu)\,T_{gR}(1,\mu)\Bigg[\mathcal{D}_{-d,g}^{\rm DGLAP}(\vec{n}_1)\mathcal{D}_{1-d,g}^{\rm DGLAP}(\vec{n}_3)\, \delta^{(d-2)}(\vec{n}_1- \vec{n}_2)\nn\\
&\hspace{3.55cm}+\mathcal{D}_{-d,g}^{\rm DGLAP}(\vec{n}_1)\mathcal{D}_{1-d,g}^{\rm DGLAP}(\vec{n}_2)\, \delta^{(d-2)}(\vec{n}_1- \vec{n}_3)\,\nn\\
&\hspace{3.55cm}+\mathcal{D}_{-d,g}^{\rm DGLAP}(\vec{n}_2)\mathcal{D}_{1-d,g}^{\rm DGLAP}(\vec{n}_1)\, \delta^{(d-2)}(\vec{n}_2- \vec{n}_3)\Bigg]\nn\\
&+\tilde{T}_{gR}(3,\mu)\, \mathcal{D}_{-1-d,g}^{\rm DGLAP}(\vec{n}_1)\, \delta^{(d-2)}(\vec{n}_1- \vec{n}_2)\,\delta^{(d-2)}(\vec{n}_2- \vec{n}_3)\,,
\end{align}
where the scale $\mu$ of the UV detectors is suppressed for brevity.

The inclusive unit-weight limit makes the cancellations in this matching explicit. For $R=\forall$, momentum conservation fixes $T_{g\forall}(z,\mu)=\delta(1-z)$, and hence all of its ordinary, undiagonalized moments satisfy $T_{g\forall}(n,\mu)=1$. The shift symmetry implies that the diagonalized second and third moments are the corresponding central moments, so that
\begin{align}
F_g^{\forall\forall}(1-d,1-d,\mu)+F_g^{\forall}(-d,\mu)
&=\tilde{T}_{g\forall}(2,\mu)\,,\\
F_g^{\forall}(-1-d,\mu)
+3F_g^{\forall\forall}(-d,1-d,\mu)
+F_g^{\forall\forall\forall}(1-d,1-d,1-d,\mu)
&=\tilde{T}_{g\forall}(3,\mu)\,.\nonumber
\end{align}
where
\begin{align}
\tilde{T}_{g\forall}(2,\mu)
&=T_{g\forall}(2,\mu)-T_{g\forall}(1,\mu)^2=1-1=0\,,\\
\tilde{T}_{g\forall}(3,\mu)
&=T_{g\forall}(3,\mu)
-3T_{g\forall}(2,\mu)T_{g\forall}(1,\mu)
+2T_{g\forall}(1,\mu)^3=1-3+2=0\,.\nonumber
\end{align}
The first equality shows that every pair-contact coefficient in~\eq{EEEmatchtrack} vanishes, while the second shows that the three-detector coefficient vanishes as well. Since $T_{g\forall}(1,\mu)=F_g^\forall(1-d,\mu)=1$, only the separately matched product of UV detectors remains. More generally, the deterministic track function $T_{g\forall}(z,\mu)=\delta(1-z)$ has no fluctuations, so the analogous higher-point combinations associated with unresolved subsets vanish. Thus products of inclusive unit-weight detectors match directly onto products of ANE detectors, without additional hadronization-matching contact corrections.

In summary, the standard track function formalism provides a convenient way to package the combinations of multi-hadron fragmentation-function moments needed for the matching of multi-point products of IR detectors of the form
\begin{align}
\label{eq:multitrackIR}
\mathcal{E}_{R}(\vec{n}_1) \mathcal{E}_{R}(\vec{n}_2) \cdots \mathcal{E}_{R}(\vec{n}_N)\,,
\end{align}
where all the IR detectors carry the same quantum number $R$ and unit energy weight $m=1$. This is not surprising. The track function formalism was necessary for the recent calculations of energy correlators measured on tracks~\cite{Chen:2020vvp,Jaarsma:2023ell}. The natural question is whether the formalism can be generalized to efficiently package the combinations of multi-hadron fragmentation-function moments required to match the general correlators of the form
\begin{align}
\label{eq:multigenIR}
\mathcal{E}_{R_1}^{m_1}(\vec{n}_1) \mathcal{E}_{R_2}^{m_2}(\vec{n}_2)\cdots \mathcal{E}_{R_N}^{m_N}(\vec{n}_N)\,.
\end{align}
For unit energy weights but distinct quantum numbers $R_i$, this generalization is provided by the ``joint track functions'' introduced in~\cite{Lee:2023npz,Lee:2023tkr}. Before discussing the fully general case~\eq{multigenIR}, we first consider in the next subsection the simpler generalization in which all detectors carry a common quantum number $R$ and a common energy weight $m$, generalized away from unity,
\begin{align}
\label{eq:samemIR}
\mathcal{E}_{R}^m(\vec{n}_1) \mathcal{E}_{R}^m(\vec{n}_2)\cdots \mathcal{E}_{R}^m(\vec{n}_N)\,.
\end{align}

\subsection{One-dimensional Generalized Track Functions $T_{iR}^m(z_m,\mu)$}
\label{sec:gentrack1d}

To describe detectors with general energy weight $m$, we must generalize the track functions to probability distributions in which the momentum fractions of the hadrons are raised to the power $m$ before being summed. This is achieved simply by replacing the measurement $\delta(z-\sum_{h\in R} p_h^-/k^-)$ in~\eqs{trackdefg}{trackdefq} with $\delta(z_m-\sum_{h\in R} (p_h^-/k^-)^m)$. In this subsection we define these one-dimensional generalized track functions, relate their moments to the multi-hadron fragmentation functions, and study their RG evolution, in particular the loss of the shift symmetry away from $m=1$ and the resulting rapid decay of their moments. We then show how their diagonalized moments reproduce the matching of Sec.~\ref{sec:matchmultiple} for IR detectors carrying a common quantum number $R$ and a common energy weight $m$.

\subsubsection{Definition and Relation to Multi-Hadron Fragmentation Functions}
\label{sec:gentrackdef}

The one-dimensional generalized track functions $T_{iR}^m(z_m,\mu)$ are defined as
\begin{align}
\label{eq:gentrackdef}
T^m_{gR}(z_m,\mu)=&\ \frac{-1}{(d-2)\left(N_c^2-1\right) k^{-}}\int \mathrm{d} y^{+} \mathrm{d}^{d-2} y_{\perp}\, e^{i k^{-} y^{+} / 2} \nn\\
&\times \sum_X \langle 0| G_{-\lambda}^a\left(y^{+}, 0, y_{\perp}\right)|X\rangle\langle X| G_{-}^{\lambda, a}(0)|0\rangle\,  \delta\left(z_m-\sum_{h\in R} (p_h^-/k^-)^m\right) \,,\\
T^m_{qR}(z_m,\mu)=&\ \int \mathrm{d} y^{+} \mathrm{d}^{d-2} y_{\perp}\, e^{i k^{-} y^{+} / 2} \nn\\
&\times\sum_X
\frac{1}{2 N_c} \operatorname{tr}\left[\frac{\gamma^{-}}{2}\langle 0| \psi\left(y^{+}, 0, y_{\perp}\right)|X\rangle\langle X| \bar{\psi}(0)|0\rangle\right]  \delta\left(z_m-\sum_{h\in R} (p_h^-/k^-)^m\right) \,,\nn
\end{align}
where, as in~\eqs{trackdefg}{trackdefq}, we work in light-cone gauge to suppress gauge links. For $m=1$ they reduce to the track functions of Sec.~\ref{sec:trackreview}. Like the track functions, the generalized track functions are probability distributions, normalized as $\int dz_m\, T^m_{iR}(z_m,\mu)=1$, with support $z_m\in[0,1]$ for $m\geq 1$. An essential difference from the $m=1$ case, however, is that the generalized track functions remain nontrivial even when the measurement includes all hadrons, $R=\forall$: momentum conservation fixes the total momentum fraction $\sum_h z_h = 1$, but the weighted sum $z_m=\sum_h z_h^m$ depends on how this momentum is apportioned among the individual hadrons. For $m\neq 1$ the generalized track functions are therefore genuinely nonperturbative distributions for every choice of $R$, including $R=\forall$.

We define the moments of the generalized track functions as
\begin{align}
\label{eq:gentrackmom}
T^{m}_{iR} (k,\mu) \equiv \int dz_m\,z_m^k\, T^{m}_{iR}(z_m,\mu)\,,
\end{align}
which are normalized as $T^m_{iR}(0,\mu)=1$. Repeating the multinomial expansion of~\eq{multinomexp} with $z_h\to z_h^m$ (and relabeling its moment index and partition indices as $n\to k$ and $m_j\to k_j$, to avoid confusion with the energy weight $m$), the moments are again expressed in terms of the multi-hadron fragmentation function moments of~\eq{multihadmom},
\begin{align}
\label{eq:gentrackNhad}
T^m_{iR}(k,\mu) = \sum_{M=1}^{k}\, \frac{1}{M!}\sum_{\substack{k_j\in\mathbb{N}^+ \\ k_1+\cdots+k_M =k}} \frac{k!}{k_1!\cdots k_M!}\; D_i^{R\cdots R}(m\,k_1,\ldots,m\,k_M,\mu)\,,
\end{align}
which reduces to~\eq{trackNhad} at $m=1$. The $k$-th moment thus again contains up to $k$-hadron fragmentation functions, now evaluated at energy weights that are multiples of $m$. The first few moments are given explicitly by
\begin{align}
\label{eq:gentracklowmom}
T^m_{iR}(1,\mu) &= D_i^{R}(m,\mu)\,,\nn\\
T^m_{iR}(2,\mu) &= D_i^{R}(2m,\mu) + D_i^{RR}(m,m,\mu)\,,\nn\\
T^m_{iR}(3,\mu) &= D_i^{R}(3m,\mu) + 3\,D_i^{RR}(2m,m,\mu) + D_i^{RRR}(m,m,m,\mu)\,.
\end{align}
In particular, the first moment,
\begin{align}
\label{eq:gentrackfirstmom}
T^m_{iR}(1,\mu) = D_i^{R}(m,\mu) = F_i^{R}(2-d-m,\mu)\,,
\end{align}
is precisely the single-hadron detector function of~\eq{F1} that matches a single IR detector $\mathcal{E}^m_R$ onto the leading-twist UV detector in~\eq{singleHad}.

\subsubsection{Renormalization Group Evolution and the Loss of Shift Symmetry}
\label{sec:gentrackRG}

The evolution kernels are properties of different perturbative $1\to M$ splittings, and are independent of the measurement performed on the hadronic final state. The RG evolution of the generalized track functions is therefore governed by the same kernels $K_{i\to i_1 i_2\cdots i_M}$ as for the track functions in~\eq{trackRGE}, known to NLO~\cite{Chen:2022muj,Chen:2022pdu},
\begin{align}
\label{eq:gentrackRGE}
\frac{d}{d \ln \mu^2} T^m_{iR}(z_m,\mu)=\sum_{\left\{i_f\right\}} \sum_{\ell=0}^{\infty} a_s^{\ell+1}\, K_{i \rightarrow\left\{i_f\right\}}^{(\ell)} \otimes \prod_{j \in\left\{i_f\right\}}  T^m_{jR}\,.
\end{align}
The only modification relative to the track function case is the convolution structure, which now builds up the measured variable of the parent parton from those of the daughter partons with the momentum fractions raised to the power $m$. For example, the convolutions from~\eq{trackconvol} becomes
\begin{align}
\label{eq:gentrackconvol}
 K_{i \rightarrow i_1 i_2} \otimes T^m_{i_1R} T^m_{i_2R}\, (z_m)
 =&\int_0^1 dx_1\, dx_2\ \delta\left(1-x_1-x_2\right) K_{i \rightarrow i_1 i_2}\left(x_1, x_2\right) \\
 &\hspace{-4cm}\times \int dy_1\,dy_2 \,\delta (z_m-x_1^m y_{1m}-x_2^m y_{2m})\,T^m_{i_1R}(y_{1m})\, T^m_{i_2R}(y_{2m})\,,\nonumber\\
  K_{i \rightarrow i_1 i_2 i_3} \otimes T_{i_1R}^mT_{i_2R}^mT_{i_3R}^m\,(z_m) 
=&\int_0^1 dx_1\, dx_2\, dx_3\ \delta\left(1-x_1-x_2-x_3\right)K_{i \rightarrow i_1 i_2 i_3}\left(x_1, x_2, x_3\right)\nonumber\\
&\hspace{-4cm} \times \int_0^1 dy_1\,dy_2\,dy_3\ \delta (z_m-x_1^m y_{1m}-x_2^m y_{2m}-x_3^m y_{3m})\, T_{i_1R}^m(y_{1m})\,T_{i_2R}^m(y_{2m})\,T_{i_3R}^m(y_{3m})\,,\nonumber
\end{align}
such that the measured variable is built up as
\begin{align}
\label{eq:gentrackzbuild}
z_m = \sum^M_{j=1} x_j^m\, y_{jm}\,,\qquad \sum_{j=1}^M x_j = 1\,.
\end{align}
Taking moments of~\eq{gentrackRGE}, the triangular structure of the RG evolution kernel persists: the evolution of the $k$-th moment involves the moment itself, together with products of lower moments whose orders sum to $k$. The linear term is again governed by the timelike DGLAP anomalous dimensions, now evaluated at arguments spaced in units of the energy weight, since the momentum fractions of the daughter partons enter raised to the power $m$,
\begin{align}
\label{eq:gentrackmomlinear}
\frac{d}{d\ln\mu^2} T^m_{iR}(k,\mu) \supset -\gamma^T_{ji}(km+1)\,  T^m_{jR}(k,\mu)\,,
\end{align}
with an implicit sum over the parton flavor $j$.

The crucial structural difference from the track function case is the lack of shift symmetry. Momentum conservation fixes $\sum_j x_j=1$ but not $\sum_j x_j^m$, so the simultaneous shift $z_m\to z_m + a$, $y_{jm}\to y_{jm}+a$ changes~\eq{gentrackzbuild} by $a\,(\sum_j x_j^m - 1)\neq 0$ for $m\neq 1$. Therefore, the shift symmetry of the evolution is lost. Correspondingly, no nontrivial moment of the generalized track functions is protected under evolution and only the normalization $T^m_{iR}(0,\mu)=1$ is preserved. In particular, the first moment now evolves with the nonvanishing anomalous dimension $\gamma^T_{ji}(m+1)$ and, for $m>1$, decays away towards the UV, in stark contrast to the usual track function case, where the difference between the quark and gluon first moments decays and they approach a common constant. In the deep UV the generalized track functions with $m>1$ therefore collapse onto a delta function at the origin, $T^m_{iR}(z_m,\mu)\to \delta(z_m)$. This implies that, unlike for track functions, the RG evolution of the generalized track functions can significantly modify the angular dependence of the associated correlators.

\subsubsection{Rapid Decay under Renormalization Group Evolution}
\label{sec:gentrackdecay}

\begin{figure}[t!]
\begin{center}
\includegraphics[width =4 in]{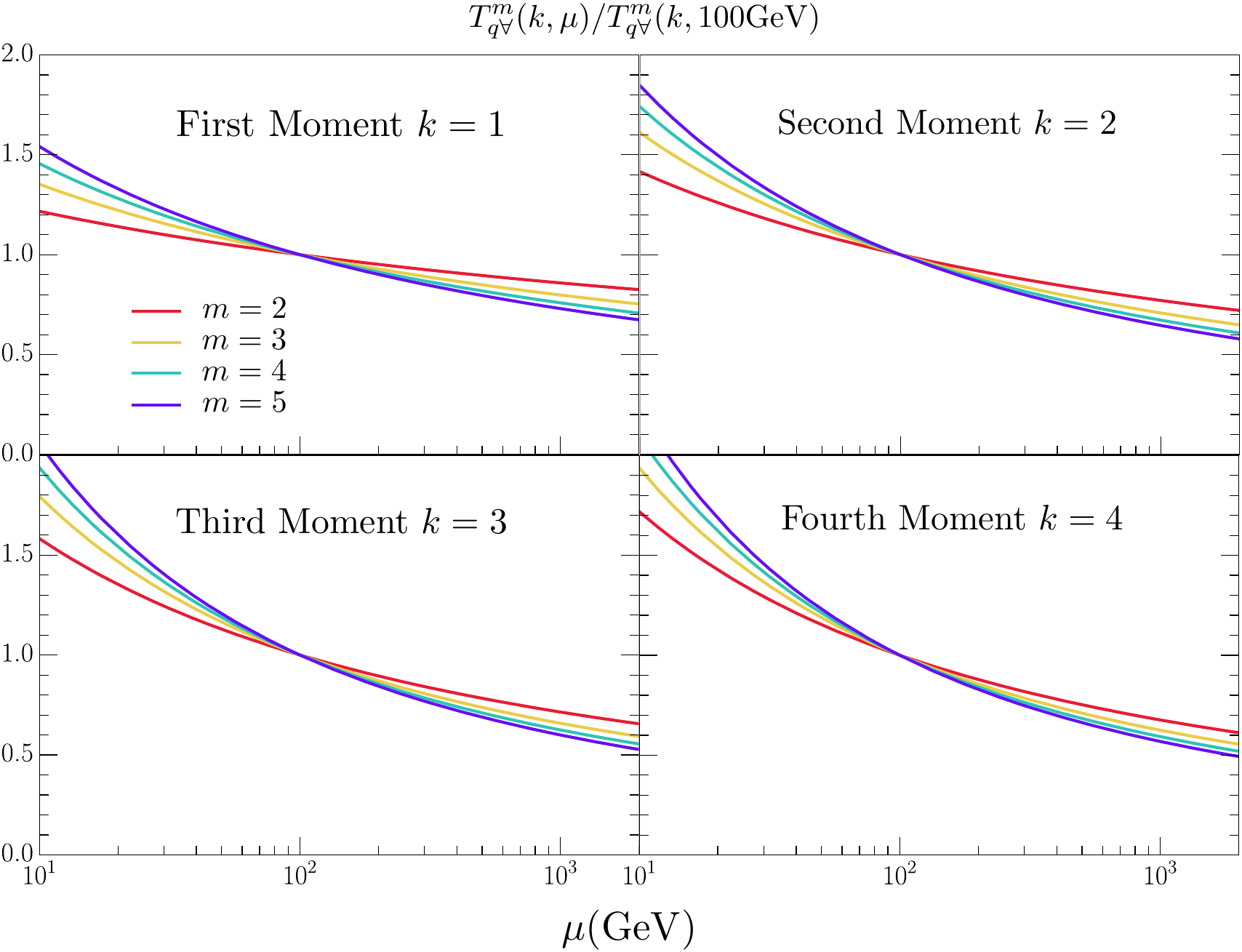}
\end{center}
\caption{The LL evolution of the first four quark moments, $k=1,2,3,4$, of the one-dimensional generalized track functions for energy weights $m=2,3,4,5$, normalized to their values at $\mu_0=100~\rm{GeV}$. The moments evolve rapidly as a function of the scale, particularly for the higher energy weights.}\label{fig:gentrackmom}
\end{figure}

To gain intuition for the size of these effects, we study the leading logarithmic (LL) evolution of the moments of the generalized track functions in full QCD. For simplicity, we display the evolution equations only for the first two moments, while the numerical study below includes the first four. Taking moments of~\eq{gentrackRGE}, the LL evolution equations read
\begin{align}
\label{eq:gentrackLL}
\frac{d}{d\ln\mu^2}T_{qR}^m(1,\mu) =& - \gamma_{qq}^{T(0)}(m+1)\, T_{qR}^m(1,\mu) - \gamma_{gq}^{T(0)}(m+1)\, T_{gR}^m(1,\mu)\,,\nn\\
\frac{d}{d\ln\mu^2}T_{gR}^m(1,\mu) =& - \gamma_{gg}^{T(0)}(m+1)\, T_{gR}^m(1,\mu) - 2n_f\, \gamma_{qg}^{T(0)}(m+1)\, T_{qR}^m(1,\mu)\,,\nn\\
\frac{d}{d\ln\mu^2}T_{qR}^m(2,\mu) =& - \gamma_{qq}^{T(0)}(2m+1)\, T_{qR}^m(2,\mu) - \gamma_{gq}^{T(0)}(2m+1)\, T_{gR}^m(2,\mu) \nn\\
&+ 2\int_0^1 dx_1 dx_2\ \delta(1-x_1-x_2)\, x_1^m\,x_2^m\, K^{(0)}_{q\to qg}(x_1,x_2)\; T_{qR}^m(1,\mu)\,T_{gR}^m(1,\mu)\,,\nn\\
\frac{d}{d\ln\mu^2}T_{gR}^m(2,\mu) =& - \gamma_{gg}^{T(0)}(2m+1)\, T_{gR}^m(2,\mu) - 2n_f\, \gamma_{qg}^{T(0)}(2m+1)\, T_{qR}^m(2,\mu) \nn\\
&+ 2\int_0^1 dx_1 dx_2\ \delta(1-x_1-x_2)\, x_1^m\,x_2^m \Big[K^{(0)}_{g\to gg}(x_1,x_2)\, T_{gR}^m(1,\mu)^2 \nn\\
&\hspace{4.9cm}+ n_f\, K^{(0)}_{g\to q\bar{q}}(x_1,x_2)\, T_{qR}^m(1,\mu)^2\Big]\,,
\end{align}
where the superscript $(0)$ denotes the leading order anomalous dimensions and kernels, $K^{(0)}_{g\to q\bar{q}}$ is defined per quark flavor, and we have taken the generalized track functions to be identical for all light quark and antiquark flavors. Higher moments take a similar form. Note that the $g\to gg$ mixing term in the second moment is precisely the kernel $\hat{\gamma}_g(m,m)$ of~\eq{mixingkernel}: the coefficient of $T_{gR}^m(1,\mu)^2$ in~\eq{gentrackLL} and the kernel $\hat{\gamma}_g(m,m)$ both equal $2\int_0^1 dx\, x^m (1-x)^m\, \hat{P}^T_{g\to gg}(x)$, with $K^{(0)}_{g\to gg}$ carrying the same identical-gluon symmetry factor as $\hat{P}^T_{g\to gg}$; in other words, at this order, $\hat{P}^T_{g\to gg}(x) = a_s\, K^{(0)}_{g\to gg}(x,1-x)$.

To study this evolution numerically, we extract the first four moments using Pythia with LO matching in the $e^+e^-$ process at $\mu_0 = Q = 100~\rm{GeV}$. For simplicity, we additionally set the gluon and quark moments equal. The extracted moments at $\mu_0=100~\rm{GeV}$ for $R=\forall$ are then given by
\begin{align}
\label{eq:gentrackpythia}
\{T^m_{\forall}(1,\mu_0),\,&T^m_{\forall}(2,\mu_0),\,T^m_{\forall}(3,\mu_0),\,T^m_{\forall}(4,\mu_0)\} \nn\\
=&\{0.173,\,0.0405,\,0.0127,\,5.12\times 10^{-3}\}\quad &\text{for }m&=2\,,\nn\\
=&\{0.0536,\,7.97\times 10^{-3},\,2.38\times 10^{-3},\,1.06\times 10^{-3}\}\quad &\text{for }m&=3\,,\nn\\
=&\{0.0226,\,3.04\times 10^{-3},\,9.53\times 10^{-4},\,4.36\times 10^{-4}\}\quad &\text{for }m&=4\,,\nn\\
=&\{0.0115,\,1.53\times 10^{-3},\,4.93\times 10^{-4},\,2.28\times 10^{-4}\}\quad &\text{for }m&=5\,,
\end{align}
where we drop the partonic label since the quark and gluon moments are taken equal in the extraction. Already at the input scale the moments are far below unity, even though no restriction is imposed on the hadrons ($R=\forall$). For $m\neq 1$ the measurement is sensitive to how the momentum is shared among the hadrons, and the sizable hadron multiplicity strongly suppresses the weighted moments.

\fig{gentrackmom} shows the LL evolution of these moments across a wide range of scales, normalized to their values at $\mu_0=100~\rm{GeV}$. Note that, while the quark and gluon moments are set equal at the input scale in~\eq{gentrackpythia}, this equality is not preserved by the evolution: away from $\mu_0$, the quark and gluon components evolve differently and mix according to~\eq{gentrackLL}. In \fig{gentrackmom}, we display the quark moments $T^m_{q\forall}(k,\mu)$. As anticipated from the loss of the shift symmetry, the moments decay rapidly as a function of the scale, with both higher energy weights $m$ and higher moments exhibiting faster decay rates. We leave a precision extraction and evolution of the generalized track functions to future work.

\subsubsection{Generalized Track Function Moments as Detector Matching Functions}
\label{sec:gentrackmatch}

We now show how the moments of the generalized track functions reproduce the combinations of detector functions that appear in the matching of Sec.~\ref{sec:matchmultiple} for products of IR detectors $\mathcal{E}_R^m$, i.e.~\eq{samemIR}. As in Sec.~\ref{sec:trackmatch}, we restrict to the pure YM case. The logic is identical to the $m=1$ case; however, in the absence of the shift symmetry, the mixing terms are no longer related to the timelike anomalous dimensions.

Applying the evolution equations \eqs{Tevol}{dihadronmomevol} to the moment relations~\eq{gentracklowmom}, the first and second moments of $T^m_{gR}(z_m,\mu)$ evolve as the closed triangular system
\begin{align}
\label{eq:gentrackmatrix2}
\hspace{-1cm}\frac{d}{d \ln \mu^2}\left(\begin{array}{c}
T^m_{gR}(2) \\
T^m_{gR}(1)^2
\end{array}\right)=
\left(\begin{array}{cc}
-\gamma_{T}(2m+1) & \hat{\gamma}_g(m,m)\\
0& -2\gamma_{T}(m+1)
\end{array}\right)
\left(\begin{array}{c}
T^m_{gR}(2) \\
T^m_{gR}(1)^2
\end{array}\right)\,,
\end{align}
generalizing~\eq{trackmatrix2}, where we again suppress the $\mu$ dependence for brevity. The diagonal entries are fixed by the timelike anomalous dimensions, in accordance with~\eq{gentrackmomlinear}, and the mixing entry is the kernel $\hat{\gamma}_g(m,m)$ of~\eq{mixingkernel}. For $m\neq 1$, however, there is no analogue of the shift symmetry relations~\eq{trackshiftvals}: $\gamma_T(m+1)\neq 0$, and $\hat{\gamma}_g(m,m)$ is an independent quantity that is not fixed by the timelike anomalous dimensions.

The linear combination that evolves purely multiplicatively is, at fixed coupling,
\begin{align}
\label{eq:gentracktildeT2}
\tilde{T}^m_{gR}(2) = T^m_{gR}(2) - \frac{\hat{\gamma}_g(m,m)}{\gamma_T(2m+1)-2\gamma_T(m+1)}\,T^m_{gR}(1)^2\,,
\end{align}
in terms of which the system~\eq{gentrackmatrix2} diagonalizes,
\begin{align}
\label{eq:gentrackdiag2}
\hspace{-1cm}\frac{d}{d \ln \mu^2}\left(\begin{array}{c}
\tilde{T}^m_{gR}(2) \vspace{0.05cm}\\
T^m_{gR}(1)^2
\end{array}\right)=
\left(\begin{array}{cc}
-\gamma_{T}(2m+1) & 0\\
0& -2\gamma_{T}(m+1)
\end{array}\right)
\left(\begin{array}{c}
\tilde{T}^m_{gR}(2) \vspace{0.05cm}\\
T^m_{gR}(1)^2
\end{array}\right)\,.
\end{align}
For $m=1$ this reduces to \eqs{trackeigen2}{trackdiag2}. For $m\neq 1$, however, $\tilde{T}^m_{gR}(2)$ is no longer a central moment, and for $m>1$ both $T^m_{gR}(1)^2$ and $\tilde{T}^m_{gR}(2)$ decay under evolution towards the UV. Substituting the moment relations~\eq{gentracklowmom} into~\eq{gentracktildeT2} and comparing with the diagonalized di-hadron moment~\eq{dihadrondiag} at $m_1=m_2=m$, we find
\begin{align}
\label{eq:gentracktildeT2frag}
\tilde{T}^m_{gR}(2,\mu) &= D_g^{R}(2m,\mu) + D_g^{RR}(m,m,\mu) - \frac{\hat{\gamma}_g(m,m)}{\gamma_T(2m+1)-2\gamma_T(m+1)}\, D_g^{R}(m,\mu)^2\nn\\
&= D_g^{R}(2m,\mu) + \tilde{D}_g^{RR}(m,m,\mu)\nn\\
&= F_g^{R}(2-d-2m,\mu) + F_g^{RR}(2-d-m,2-d-m,\mu)\,,
\end{align}
which is exactly the matching coefficient of the contact term in~\eq{twoHad} for $m_1=m_2=m$ and $R_1=R_2=R$. Together with the first moment~\eq{gentrackfirstmom} for the well-separated matching, the two-detector matching~\eq{twoHad} can thus be written compactly in terms of generalized track function moments as
\begin{align}
\label{eq:EmEmmatch}
\mathcal{E}^m_{R}(\vec{n}_1)\, \mathcal{E}^m_{R}(\vec{n}_2) =&\ T^m_{gR}(1,\mu)\,T^m_{gR}(1,\mu)\, \mathcal{D}_{2-d-m,g}^{\rm DGLAP}(\vec{n}_1,\mu)\,\mathcal{D}_{2-d-m,g}^{\rm DGLAP}(\vec{n}_2,\mu) \nn\\
&+ \tilde{T}^m_{gR}(2,\mu)\,
\mathcal{D}_{2-d-2m,g}^{\rm DGLAP}(\vec{n}_1,\mu)\, \delta^{(d-2)}(\vec{n}_1- \vec{n}_2)\,,
\end{align}
generalizing~\eq{EEmatchtrack} to arbitrary energy weight, where the boost weights $2-d-m$ and $2-d-2m$ follow from \eqs{JLmrel}{Jtilde12}.

For the matching of three IR detectors, we need the third moments. Applying the evolution of the fragmentation function moments to the third-moment relation of~\eq{gentracklowmom}, the evolution closes onto the triangular system
\begin{align}
\label{eq:gentrackmatrix3}
\frac{d}{d \ln \mu^2}&\left(\begin{array}{c}
T_{gR}^{m}(3) \\
T_{gR}^{m}(2)\, T_{gR}^{m}(1)\\
T_{gR}^{m}(1)^3
\end{array}\right)\nn\\
&=
\left(\begin{array}{ccc}
-\gamma_{T}(3m+1) & \gamma_c^m& \gamma_b^m \\
0& -\gamma_{T}(m+1)-\gamma_{T}(2m+1) & \hat{\gamma}_g(m,m)  \\
0& 0 &-3\gamma_{T}(m+1)
\end{array}\right)
\left(\begin{array}{c}
T_{gR}^{m}(3) \\
T_{gR}^{m}(2)\, T_{gR}^{m}(1)\\
T_{gR}^{m}(1)^3
\end{array}\right)\,.
\end{align}
At leading order the mixing entries are again fixed by moments of the splitting functions: $\gamma_c^m = 3\,\hat{\gamma}_g(2m,m)$, while $\gamma_b^m$ vanishes, since a $1\to 2$ splitting cannot produce three independent factors of $T^m_{gR}(1)$; a nonzero $\gamma_b^m$ first arises at NLO through the $1\to 3$ kernels.\footnote{For $m=1$, the shift symmetry fixes $\gamma_b^{m=1} = 3\gamma_T(3)-2\gamma_T(4)$ and $\gamma_c^{m=1} = 3\gamma_T(4)-3\gamma_T(3)$ in terms of the timelike anomalous dimensions of pure YM~\cite{Jaarsma:2022kdd,Li:2021zcf}. This is consistent with the leading order identifications: explicitly, $\hat{\gamma}_g(2,1)=\gamma_T(4)-\gamma_T(3)$ at LO, while the vanishing of $\gamma_b^{m=1}$ at this order is ensured by the nontrivial relation $2\gamma_T(4)=3\gamma_T(3)$ satisfied by the LO timelike anomalous dimensions of pure YM.}
Diagonalizing~\eq{gentrackmatrix3}, again at fixed coupling, the combination
\begin{align}
\label{eq:gentracktildeT3}
\tilde{T}^m_{gR}(3) = T^m_{gR}(3) + a_m\, T^m_{gR}(2)\,T^m_{gR}(1) + b_m\, T^m_{gR}(1)^3\,,
\end{align}
with
\begin{align}
\label{eq:gentracktildeT3coeffs}
a_m = \frac{\gamma_c^m}{\gamma_T(m+1)+\gamma_T(2m+1)-\gamma_T(3m+1)}\,,\qquad
b_m = \frac{\gamma_b^m + a_m\, \hat{\gamma}_g(m,m)}{3\gamma_T(m+1)-\gamma_T(3m+1)}\,,
\end{align}
evolves purely multiplicatively with $-\gamma_T(3m+1)$. For $m=1$, inserting the shift symmetry values gives $a_1=-3$ and $b_1=2$, so that $\tilde{T}_{gR}(3) = T_{gR}(3) - 3\,T_{gR}(2)T_{gR}(1) + 2\,T_{gR}(1)^3$ is precisely the third central moment of the track function, extending the pattern found for $\tilde{T}_{gR}(2)$ in Sec.~\ref{sec:trackmatch}. Plugging the multi-hadron moment relations~\eq{gentracklowmom} into the right-hand side of~\eq{gentracktildeT3}, we find
\begin{align}
\label{eq:gentracktildeT3frag}
\tilde{T}^m_{gR}(3,\mu) &= D_g^{R}(3m,\mu) + 3\,\tilde{D}_g^{RR}(2m,m,\mu) + \tilde{D}_g^{RRR}(m,m,m,\mu)\nn\\
&= F_g^{R}(2-d-3m,\mu) + 3\,F_g^{RR}(2-d-2m,2-d-m,\mu) \nn\\
&\hspace{0.42cm}+ F_g^{RRR}(2-d-m,2-d-m,2-d-m,\mu)\,,
\end{align}
which is precisely the matching coefficient of the fully unresolved contact term in~\eq{threeHad} for $m_i=m$ and $R_i=R$. The pair contact terms similarly carry the coefficient $\tilde{T}^m_{gR}(2)\,T^m_{gR}(1)$, and the three-detector matching takes exactly the form of~\eq{EEEmatchtrack}, with the replacements $T_{gR}\to T^m_{gR}$, $\tilde{T}_{gR}\to \tilde{T}^m_{gR}$, and the boost weights $1-d \to 2-d-m$, $-d \to 2-d-2m$, and $-1-d \to 2-d-3m$.

In summary, the moments of the one-dimensional generalized track functions, organized into the diagonalized combinations $\{T^m_{gR}(1),\, \tilde{T}^m_{gR}(2),\, \tilde{T}^m_{gR}(3),\,\ldots\}$, package precisely the combinations of multi-hadron fragmentation-function moments required for the matching of the correlators~\eq{samemIR}, generalizing the track function formalism of Sec.~\ref{sec:trackreview} to arbitrary energy weight. The remaining generalization, to products of detectors carrying different energy weights $m_a$ and different quantum numbers $R_a$ as in~\eq{multigenIR}, requires multi-dimensional generalized track functions, which we introduce next.

\subsection{Multi-dimensional Generalized Track Functions}
\label{sec:gentrackmulti}

The one-dimensional generalized track functions of Sec.~\ref{sec:gentrack1d} package the detector-function combinations needed to match products of \emph{identical} detectors $\mathcal{E}_R^m$. Our ultimate interest, however, is in the most general correlators of~\eq{generalcorr}, in which each detector carries its own energy weight $m_a$ and quantum number $R_a$. When an unresolved cluster contains detectors with different weights or quantum numbers, the corresponding contact terms in \eqs{twoHad}{threeHad} involve correlations \emph{between} the differently weighted momentum fractions, which cannot be captured by a distribution in a single variable. The required generalization is nevertheless immediate: we promote the generalized track function to a joint probability distribution, with one weighted momentum fraction for each pair $(R_a,m_a)$ appearing in the correlator.

The multi-dimensional generalized track functions $T^{(R_1,m_1)\cdots(R_N,m_N)}_i(z_{m_1},\ldots,z_{m_N},\mu)$ are defined from~\eq{gentrackdef}, for both quarks and gluons, simply by replacing the single measurement with a product of measurements,
\begin{align}
\label{eq:multidimdef}
\delta\bigg(z_m-\sum_{h\in R} (p_h^-/k^-)^{m}\bigg)\ \longrightarrow\ \prod_{a=1}^{N}\,\delta\bigg(z_{m_a}-\sum_{h\in R_a} (p_h^-/k^-)^{m_a}\bigg)\,.
\end{align}
They describe the joint probability distribution of the $N$ weighted momentum fractions $z_{m_a}=\sum_{h\in R_a} z_h^{m_a}$ in the fragmentation of a parton $i$, where $z_{m_a}$ denotes the variable of the $a$-th measurement even when several weights coincide. Integrating over any one variable reduces the distribution to the $(N-1)$-dimensional one with that measurement omitted, and integrating over all variables gives unity, as required of a joint probability distribution. For unit energy weights, $m_a=1$, they reduce to the joint track functions introduced in~\cite{Lee:2023npz,Lee:2023tkr}, while for $N=1$ they reduce to the one-dimensional generalized track functions of Sec.~\ref{sec:gentrack1d}. When all detectors carry a common quantum number $R$, we abbreviate
\begin{align}
\label{eq:gentrackReqT}
T_{iR}^{m_1 m_2\cdots m_N}(z_{m_1},z_{m_2},\ldots,z_{m_N},\mu)\equiv T^{(R,m_1)(R,m_2)\cdots(R, m_N)}_i(z_{m_1},z_{m_2},\ldots,z_{m_N},\mu)\,,
\end{align}
and when additionally all the energy weights coincide, the measurements all act on the same set of hadrons, so that the joint distribution collapses onto the diagonal,
\begin{align}
\label{eq:gentrackReqn}
T_{iR}^{m\,m\, \cdots\, m}(z_{m_1},z_{m_2},\ldots,z_{m_N},\mu) &= T^{m}_{iR}(z_{m_1},\mu)\,\delta(z_{m_1}-z_{m_2})\,\delta(z_{m_1}-z_{m_3})\cdots \delta(z_{m_1}-z_{m_N})\,.
\end{align}

We define the mixed moments of the multi-dimensional generalized track functions as
\begin{align}
\label{eq:multidimmom}
T^{(R_1,m_1)\cdots (R_N,m_N)}_i (k_1,\ldots, k_N,\mu) \equiv \int \prod_{a=1}^N \Big[dz_{m_a}\, z_{m_a}^{k_a}\Big]\; T^{(R_1,m_1)\cdots(R_N, m_N)}_i(z_{m_1},\ldots,z_{m_N},\mu)\,,
\end{align}
normalized as $T^{(R_1,m_1)\cdots (R_N,m_N)}_i(0,\ldots,0,\mu)=1$. Setting any $k_a=0$ removes the corresponding measurement, while on the diagonal, \eq{gentrackReqn} implies $T^{m\,m\,\cdots\, m}_{iR}(k_1,\ldots,k_N,\mu) = T^m_{iR}(k_1+\cdots+k_N,\mu)$. As in the one-dimensional case, expanding the powers of the weighted sums multinomially expresses the mixed moments in terms of the multi-hadron fragmentation function moments of~\eq{multihadmom}, now with independent energy weights distributed among the identified hadrons; the general relation is a straightforward, though notationally heavy, generalization of~\eq{gentrackNhad}, and we do not write it out. The moment needed for the matching of two detectors is
\begin{align}
\label{eq:multidimT11}
T^{(R_1,m_1)(R_2,m_2)}_i(1,1,\mu) = D_i^{R_1 R_2}(m_1,m_2,\mu) + \delta_{R_1 R_2}\, D_i^{R_1}(m_1+m_2,\mu)\,,
\end{align}
where the two terms arise from the two measurements acting on distinct hadrons or on the same hadron, respectively.\footnote{As implicitly assumed in~\eq{twoHad}, we take the quantum numbers $R_a$ to be either identical or disjoint. More generally, the same-hadron term is supported on the overlap of the two classes, $\delta_{R_1 R_2}\,D_i^{R_1}(m_1+m_2,\mu)\to D_i^{R_1\cap R_2}(m_1+m_2,\mu)$.}

The RG evolution follows the same pattern as in the one-dimensional case. The evolution kernels are again those of~\eq{trackRGE}, with the convolutions building up each measured variable additively from the daughter partons, $z_{m_a}=\sum_j x_j^{m_a}\, y_{j,a}$, in direct analogy with~\eq{gentrackzbuild}. Taking mixed moments, the evolution of a given moment involves products of lower mixed moments whose orders sum, in each slot, to those of the original moment, with the linear term governed by the timelike anomalous dimension at the total energy weight,
\begin{align}
\label{eq:multidimlinear}
\hspace{-0.7cm}\frac{d}{d\ln\mu^2}\, T^{(R_1,m_1)\cdots(R_N,m_N)}_{i}(k_1,\ldots, k_N,\mu) \supset -\gamma^T_{ji}\bigg(\sum_{a=1}^N k_a m_a+1\bigg)\,  T^{(R_1,m_1)\cdots(R_N,m_N)}_{j}(k_1,\ldots,k_N,\mu)\,,
\end{align}
and the shift symmetry is broken in every slot with $m_a\neq 1$; only when all $m_a=1$ does the full shift symmetry of the joint track functions survive. In particular, restricting to the pure YM case, the mixed moment~\eq{multidimT11} mixes only into the product of first moments $T^{m_1}_{gR_1}(1)\,T^{m_2}_{gR_2}(1)$, with the mixing entry given by the kernel $\hat{\gamma}_g(m_1,m_2)$ of~\eq{mixingkernel}, and the diagonalized combination, at fixed coupling, is given as
\begin{align}
\label{eq:tildeT11}
\tilde{T}^{(R_1,m_1)(R_2,m_2)}_g(1,1) &= T^{(R_1,m_1)(R_2,m_2)}_g(1,1) + \frac{\hat{\gamma}_g(m_1,m_2)\; T^{m_1}_{gR_1}(1)\,T^{m_2}_{gR_2}(1)}{\gamma_T(m_1+1)+\gamma_T(m_2+1)-\gamma_T(m_1+m_2+1)}\nn\\
&= \tilde{D}_g^{R_1 R_2}(m_1,m_2) + \delta_{R_1R_2}\, D_g^{R_1}(m_1+m_2)\nn\\
&= F_g^{R_1 R_2}(J_{L_1},J_{L_2}) + \delta_{R_1R_2}\, F_g^{R_1}(\tilde{J}_{L_{12}})\,,
\end{align}
This combination evolves purely multiplicatively with $-\gamma_T(m_1+m_2+1)$. In the second line we recognized the diagonalized di-hadron moment of~\eq{dihadrondiag} and used the map of Tab.~\ref{tab:map} with $J_{L_a}=2-d-m_a$ and $\tilde{J}_{L_{12}}=2-d-m_1-m_2$: this is precisely the full contact bracket of~\eq{twoHad}. In the common-$R$ shorthand of~\eq{gentrackReqT} we write $\tilde{T}^{m_1 m_2}_{gR}(1,1,\mu)\equiv \tilde{T}^{(R,m_1)(R,m_2)}_g(1,1,\mu)$, which for $m_1=m_2=m$ reduces to $\tilde{T}^{m}_{gR}(2,\mu)$ of~\eq{gentracktildeT2}. The matching of two arbitrary IR detectors can therefore be written compactly in terms of generalized track function moments as
\begin{align}
\label{eq:genEEmatch}
\mathcal{E}^{m_1}_{R_1}(\vec{n}_1)\, \mathcal{E}^{m_2}_{R_2}(\vec{n}_2) =&\ T^{m_1}_{gR_1}(1,\mu)\, T^{m_2}_{gR_2}(1,\mu)\, \mathcal{D}_{2-d-m_1,g}^{\rm DGLAP}(\vec{n}_1,\mu)\,\mathcal{D}_{2-d-m_2,g}^{\rm DGLAP}(\vec{n}_2,\mu) \nn\\
&+ \tilde{T}^{(R_1,m_1)(R_2,m_2)}_g(1,1,\mu)\,
\mathcal{D}_{2-d-m_1-m_2,g}^{\rm DGLAP}(\vec{n}_1,\mu)\, \delta^{(d-2)}(\vec{n}_1- \vec{n}_2)\,,
\end{align}
generalizing~\eq{EmEmmatch} to an arbitrary pair of detectors.

The same pattern extends to any number of detectors: every contact bracket in the matching of \eqs{twoHad}{threeHad}, and in their higher-multiplicity generalizations, is a diagonalized mixed moment of the multi-dimensional generalized track function, with the moment index set to one in each detector's slot. For instance, the fully unresolved bracket of~\eq{threeHad} is $\tilde{T}^{(R_1,m_1)(R_2,m_2)(R_3,m_3)}_g(1,1,1,\mu)$, while the pair contacts carry the coefficients $\tilde{T}^{(R_i,m_i)(R_j,m_j)}_g(1,1,\mu)\,T^{m_k}_{gR_k}(1,\mu)$. The multi-dimensional generalized track functions therefore supply, through their diagonalized mixed moments, the complete combinations of detector functions required for the matching of the most general correlators~\eq{generalcorr} considered in this paper. In Sec.~\ref{sec:fact}, we use these objects to present collinear factorization theorems for the $x_L$ projections of such correlators, and to compute the associated jet functions to NLO.

\section{Collinear Factorization of General Correlators}\label{sec:fact}
In this section, we derive QCD factorization theorems in the collinear limit for the projected correlators
\begin{align}
\label{eq:multixLcorr}
\langle \mathcal{E}_{R_1}^{m_1}(\vec{n}_1) \mathcal{E}_{R_2}^{m_2}(\vec{n}_2)\cdots \mathcal{E}_{R_N}^{m_N}(\vec{n}_N) \rangle_{x_L}\,,
\end{align}
where the $x_L$ projection onto the largest pairwise angle was defined in~\eq{xLmeas}. These factorization theorems fully incorporate quark-gluon mixing and the scale violation induced by the running of the coupling. The generalized track functions of Sec.~\ref{sec:GenTrack} enter as the nonperturbative boundary data of the factorization, and we compute the corresponding jet functions to NLO, providing the ingredients required for NLL resummation. For notational simplicity, we will always consider the case $R_1=R_2 = \cdots = R_N =R$. The more general case is easily restored using the map~\eq{gentrackReqT} and promoting each $(R,m_i)$ of the right-hand side to $(R_i,m_i)$.

In Sec.~\ref{sec:twopointfact}, we begin with the two-point correlators $\langle \mathcal{E}_R^n\, \mathcal{E}_R^m\rangle_{x_L}$. We present the factorization theorem, compute the NLO jet functions, and show, through a reciprocity argument, that when quark-gluon mixing and the running of the coupling are neglected, the resulting anomalous $x_L$ scaling exactly reproduces the predictions of the light-ray OPE analysis of Sec.~\ref{sec:LROPE}. In Sec.~\ref{sec:higherfact}, we generalize to the higher-point projected correlators and again compare with the light-ray OPE, presenting the NLO jet functions up to the four-point case.

\subsection{General Two-Point Energy Correlators}
\label{sec:twopointfact}

We begin by considering the two-point energy correlator of generalized detector operators, $\langle \mathcal{E}_R^n(\vec{n}_1) \mathcal{E}_R^m(\vec{n}_2) \rangle_{x_L}$, in a state produced by an $e^+e^-$ collision. Since our focus is on the modifications due to the generalized detectors, it is sufficient to consider the simplest case of $e^+e^-$ collisions to avoid irrelevant complications; the factorization presented here is easily generalized to $pp$ collisions following~\cite{Lee:2022uwt,Lee:2024icn}. 

The $x_L$ projection of the two-point correlator for $e^+e^-$ collisions can be written explicitly in terms of the four-point function involving the electromagnetic current $J^{\mu}=\bar{\psi}\gamma^\mu\psi$, sourcing a state with momentum $q=l+l'$, where $l^\mu$ and ${l'}^\mu$ are the incoming lepton momenta, as
\begin{align}
\label{eq:EnEmC}
\frac{1}{\sigma_{\rm tot}}\frac{d \sigma^{\langle\mathcal{E}_R^n \mathcal{E}_R^{m}\rangle}}{dx_L}
=& \frac{1}{\sigma}\int d^4 x \frac{e^{i q \cdot x}}{Q^{n+m}}\int d^2 \Omega_{\vec{n}_1}d^2 \Omega_{\vec{n}_2} \delta\left(x_L-\frac{1-\vec{n}_1 \cdot \vec{n}_2}{2}\right) \nn\\
\times&L_{\mu\nu}\langle 0|J^{\mu\dagger}(x) \mathcal{E}_R^{n}\left(\vec{n}_1\right) \mathcal{E}_R^{m}\left(\vec{n}_2\right) J^\nu(0)| 0\rangle\,,
\end{align}
where $Q=\sqrt{q^2}$ is the center-of-mass energy, $L^{\mu \nu}=l^\mu l^{\prime \nu}+l^{\prime \mu} l^\nu-\left(l \cdot l^{\prime}\right) g^{\mu \nu}$ is the leptonic tensor, $\sigma = L_{\mu \nu} \times \int d^4 x\, e^{i q \cdot x} \langle 0| J^{\mu \dagger}(x) J^\nu(0)|0\rangle$, and $\sigma_{\rm tot}=\sigma (4\pi\alpha_{\rm EM})^2/2Q^6$ is the total cross section.

Integrated over the full angular range, the correlator~\eq{EnEmC} yields
 \begin{align}
 \label{eq:enemnorm}
 \int dx_L\, \frac{1}{\sigma_{\rm tot}}\frac{d \sigma^{\langle\mathcal{E}_R^n \mathcal{E}_R^{m}\rangle}}{dx_L} = T_{R}^{n\,m}(1,1)\,,
 \end{align}
where the right-hand side is the event-level mixed moment, obtained by convolving the partonic generalized track function moments of Sec.~\ref{sec:gentrackmulti} with the hard function, and hence carrying no partonic label. For $n=m=1$ and $R=\forall$, energy conservation fixes $T_{\forall}^{1\,1}(1,1)=1$, while for generic energy weights or a restricted hadron class $R$, the integral is instead a nontrivial nonperturbative quantity.

To study the all-orders factorization in the collinear limit, it is convenient to work in terms of the cumulant distribution
\begin{align}
\Sigma^{\langle\mathcal{E}_R^n \mathcal{E}_R^{m}\rangle}\left(x_L, \frac{Q^2}{\mu^2}, \mu\right) &\equiv\frac{1}{\sigma_{0}} \int_0^{x_L} dx_L^{\prime} \frac{d \sigma^{\langle\mathcal{E}_R^n \mathcal{E}_R^{m}\rangle}}{dx_L^{\prime}}\,,
\end{align}
where $\sigma_0$ is the Born-level total cross section. We emphasize that the distribution~\eq{EnEmC} and its cumulant are defined and measurable over the entire angular range $0< x_L< 1$ in $e^+e^-$ collisions. In the collinear limit $x_L\to 0$, the cumulant can be factorized as
\begin{align}
\label{eq:Fact}
\Sigma^{\langle\mathcal{E}_R^n \mathcal{E}_R^{m}\rangle}\left(x_L, \frac{Q^2}{\mu^2}, \mu\right) 
&=\int_0^1 d x\, x^{n+m} \vec{J}^{\langle\mathcal{E}_R^n \mathcal{E}_R^{m}\rangle}\left(\frac{x_Lx^2Q^2}{\mu^2}, \mu\right) \cdot \vec{H}\left(x, \frac{Q^2}{\mu^2}, \mu\right)\,,
\end{align}
where $\vec{H}$ denotes the hard function that describes the production of the energetic partons in the underlying hard process, which is known to N$^3$LO~\cite{He:2025hin} for $e^+e^-$ collisions, and $\vec{J}^{\langle\mathcal{E}_R^n \mathcal{E}_R^{m}\rangle}$ denotes the new two-point jet function for the $\langle\mathcal{E}_R^n \mathcal{E}_R^{m}\rangle_{x_L}$ correlator. Both functions are vectors in flavor space. 

The hard function satisfies the standard timelike DGLAP equation
\begin{align}
\frac{\mathrm{d} \vec{H}\left(x, \frac{Q^2}{\mu^2}, \mu\right)}{\mathrm{d} \ln \mu^2}=-\int_x^1 \frac{\mathrm{~d} y}{y} \widehat{P}\left(y\right) \cdot \vec{H}\left(\frac{x}{y}, \frac{Q^2}{\mu^2}, \mu\right)\,,
\end{align}
where $\widehat{P}(y)$ is the singlet timelike splitting matrix, which is known to three loops~\cite{Mitov:2006ic,Mitov:2006wy,Moch:2007tx,Chen:2020uvt}. It is given by
\begin{align}
  \label{eq:splitK}
  \widehat P = 
  \begin{pmatrix}
    P_{qq} &\hspace{0.15cm}  P_{qg}
\\
    P_{gq} &\hspace{0.15cm} P_{gg}
  \end{pmatrix} \,.
\end{align}
Taking moments, the timelike DGLAP anomalous dimensions are defined as
\begin{equation}
\label{eq:gammaPTrelation}
\gamma_T(k)\ \equiv\ - \int_0^1 dy \, y^{k-1} \, \widehat{P}(y)\,,
\end{equation}
which is the flavor-matrix version of the timelike anomalous dimension of~\eqs{gammaTfromP}{Devol}, whose entries $\gamma^T_{ji}$ governed the evolution of the generalized track function moments in Sec.~\ref{sec:GenTrack}. 

The dependence on the specific choice of generalized detector lies entirely in the jet functions $\vec{J}^{\langle\mathcal{E}_R^n \mathcal{E}_R^{m}\rangle}$. These jet functions are universal in the sense that they also apply at hadron colliders, where the dependence on the different initial state is encoded in the hard function. We now discuss the jet functions for the correlator $\langle\mathcal{E}_R^n \mathcal{E}_R^{m}\rangle_{x_L}$ in detail.

\subsubsection{Jet Function for the Two-Point Correlators $\langle\mathcal{E}_R^n \mathcal{E}_R^{m}\rangle_{x_L}$}
\label{sec:TwoPTJet}
While the diagonalized combinations of Sec.~\ref{sec:GenTrack} make the anomalous scaling of the individual light-ray OPE structures transparent, for fixed-order calculations it is more convenient to present the jet functions directly in terms of the non-diagonalized moments of the generalized track functions, which we do in the following.

Using RG consistency, the jet function $\vec{J}^{\langle\mathcal{E}_R^n \mathcal{E}_R^{m}\rangle}$ satisfies the RG equation
\begin{align}
\label{eq:twojetRG}
\frac{d \vec{J}^{\langle\mathcal{E}_R^n \mathcal{E}_R^{m}\rangle}\left(\frac{x_L\, Q^2}{\mu^2}\right)}{d \ln \mu^2}=\int_0^1 dy\,y^{n+m}\, \vec{J}^{\langle\mathcal{E}_R^n \mathcal{E}_R^{m}\rangle}\left(\frac{x_L\,y^2 Q^2}{\mu^2}\right) \cdot \widehat{P}(y)\,.
\end{align}
 For each given flavor, we write the jet function as $J_i^{\langle\mathcal{E}_R^n \mathcal{E}_R^{m}\rangle} = \hat{J}_i^{\langle\mathcal{E}_R^n \mathcal{E}_R^{m}\rangle}\, +\, $p.c., where the additional term `$+$p.c.' represents nonperturbative power corrections, which will be discussed in Sec.~\ref{sec:NP}. 
 
To all orders, the perturbative jet functions take the form
\begin{align}
\label{eq:JetTwoAnsatz}
\hat{J}_{i}^{\langle\mathcal{E}_R^n \mathcal{E}_R^{m}\rangle}\left(\frac{x_L\, x^2 Q^2}{\mu^2}, \mu\right)=\sum_{a=0}^{\infty} \sum_{b=0}^a\left(\frac{\alpha_s}{4 \pi}\right)^a \ln^b \frac{\mu^2}{x_L x^2 Q^2} J_{i(a, b)}^{\langle\mathcal{E}_R^n \mathcal{E}_R^{m}\rangle}\,.
\end{align}
The log-enhanced terms are iteratively predicted by the RG evolution equation given in~\eq{twojetRG}. Each coefficient $J_{i(a, b)}^{\langle\mathcal{E}_R^n \mathcal{E}_R^{m}\rangle}$ will in general involve terms proportional to \{$T_{jR}^{n\,m}(1,1,\mu), T_{kR}^{n}(1,\mu) T_{lR}^{m}(1,\mu)$\}, where $j,k,l$ are some parton flavors. 

To achieve N$^r$LL resummation, we need the non-logarithmic jet function coefficients through $J_{i(r, 0)}^{\langle\mathcal{E}_R^n \mathcal{E}_R^{m}\rangle}$. It is straightforward to compute them to next-to-leading order (NLO).

The jet function at the tree level is given by\footnote{The overall normalization $2^{-(n+m)}$ follows from normalizing the energy weights by $Q^{n+m}$ in \eq{EnEmC}, rather than by $(Q/2)^{n+m}$.}
\begin{align}
\label{eq:Jzeroth}
2^{(n+m)}J_{i(0,0)}^{\langle\mathcal{E}_R^n \mathcal{E}_R^{m}\rangle}  =\,& T_{iR}^{n\,m}(1,1,\mu)\,,
\end{align}
and at one loop by
\begin{align}
\label{eq:JgTwo}
2^{(n+m)}J_{g(1,0)}^{\langle \cE_R^n \cE_R^m \rangle}& = \Bigg[\left(m^2+3 m+n^2+3 n+4\right) \psi ^{(0)}(n+1)-\left(2 n^2+6 n+4\right) \psi^{(0)}(m+n+4)  \\
&+\frac{m n}{2}+3 n+\frac{7}{2}\Bigg] \frac{8  n_f T_f\, n\, m\, \Gamma(n) \Gamma(m)}{\Gamma(n+m+4)} T_{qR}^n(1) T_{qR}^m(1) \nn \\
&+ \Bigg[\left(m^4+6 \left(m^3+n^3\right)+3 m^2 n^2+(2 m n+11) \left(m^2+n^2\right)+11 mn \right. \nn \\
&+ \left. (m+n) (9 m n+6)+n^4\right) (\psi ^{(0)}(n)-\psi ^{(0)}(m+n+4)) \nn \\
&+3 \left((6+2 (m+n))m n+2 n^3+9 n^2+11 n+2\right)\Bigg] \frac{8  C_A \Gamma(n) \Gamma(m)}{\Gamma(n+m+4)} T_{gR}^n(1) T_{gR}^m(1) \nn \\
&+ (n \leftrightarrow m)\,,\nn
\end{align}
for the gluon jet function and
\begin{align}
\label{eq:JqTwo}
&2^{(n+m)}J_{q(1,0)}^{\langle \cE_R^n \cE_R^m \rangle} = \Bigg[2 m \left(2 m^2+2 m (n+3)+n (n+3)+4\right) (-2 \psi ^{(0)}(m+n+3)+\psi^{(0)}(m)+\psi ^{(0)}(n)) \nn \\
&+16 m^2+m (n+3) (n+10)+2 n (n+3)+8\Bigg] \frac{2 C_F \Gamma(n) \Gamma(m)}{\Gamma(n+m+3)} T_{gR}^n(1) T_{qR}^m(1) + \left(n \leftrightarrow m\right)\,,
\end{align}
for the quark jet function. In these expressions, we suppressed the $\mu$ dependence of the $T_{iR}^n$ for convenience. When $n=m$, one can simply replace $T_{iR}^{n\,m}(1,1) \to T^{n}_{iR}(2)$ in the LO constant jet function, following~\eq{gentrackReqn}. The case $n=m=1$ was already calculated in~\cite{Chen:2020vvp}, and the results presented here reproduce it as a special case. Lastly, we emphasize that the explicit NLO calculation demonstrates that the IR poles are indeed absorbed by the one-loop renormalization of the generalized track functions.

\subsubsection{Reciprocity and Anomalous Scaling for the Two-Point Correlators $\langle\mathcal{E}_R^n \mathcal{E}_R^{m}\rangle_{x_L}$}
\label{sec:reciprocityTwo}
Combining the logarithmic terms iteratively determined from~\eq{JetTwoAnsatz} with the NLO jet functions of~\eqs{JgTwo}{JqTwo} and the NLO evolution of the multi-dimensional generalized track functions discussed in Sec.~\ref{sec:gentrackmulti}, we can carry out the NLL resummation of $\langle\mathcal{E}_R^n \mathcal{E}_R^{m}\rangle_{x_L}$. 

Using the reciprocity argument presented in~\cite{Dixon:2019uzg,Lee:2024icn}, we can gain intuition for what this resummation achieves. For simplicity, we restrict to a single flavor by considering pure YM at fixed coupling. We now parameterize our jet function as
\begin{align}
\label{eq:JetParamTwo}
\hat{J}_g^{\langle\mathcal{E}_R^n \mathcal{E}_R^{m}\rangle}(x_L Q^2,\mu) =&  T_{gR}^{n}(1,Q \sqrt{x_L})T_{gR}^{m}(1,Q \sqrt{x_L})C(\alpha_s) \left(\frac{x_L Q^2}{\mu^2}\right)^{\gamma_1}\nn\\
&+ \tilde{T}_{gR}^{n\,m}(1,1,Q \sqrt{x_L}) D(\alpha_s) \left(\frac{x_L Q^2}{\mu^2}\right)^{\gamma_2}\,,
\end{align}
where $\gamma_1$ and $\gamma_2$ are anomalous scaling exponents to be determined, and $\tilde{T}_{gR}^{n\,m}(1,1,Q \sqrt{x_L})$ is the diagonalized mixed moment of~\eq{tildeT11}, written in the common-$R$ shorthand of~\eq{gentrackReqT}, which evolves without non-linear mixing. This ansatz is motivated by the structure expected from the light-ray OPE, but it is not an input: the exponents follow from RG consistency alone. Note that we also evaluate the track moments at $\mu\sim Q\sqrt{x_L}$ to be consistent with the choice in \Sec{sec:LROPE}. Indeed, if we take $n=m$, then $\tilde{T}_{gR}^{n\,n}(1,1,Q \sqrt{x_L})\to \tilde{T}_{gR}^{n}(2,Q \sqrt{x_L})$ of~\eq{gentracktildeT2}. Note that these diagonalized moments evolve via
\begin{align}
\label{eq:Tgnm11Tgn1Tgm1}
\hspace{-1.6cm}\frac{d}{d \ln \mu^2}\left(\begin{array}{c}
T_{gR}^{n}(1)T_{gR}^{m}(1) \vspace{0.05cm}\\ 
\tilde{T}_{gR}^{n\,m}(1,1) 
\end{array}\right)=
\left(\begin{array}{ccccc}
-\gamma_{T}(n+1)-\gamma_{T}(m+1) & 0\\
0& -\gamma_{T}(n+m+1) 
\end{array}\right)
\left(\begin{array}{c}
T_{gR}^{n}(1)T_{gR}^{m}(1) \vspace{0.05cm}\\ 
\tilde{T}_{gR}^{n\,m}(1,1) 
\end{array}\right)\,,
\end{align}
where we suppressed the $\mu$ dependence for convenience. Then substituting~\eq{JetParamTwo} into~\eq{twojetRG}, we arrive at two consistency equations
\begin{align}
2\gamma_1 &= 2\gamma_T (n+m+1+2\gamma_1-2\gamma_{T}(n+1)-2\gamma_{T}(m+1))\,,\\
2\gamma_2 &= 2\gamma_T (n+m+1+2\gamma_2-2\gamma_{T}(n+m+1))\,,\nn
\end{align}
which imply 
\begin{align}
\gamma_1 &= \gamma_S(n+m+1-2\gamma_{T}(n+1)-2\gamma_{T}(m+1))\,,\\
\gamma_2 &= \gamma_{T}(n+m+1)\,.\nn
\end{align}
where we used the reciprocity relation $2\gamma_S(J) = 2\gamma_T(J+2\gamma_S(J))$. In particular, the term proportional to $T_{gR}^{n}(1,Q \sqrt{x_L})T_{gR}^{m}(1,Q \sqrt{x_L})$ has the anomalous scaling given by the light-ray operator with the spin `$n+m+1-2\gamma_{T}(n+1)-2\gamma_{T}(m+1)$' and the term proportional to $ \tilde{T}_{gR}^{n\,m}(1,1,Q \sqrt{x_L})$ has its $x_L$ scaling exactly cancelled between the diagonalized moment $\tilde{T}_{gR}^{n\,m}(1,1,Q \sqrt{x_L})$ and $\left(\frac{x_L Q^2}{\mu^2}\right)^{\gamma_2}$, implying that it is actually a contact term. The contact term is not explicitly proportional to $\delta(x_L)$, since our jet function arises from the factorization of the cumulant in~\eq{Fact}. This conclusion is consistent with the light-ray OPE picture for the two-point correlator: the anomalous scaling agrees with~\eq{gammaSJ'} under the identification $(m_1,m_2)=(n,m)$, and the contact structure agrees with the contact bracket of~\eq{twopointOPE}, whose coefficient is precisely the diagonalized moment $\tilde{T}_{gR}^{n\,m}(1,1)$ of~\eq{tildeT11}.

\subsection{General Higher-Point Projected Energy Correlators}
\label{sec:higherfact}

We now discuss the higher-point $x_L$-projected energy correlators of~\eq{multixLcorr}. Allowing for the possibility that several of the energy weights coincide, we consider
\begin{align}
\label{eq:gco}
\gco\equiv\langle\underbrace{\mathcal{E}_R^{n_1} \cdots \mathcal{E}_R^{n_1}}_{N_1\text{ times}}\underbrace{\mathcal{E}_R^{n_2} \cdots \mathcal{E}_R^{n_2}}_{N_2\text{ times}} \cdots \underbrace{\mathcal{E}_R^{n_M} \cdots \mathcal{E}_R^{n_M}}_{N_M\text{ times}}\rangle_{x_L}\,,
\end{align}
where we use $\gco$ as a shorthand notation for the general correlators involving $M$ distinct types of detectors $\mathcal{E}_R^{n_i}$, each appearing $N_i$ times ($1 \leq i \leq M$), with $N=N_1+N_2+\cdots +N_M$.

As in the two-point case, the cumulant of $\gco$ factorizes in the collinear limit into a jet function and a hard function,
\begin{align}
\label{eq:gcoFact}
\Sigma_\gco\left(x_L, \frac{Q^2}{\mu^2}, \mu\right) &\equiv\frac{1}{\sigma_0} \int_0^{x_L} dx_L^{\prime} \frac{d \sigma^\gco}{d x_L^{\prime}}\nonumber\\
&=\int_0^1 d x\, x^{n_1 N_1 + \cdots + n_M N_M} \vec{J}^{\,\gco}\left(\frac{x_Lx^2Q^2}{\mu^2}, \mu\right) \cdot \vec{H}\left(x, \frac{Q^2}{\mu^2}, \mu\right)\,,
\end{align}
where $\vec{H}$ denotes the same hard function as in~\eq{Fact}, and $\vec{J}^{\,\gco}$ is the jet function describing the collinear limit of the largest angle $x_L$ in the correlator $\gco$ of~\eq{gco}. The projected jet function satisfies the evolution equation
\begin{align}
\label{eq:multiJetRG}
\frac{d \vec{J}^{\,\gco}\left(\frac{x_L\, Q^2}{\mu^2}\right)}{d \ln \mu^2}=\int_0^1 dy\,y^{n_1 N_1 + \cdots + n_M N_M} \, \vec{J}^{\,\gco}\left(\frac{x_L\,y^2 Q^2}{\mu^2}\right) \cdot \widehat{P}(y)\,,
\end{align}
where $\widehat{P}(y)$ is again the singlet timelike splitting matrix of~\eq{splitK}.

\subsubsection{Jet Functions for the Higher-Point Correlators}
As for the two-point correlators, we decompose the projected jet function into perturbative and nonperturbative parts as $J_i^{\gco} = \hat{J}_i^{\gco}\, +\, $p.c., where `$+$p.c.' represents the nonperturbative power corrections. The perturbative jet functions for the projected general correlators can be organized as
\begin{align}
\hat{J}_{i}^{\gco}\left(\frac{x_L\, x^2 Q^2}{\mu^2}, \mu\right)=\sum_{a=0}^{\infty} \sum_{b=0}^a\left(\frac{\alpha_s}{4 \pi}\right)^a \ln^b \frac{\mu^2}{x_L\, x^2 Q^2} J_{i(a, b)}^{\gco}\,.
\end{align}
The log-enhanced terms are again iteratively predicted by the RG evolution equation~\eq{multiJetRG}. For the $N$-point correlator $\gco$ of~\eq{gco}, each coefficient $J_{i(a, b)}^\gco$ will in general involve terms proportional to products of generalized track function moments, $\prod_c T_{j_c R}^{n_1\cdots n_M}(k_{1_c},\ldots,k_{M_c},\mu)$, where the moment indices in each slot sum to the corresponding detector multiplicity, $\sum_c k_{l_c} = N_l$.

The LO jet functions are given as
\begin{align}
J_{i(0,0)}^{\gco}  =&\ 2^{-(n_1 N_1 + \cdots + n_M N_M)}\,T^{n_1\,n_2\,\cdots\,n_M}_{iR}(N_1,N_2,\cdots,N_M,\mu)\,.
\end{align}
The NLO coefficients cannot be given without specifying $\gco$ explicitly. In Appendix~\ref{Appendix}, we present the NLO constants for the projected correlators with arbitrary distinct weights for the three- and four-point cases, $\langle\mathcal{E}_R^{n_1} \mathcal{E}_R^{n_2}\mathcal{E}_R^{n_3}\rangle_{x_L}$ and $\langle\mathcal{E}_R^{n_1} \mathcal{E}_R^{n_2}\mathcal{E}_R^{n_3}\mathcal{E}_R^{n_4}\rangle_{x_L}$; higher-point cases can be computed in the same manner. Those results also contain the cases involving multiple identical detectors as special instances: when $n_a = n_b$, one simply merges the moments as $T_{iR}^{\cdots\,n_a\,n_b\,\cdots}(\cdots,A,B,\cdots) \to T_{iR}^{\cdots\,n_a\,\cdots}(\cdots,A+B,\cdots)$, following~\eq{gentrackReqn}. Here, we present the cases of three and four identical detectors. For the three-point case, the NLO constants are given by
\begin{align}
2^{3n}&J_{g(1,0)}^{\langle \cE_R^n \cE_R^n \cE_R^n \rangle} = \Bigg[(5 n+4) \left(\psi ^{(0)}(n+1)+\psi ^{(0)}(2 n+1)-2 \psi ^{(0)}(3 n+4)\right)+2 n+7\Bigg]  \\
&\times \frac{48 n_f T_F n^2 (n+1) \Gamma(n) \Gamma(2n)}{\Gamma(3n+4)} T_{qR}^n(1) T_{qR}^n(2) - \Bigg[110+n (n (9 n (14 n+61)+806)+497) \nn \\
&-3 (n+1) (3 n+1) (3 n+2) (n (49 n+59)+18) \left(\psi ^{(0)}(n)+\psi ^{(0)}(2 n)-2 \psi ^{(0)}(3 n)\right)\Bigg] \nn \\
&\times \frac{8 C_A \Gamma(n) \Gamma(2n)}{9(n+1)\left(9n(n+1)+2\right)^2\, \Gamma(3n)} T_{gR}^n(1) T_{gR}^n(2)\,, \nn
\end{align}
for the gluon jet function and
\begin{align}
2^{3n}J_{q(1,0)}^{\langle \cE_R^n \cE_R^n \cE_R^n \rangle} &= \Bigg[2 \left(5 n^2+6 n+2\right) \left(\psi ^{(0)}(2 n)+\psi ^{(0)}(n+1)-2 \psi ^{(0)}(3 n+3)\right)+2 n^2+ 15 n+9\Bigg]  \\
&\times \frac{12 C_F \Gamma(2n) \Gamma(n+1)}{\Gamma(3n + 3)} T_{qR}^n(1) T_{gR}^n(2) + \Bigg[n^2+33 n+18+\left(26 n^2+30 n+8\right) \left(\psi ^{(0)}(n) \right. \nn \\
&+ \left. \psi ^{(0)}(2 n+1)-2 \psi ^{(0)}(3 n+3)\right)\Bigg] \frac{6  C_F \Gamma(n) \Gamma(2n+1)}{\Gamma(3n + 3)} T_{gR}^n(1) T_{qR}^n(2)\,, \nn
\end{align}
for the quark jet function. For the four-point case, they are given as
\begin{align}
&2^{4n}J_{g,(1,0)}^{\langle \cE_R^n \cE_R^n \cE_R^n \cE_R^n \rangle} = \Bigg[8 \left(2 n^2+3 n+1\right) \left(\psi ^{(0)}(2 n+1)-\psi ^{(0)}(4 n+4)\right)+4 n^2+12 n+7\Bigg]  \\
&\times \frac{96 n_f\, T_F\, n^2\, \left(\Gamma(2n)\right)^2}{\Gamma(4n + 4)} \left(T_{qR}^n(2)\right)^2 + \Bigg[2 \left(5 n^2+6 n+2\right) \left(\psi ^{(0)}(n+1)+\psi ^{(0)}(3 n+1) \right. \nn \\
&- \left. 2 \psi ^{(0)}(4 n+4)\right)+3 n^2+12 n+7\Bigg]\frac{96 n_f\, T_F\, n^2\, \Gamma(n) \Gamma(2n)}{\Gamma(4n + 4)} T_{qR}^n(1) T_{qR}^n(3) \nn \\
&+ \Bigg[2 (4n+3)^2 \Gamma (2 n) \Gamma (4 n+2) \left(\psi ^{(0)}(2 n)-\psi^{(0)}(4 n)\right) - \Gamma (4 n) \Gamma (2 n+1)\Big(33+n(85+56 n) \nn \\
&+ 2n(3+4n)(11+14 n) \left(\psi ^{(0)}(2n)-\psi ^{(0)}(4 n+2)\right)\Big)\Bigg]\frac{24 C_A \Gamma(2n)}{\Gamma(4n) \Gamma(4n+2) (4n + 3)^2} \left(T_{gR}^n(2)\right)^2 \nn \\
&+ \Bigg[(2n+1)(4n+3)\left(48+2n(239+n(848+n(1273+676n)))\right) \left(\psi ^{(0)}(n)+\psi ^{(0)}(3 n) - 2 \psi ^{(0)}(4 n)\right) \nn \\
&- 165 - n(1348+n(4371+4n(1755+4n(345+104n))))\Bigg] \nn\\
&\times\frac{16\, C_A\, \Gamma(3n) \Gamma(n+1) \Gamma(4n+1)}{(2n+1) (4n+3)\Gamma(4n+2) \Gamma(4n + 4)} T_{gR}^n(1) T_{gR}^n(3)\,,\nn
\end{align}
for the gluon jet function and
\begin{align}
&2^{4n}J_{q,(1,0)}^{\langle \cE_R^n \cE_R^n \cE_R^n \cE_R^n \rangle} = \Bigg[\left(34 n^2+30 n+8\right) \left(\psi ^{(0)}(3 n)+\psi ^{(0)}(n+1)-2\psi ^{(0)}(4 n+3)\right)+9 n^2+39 n+18\Bigg]\nn \\
&\times \frac{8  C_F \Gamma(3n) \Gamma(n+1)}{\Gamma(4n+3)} T_{qR}^n(1) T_{gR}^n(3) +\Bigg[\left(50 n^2+42 n+8\right) \left(\psi ^{(0)}(n)+\psi ^{(0)}(3 n+1) \right.  \\
&- \left. 2 \psi ^{(0)}(4 n+3)\right)+n^2+45 n+18\Bigg] \frac{8 C_F \Gamma(n) \Gamma(3n+1)}{\Gamma(4n+3)} T_{qR}^n(3) T_{gR}^n(1) +\Bigg[2 n^2+21 n+9 \nn \\
&+2 \left(10 n^2+9 n+2\right) (\psi ^{(0)}(2 n)+\psi ^{(0)}(2 n+1)-2 \psi ^{(0)}(4 n+3))\Bigg] \frac{24 C_F \Gamma(2n) \Gamma(2n+1)}{\Gamma(4n+3)} T_{qR}^n(2) T_{gR}^n(2)\,,\nn
\end{align}
for the quark jet function. In these expressions, we suppressed the $\mu$ dependence of the $T_{iR}^n$ for convenience. The three- and four-point cases with $n=1$ were already calculated in~\cite{Chen:2020vvp}, and the results presented here reproduce them as special cases.

\subsubsection{Reciprocity and Anomalous Scaling for the Higher-Point Correlators}
\label{sec:reciprocityHigher}
In this section, we repeat the reciprocity argument of Sec.~\ref{sec:reciprocityTwo} to study the scaling of the higher-point correlators. Conceptually, the higher-point case shows how contact terms like the one in~\eq{JetParamTwo}, proportional to $\tilde{T}_{gR}^{n\,m}(1,1)$, produce finite-$x_L$ contributions. 

To be concrete, we consider the $x_L$ projection of the three-point correlator $\langle\mathcal{E}_R^m \mathcal{E}_R^{m}\mathcal{E}_R^{m}\rangle_{x_L}$. Again working in pure YM at fixed coupling, we parameterize our jet function as
\begin{align}
\label{eq:JetParamThree}
\hat{J}_g^{\langle\mathcal{E}_R^m \mathcal{E}_R^{m}\mathcal{E}_R^{m}\rangle}(x_L Q^2,\mu) =&  T_{gR}^{m}(1,Q \sqrt{x_L})^3C(\alpha_s) \left(\frac{x_L Q^2}{\mu^2}\right)^{\gamma_1}\\
&+ \tilde{T}_{gR}^{m}(2,Q \sqrt{x_L}) T_{gR}^{m}(1,Q \sqrt{x_L})D(\alpha_s) \left(\frac{x_L Q^2}{\mu^2}\right)^{\gamma_2}\nn\\
&+ \tilde{T}_{gR}^{m}(3,Q \sqrt{x_L}) E(\alpha_s) \left(\frac{x_L Q^2}{\mu^2}\right)^{\gamma_3}\,,\nn
\end{align}
where $\tilde{T}_{gR}^{m}(2,Q \sqrt{x_L})$ and  $\tilde{T}_{gR}^{m}(3,Q \sqrt{x_L})$ are the diagonalized moments given in~\eqs{gentracktildeT2}{gentracktildeT3}, respectively. Their RG evolution equations were given in Sec.~\ref{sec:gentrackmatch}. Then substituting~\eq{JetParamThree} into~\eq{multiJetRG}, we arrive at the following consistency equations
\begin{align}
2\gamma_1 &= 2\gamma_T (3m+1+2\gamma_1-6\gamma_{T}(m+1))\,,\\
2\gamma_2 &= 2\gamma_T (3m+1+2\gamma_2-2\gamma_{T}(2m+1)-2\gamma_{T}(m+1))\,,\nn\\
2\gamma_3 &= 2\gamma_T (3m+1+2\gamma_3-2\gamma_{T}(3m+1))\,,\nn
\end{align}
which imply
\begin{align}
\gamma_1 &= \gamma_S(3m+1-6\gamma_{T}(m+1))\,,\\
\gamma_2 &= \gamma_S(3m+1-2\gamma_{T}(2m+1)-2\gamma_{T}(m+1))\,,\nn\\
\gamma_3 &= \gamma_T(3m+1)\,,\nn
\end{align}
where we again used the reciprocity relation $2\gamma_S(J) = 2\gamma_T(J+2\gamma_S(J))$. This result is consistent with the scaling derived from the light-ray OPE picture for the three-point case in~\eq{gammaSJThree} when $m_1=m_2=m_3=m$. In particular, we identify the contribution from the OPE of the two-detector contact term with the third detector, governed by the light-ray operator with the spin `$3m+1-2\gamma_{T}(2m+1)-2\gamma_{T}(m+1)$'.

\section{Nonperturbative Power Corrections}\label{sec:NP}
In QCD, nonperturbative corrections grow in significance as one approaches the kinematic endpoints, $x_L\to 0,1$, where the angular scale becomes nonperturbative, $Q\sqrt{x_L(1-x_L)}\sim \Lambda_{\rm QCD}$. For the two-point energy correlator, the leading power correction was first identified in~\cite{Korchemsky:1999kt}, and is described by a universal matrix element sourced by back-to-back Wilson lines $Y_{n(\bar{n})}^a$,
\begin{align}
\label{eq:NP2}
\Omega_{1a}\equiv& \frac{1}{N_a}\langle 0|\operatorname{tr} \bar{Y}_{\bar{n}}^{\dagger a} Y_n^{\dagger a} \mathcal{E}_{T}(0) Y_n^a \bar{Y}_{\bar{n}}^a| 0\rangle\,,
\end{align}
where $a = q,g$ denotes the color channel, $N_q = N_c$ and $N_g=N_c^2-1$, and $\mathcal{E}_T(\eta)$ is the transverse energy flow operator defined below. The same matrix element was recently shown to govern the leading power corrections of the $N$-point projected correlators $\langle\mathcal{E}(\vec{n}_1)\cdots \mathcal{E}(\vec{n}_N)\rangle_{x_L}$~\cite{Lee:2024esz}. In this section, we present the corresponding results for the correlators of generalized detectors, focusing for notational simplicity on the two-point case $\langle\mathcal{E}_R^n \mathcal{E}_R^{m}\rangle_{x_L}$. The generalization to $N$-point projected correlators proceeds as in~\cite{Lee:2024esz,Chen:2024nyc}.

\subsection{Universal Power Corrections from Soft Radiation}
\label{sec:NPuniversal}
The leading power corrections at lowest order in $\alpha_s$ arise from configurations in which one of the two detectors is triggered by soft hadrons, while the other detector sits on one of the energetic jets. Since the detector $\mathcal{E}_R^n$ weights the soft hadrons by the $n$-th power of their energy, this contribution is of order $\Lambda_{\rm QCD}^n$, and is governed by the generalization of~\eq{NP2},
\begin{align}
\label{eq:NP}
\Omega_{ka}\equiv& \frac{1}{N_a}\langle 0|\operatorname{tr} \bar{Y}_{\bar{n}}^{\dagger a} Y_n^{\dagger a} \mathcal{E}_{T}^k(0) Y_n^a \bar{Y}_{\bar{n}}^a| 0\rangle\,,
\end{align}
where the weighted transverse energy flow operator is defined through its action on hadronic states,
\begin{align}
\label{eq:ETk}
\mathcal{E}_{T}^k(\eta)|X\rangle = \sum_{i \in X}\left|\mathbf{p}_i^T\right|^k \delta\left(\eta-\eta_i\right)|X\rangle\,,
\end{align}
with $\eta_i$ and $\mathbf{p}_i^T$ the rapidity and transverse momentum of hadron $i$ with respect to the dijet axis, related to the hadron energy by $E_i = |\mathbf{p}_i^T|\cosh\eta_i$. Since the transverse momenta are invariant under boosts along the axis, the matrix element~\eq{NP} is independent of the rapidity at which the operator is inserted, and $\Omega_{ka}\sim \Lambda_{\rm QCD}^k$ is a single universal nonperturbative number for each color channel. When the detectors are restricted to hadrons carrying quantum number $R$, the transverse energy flow operator is restricted in the same way,
\begin{align}
\label{eq:ETkR}
\mathcal{E}_{TR}^k(\eta)|X\rangle = \sum_{i\in R \subset X}\left|\mathbf{p}_i^T\right|^k \delta\left(\eta-\eta_i\right)|X\rangle\,,
\end{align}
defining the restricted matrix elements $\Omega_{ka}^R$.

The resulting power corrections can be stated compactly at the level of the fixed-order cross section. Generalizing the $n=m=1$ result of~\cite{Lee:2024esz}, we have
\begin{align}
\label{eq:NPFO}
\frac{1}{\sigma_0}\frac{d \sigma^{\langle\mathcal{E}_R^n \mathcal{E}_R^{m}\rangle}}{dx_L}
=&\ \frac{1}{\sigma_0}\frac{d \hat{\sigma}^{\langle\mathcal{E}_R^n \mathcal{E}_R^{m}\rangle}}{dx_L}
+ \frac{T^{m}_{qR}(1)\; \Omega^{R}_{nq}}{2^{n+m}\, Q^n\, \left[x_L(1-x_L)\right]^{\frac{2+n}{2}}}
+ \frac{T^{n}_{qR}(1)\; \Omega^{R}_{mq}}{2^{n+m}\, Q^m\, \left[x_L(1-x_L)\right]^{\frac{2+m}{2}}}\nn\\
&+ \mathcal{O}(\alpha_s)\times\frac{\mathcal{O}(\Lambda_{\rm QCD})}{Q\, \left[x_L(1-x_L)\right]^{3/2}}\,,
\end{align}
where $d\hat{\sigma}$ denotes the perturbative contribution discussed in~\Sec{sec:LROPE} and~\Sec{sec:fact}. The structure of the $\Omega$ terms is simple to understand. A soft hadron at rapidity $\eta$ with respect to the dijet axis is measured at the angle $x_L = (1-\tanh\eta)/2$, so that $\cosh\eta = 1/(2\sqrt{x_L(1-x_L)})$; the energy weight of the soft detector contributes $(E_i/Q)^n = (|\mathbf{p}^T_i|\cosh\eta/Q)^n$, which combines with the Jacobian from the rapidity integral into the characteristic kinematic factor $[x_L(1-x_L)]^{-(2+n)/2}$. The accompanying factor $T^{m}_{qR}(1)$, evaluated at $\mu \simeq Q$, is the first moment of the quark generalized track function, arising from the remaining detector acting on the energetic jets, which at this order are the two quark jets; the overall $2^{-(n+m)}$ follows from the $Q^{n+m}$ normalization of~\eq{EnEmC}. For $n=m=1$ and $R=\forall$, the moments reduce to unity and~\eq{NPFO} reproduces the known result for the energy correlator, with coefficient $N/2^N$ at $N=2$~\cite{Lee:2024esz}. The scaling given in~\eq{NPFO} has also been confirmed by generalizing the renormalon analyses of~\cite{Schindler:2023cww,Lee:2024esz}.

The power corrections displayed in~\eq{NPFO} are those identified by a leading-order analysis in $\alpha_s$. At higher orders, there is a priori no reason for nonperturbative corrections with fewer powers of $\Lambda_{\rm QCD}$ to be absent: for $n,m>1$, the most important such correction would be of first order in $\Lambda_{\rm QCD}$: while suppressed by $\alpha_s$, it is less suppressed in $\Lambda_{\rm QCD}/Q$ than the $\Omega$ terms. The last term of~\eq{NPFO} represents such contributions schematically, with the characteristic angular enhancement of a unit energy weight placed on soft hadrons. From here on out, we denote the corresponding nonperturbative matrix elements by $\Delta^R_{1i}\sim\Lambda_{\rm QCD}$; their operator definitions are not yet known, and while their universality and precise structure would be interesting to explore, we do not consider them further here beyond keeping track of their contribution.

In the collinear factorization of Sec.~\ref{sec:fact}, these power corrections reside in the jet functions~\cite{Lee:2024esz,Chen:2024nyc}. Including them in the two-point jet function $J_i^{\langle\mathcal{E}_R^n \mathcal{E}_R^{m}\rangle}$ of~\eq{Fact}, we have
\begin{align} \label{eq:Jpc}
\hspace{-0.1cm}2^{n+m}\,J_i^{\langle\mathcal{E}_R^n \mathcal{E}_R^{m}\rangle}\left(\frac{x_Lx^2 Q^2}{\mu^2}, \mu\right) =\ &2^{n+m}\,\hat{J}^{\langle\mathcal{E}_R^n \mathcal{E}_R^{m}\rangle}_i\left(\frac{x_L x^2 Q^2}{\mu^2}, \mu\right)- \mathcal{J}_i^{\langle\mathcal{E}_R^{n}\mathcal{E}_R^{m}\rangle}\left(\frac{x_L x^2 Q^2}{\mu^2}, \mu\right) \,,
\end{align}
where $\hat{J}_i$ is the perturbative jet function discussed at length in Sec.~\ref{sec:TwoPTJet}, and the nonperturbative coefficient $\mathcal{J}_i$ collects all the power corrections; in contrast to $\hat{J}_i$, it depends on its first argument not only through logarithms, but also through inverse powers. Expanding in $\alpha_s$ and logarithms as in~\eq{JetTwoAnsatz}, its leading terms are given by the universal matrix elements of~\eq{NP} multiplying the single-detector structures left behind once the corresponding detector is placed on the soft radiation, together with the $\alpha_s$-suppressed contribution of first order in $\Lambda_{\rm QCD}$ discussed above,
\begin{align}
\label{eq:JcalLO}
\mathcal{J}_{i}^{\langle\mathcal{E}_R^{n}\mathcal{E}_R^{m}\rangle} \supset \frac{1}{n}\,\frac{\Omega^R_{ni}\; T_{iR}^{m}(1,\mu)}{(\sqrt{x_L}\,x\,Q)^{n}} + \frac{1}{m}\,\frac{\Omega^R_{mi}\; T_{iR}^{n}(1,\mu)}{(\sqrt{x_L}\,x\,Q)^{m}} + \frac{\mathcal{O}(\alpha_s)\,\Delta^R_{1i}}{\sqrt{x_L}\,x\,Q}\,.
\end{align}
The explicit factors $1/n$ and $1/m$ compensate the differentiation of the powers $x_L^{-n/2}$ and $x_L^{-m/2}$ in passing from the cumulant to the differential distribution, ensuring consistency with~\eq{NPFO}; at $n=m=1$ the two $\Omega$ terms combine to reproduce the factor $N=2$ of the corresponding result in~\cite{Lee:2024esz}. For the purposes of RG evolution, each power $1/(\sqrt{x_L}\,x\,Q)^{n}$ in $\mathcal{J}_i$ constitutes an independent power-suppressed component of the jet function: due to the explicit factor $x^{-n}$, the $\Omega^R_{ni}$ component satisfies the RG equation~\eq{twojetRG}, with the moment $y^{n+m}$ replaced by $y^{m}$, and the product $\Omega^R_{ni}\,T^m_{iR}(1,\mu)$ in~\eq{JcalLO} is precisely its zeroth-order boundary constant. Importantly, while the $\Omega^R_{ni}$ term is suppressed by $n$ powers of $\Lambda_{\rm QCD}$, it is enhanced at small angles by the relative factor $(1/\sqrt{x_L})^n$ and eventually dominates at sufficiently small $x_L$, driving the differential distribution towards the scaling $d\sigma/d\theta \sim 1/\theta^{1+n}$. This is precisely the behavior observed in the parton shower studies of Sec.~\ref{sec:MC}.

\subsection{Light-Ray Operator Product Expansion Picture}
\label{sec:NPLROPE}
The power corrections of~\eq{Jpc} have a transparent interpretation in the light-ray OPE of Sec.~\ref{sec:LROPE}. Working again in pure YM at fixed coupling and taking equal energy weights $m_1=m_2=m$ for simplicity, the perturbative OPE channels of~\eq{twopointOPE} are supplemented by a new nonperturbative channel, which we write schematically as
\begin{align}
\label{eq:LROPEwNP}
\mathcal{E}_{R}^{m}(\vec{n}_1)\, \mathcal{E}_{R}^{m}(\vec{n}_2)\ \supset\ &\ \Omega^R_{mg}\; C^{\Omega}_{12}(x_{12},\mu)\; F_g^{R}(J_L,\mu)\,\mathcal{D}^{\rm DGLAP}_{J_{L},g}(\vec{n}_1,\mu)\,,
\end{align}
where $J_L = 2-d-m$ is the boost weight of a single hadronic detector, \eq{JLmrel}. Here we consider only the channel generated by the universal matrix element $\Omega^R_{mg}$. The $\alpha_s$-suppressed contributions associated with $\Delta^R_{1i}$ in~\eq{Jpc} would appear as an additional channel with its own matching coefficient. Lorentz symmetry fixes the classical scaling of the OPE coefficient as
\begin{align}
\label{eq:NPLRscalings}
C^{\Omega}_{12}(x_L,\mu)\sim x_L^{-\frac{2+m}{2}}\,,
\end{align}
which matches the kinematic factor of the fixed-order result~\eq{NPFO}, and is enhanced at small angles relative to the classical $1/x_L$ scaling of the perturbative channel.

In~\eq{LROPEwNP}, one of the two detectors is placed on the soft radiation, producing the matrix element $\Omega^R_{mg}\sim\Lambda_{\rm QCD}^m$, while the remaining detector undergoes the single-detector matching of~\eq{singleHad}. The light-ray operator produced in this channel is therefore the \emph{same} leading-twist DGLAP detector $\mathcal{D}^{\rm DGLAP}_{J_{L},g}$ that matches a single $\mathcal{E}_R^m$, whose spin was given in~\eq{JdglapJL}, $J = m+1-2\gamma_T(m+1)$. Classically, this spin is lower by $m$ units than the perturbative-channel spin $J_{12} = 2m+1-4\gamma_T(m+1)$ of~\eq{J12}; for $m=1$ this recovers the known observation that the spin of the leading power correction to the energy correlator is lowered by one unit~\cite{Chen:2024nyc}.

Exactly as for $C_{12}$ in Sec.~\ref{sec:LROPE}, RG invariance of~\eq{LROPEwNP} fixes the evolution of the OPE coefficient, now with a single detector function. Here, however, the evolution~\eq{Devol} of the produced detector $\mathcal{D}^{\rm DGLAP}_{J_{L},g}$ is compensated entirely by the evolution~\eq{Tevol} of the detector function $F_g^{R}(J_L,\mu)$, since the operator produced in this channel carries the \emph{same} boost weight $J_L$ as the detector function that matches it: the combination $F_g^{R}(J_L,\mu)\,\mathcal{D}^{\rm DGLAP}_{J_{L},g}(\vec{n}_1,\mu)$ is precisely the RG-invariant single-detector matching of~\eq{singleHad}, and since $\Omega^R_{mg}$ is $\mu$-independent at this order, the coefficient $C^{\Omega}_{12}$ is $\mu$-independent, in contrast to its perturbative counterpart. 

Beyond pure YM, this compensation is no longer automatic. The channel then involves the flavor sum $\sum_{\kappa=q,g}\Omega^R_{m\kappa}\, C^{\Omega}_{12,\kappa}\, F_\kappa^{R}\,\mathcal{D}^{\rm DGLAP}_{J_{L},\kappa}$. While $\sum_\kappa F_\kappa^{R}\,\mathcal{D}^{\rm DGLAP}_{J_L,\kappa}$ remains RG-invariant, the $\Omega$-weighted combination is not, and its $\mu$-derivative is proportional to the difference of the soft matrix elements in the two color channels, $\Omega^R_{mq}-\Omega^R_{mg}$. Since the quark and gluon soft matrix elements differ in general, quark-gluon mixing induces a nontrivial evolution of the OPE coefficients $C^{\Omega}_{12,\kappa}$, which, together with the running of the coupling, produces the calculable scaling violations of the power corrections.

\section{Numerical Study with Monte Carlo Simulations}\label{sec:MC}
As an illustration of the formalism developed in this paper, in this section we study the behavior of the two-point correlators $\langle \mathcal{E}^n \mathcal{E}^m\rangle_{x_L}$ and the three-point correlators $\langle \mathcal{E}^n \mathcal{E}^n\mathcal{E}^n\rangle_{x_L}$ in $e^+e^-$ collisions in the collinear limit, using Monte Carlo simulations. Beyond their theoretical interest, such correlators are already being measured: the two-point correlator with squared energy weights, $\langle \mathcal{E}^2 \mathcal{E}^2\rangle_{x_L}$, has recently been measured inside high-energy jets by the CMS collaboration in both $pp$ and Pb-Pb collisions, where the higher energy weight suppresses the underlying-event background, and was used to study the nuclear modification of jet substructure in the quark-gluon plasma~\cite{CMS:2025ydi}.

We emphasize that the preceding sections provide all the ingredients required to compute these correlators to NLL accuracy in the collinear limit: the factorization theorems of \Sec{sec:fact}, the NLO jet functions computed in \Sec{sec:TwoPTJet} and App.~\ref{Appendix}, and the NLO evolution of the generalized track function moments of \Sec{sec:GenTrack}. Nevertheless, we refrain from carrying out these calculations in this paper. A rigorous prediction requires substantial nonperturbative input that is not yet available. First, one must extract the many generalized track functions entering the factorization, with their distinct quark and gluon components. Second, one must extract the nonperturbative matrix elements governing the power corrections for each energy weight $n$, including not only the universal matrix elements $\Omega^R_{n\kappa}$ of \Sec{sec:NP}, but also the $\Delta$-type contributions discussed there, whose precise operator definitions are not yet known. Our aim in this section is instead qualitative: we verify that the scaling behaviors observed in the simulations follow the expectations of the theoretical framework we have presented, and we leave precision studies to the future. Throughout this section, we use the angular variable $\theta =\arccos(1-2x_L)$; at small angles, $x_L \approx \theta^2/4$.

\begin{figure}[t!]
\begin{center}
	\includegraphics[width =5 in]{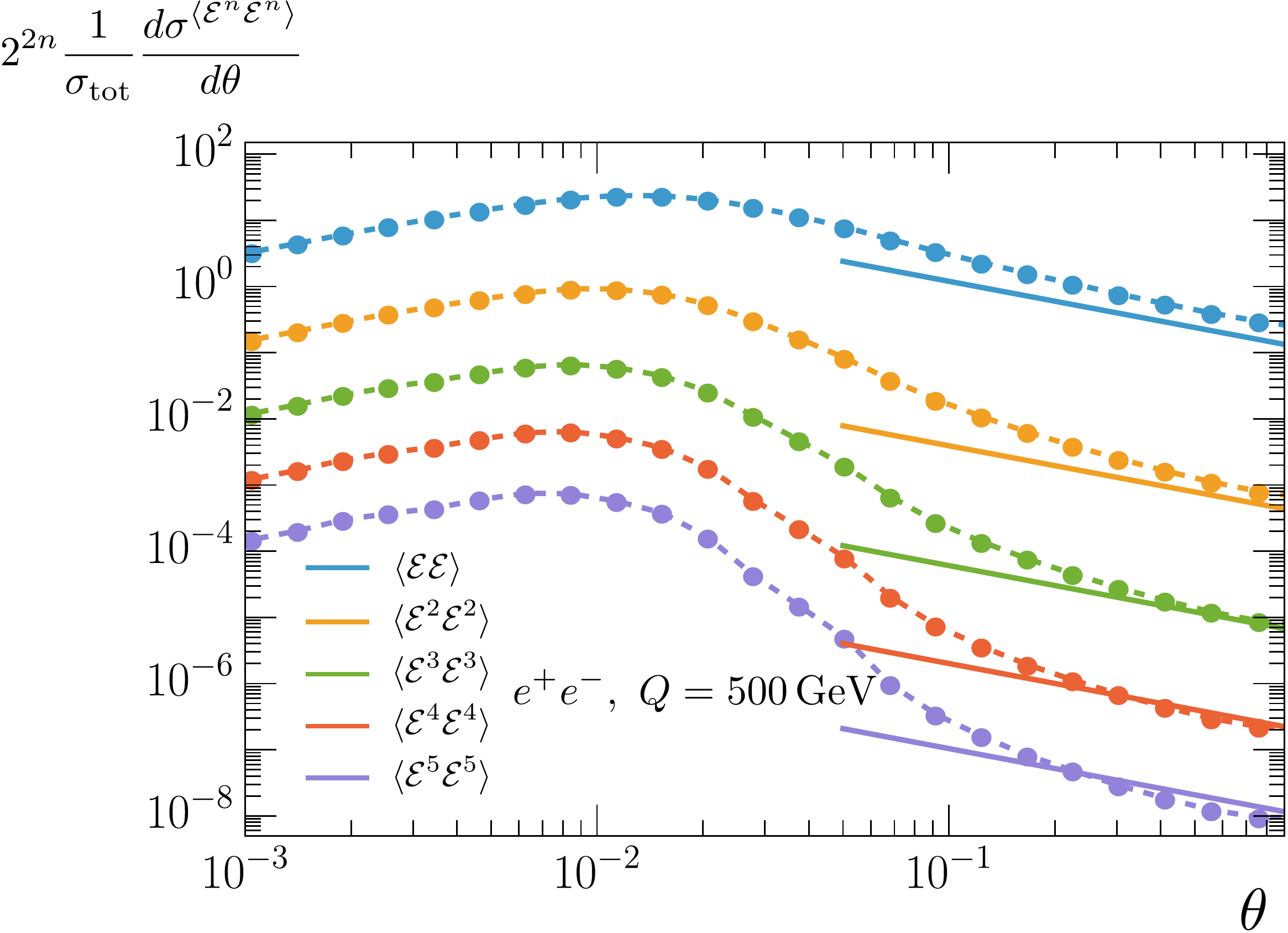}
\end{center}
\caption{Pythia simulations of $\langle\mathcal{E}^n(\vec{n}_1)\mathcal{E}^n(\vec{n}_2)\rangle_{x_L}$ at $Q=500~\rm{GeV}$ in $e^+e^-$ collisions, for energy weights $n=1,\dots,5$ (data points). The solid lines show the leading fixed-order perturbative predictions, obtained by convolving the NLO jet functions of \eqs{JgTwo}{JqTwo} with the LO hard function; their normalizations are set by the moments of the generalized track functions. For these generalized detectors, nonperturbative power corrections significantly modify the perturbative scaling, giving rise to a scaling approaching $1/\theta^{1+n}$ in the small-angle region. The linear scaling in the far-left region occurs at the hadronization scale, as the two-point function becomes fully nonperturbative.}\label{fig:pythia}
\end{figure}

As discussed in \Sec{sec:NP}, the $\mathcal{E}^n$ detectors give rise to nonperturbative power corrections that modify the classical angular scaling in the OPE. In \fig{pythia}, we observe a strong modification of the $\theta$ scaling of $\langle\mathcal{E}^n(\vec{n}_1)\mathcal{E}^n(\vec{n}_2)\rangle_{x_L}$ in the simulations as the weight $n$ is varied.\footnote{The distributions are rescaled by $2^{2n}$ for convenience, since we weight by $1/Q^{n+m}$ in~\eq{EnEmC}, rather than $1/(Q/2)^{n+m}$, while each of the two back-to-back jets carries energy $\simeq Q/2$.} In the large-angle region, the power corrections are small, and all curves exhibit the expected classical $1/\theta$ scaling. In this perturbative region, we also show the leading fixed-order perturbative predictions (solid lines), obtained by convolving the NLO jet functions of \eqs{JgTwo}{JqTwo} with the LO hard function; the moments of the generalized track functions extracted in~\eq{gentrackpythia} provide the nonperturbative input. While these fixed-order predictions do not include resummation, they already reproduce the rough normalization of the simulations for each $n$, demonstrating that the overall scale of the generalized correlators is indeed set by the moments of the generalized track functions. As the angle decreases, the distributions steepen due to the nonperturbative power corrections, with the scaling increasing all the way to $1/\theta^{1+n}$, until one reaches the transition to the hadronization region, where the distributions approach a linear scaling in $\theta$, corresponding to a flat distribution in $x_L$.

\begin{figure}[t!]
\begin{center}
	\includegraphics[width =6 in]{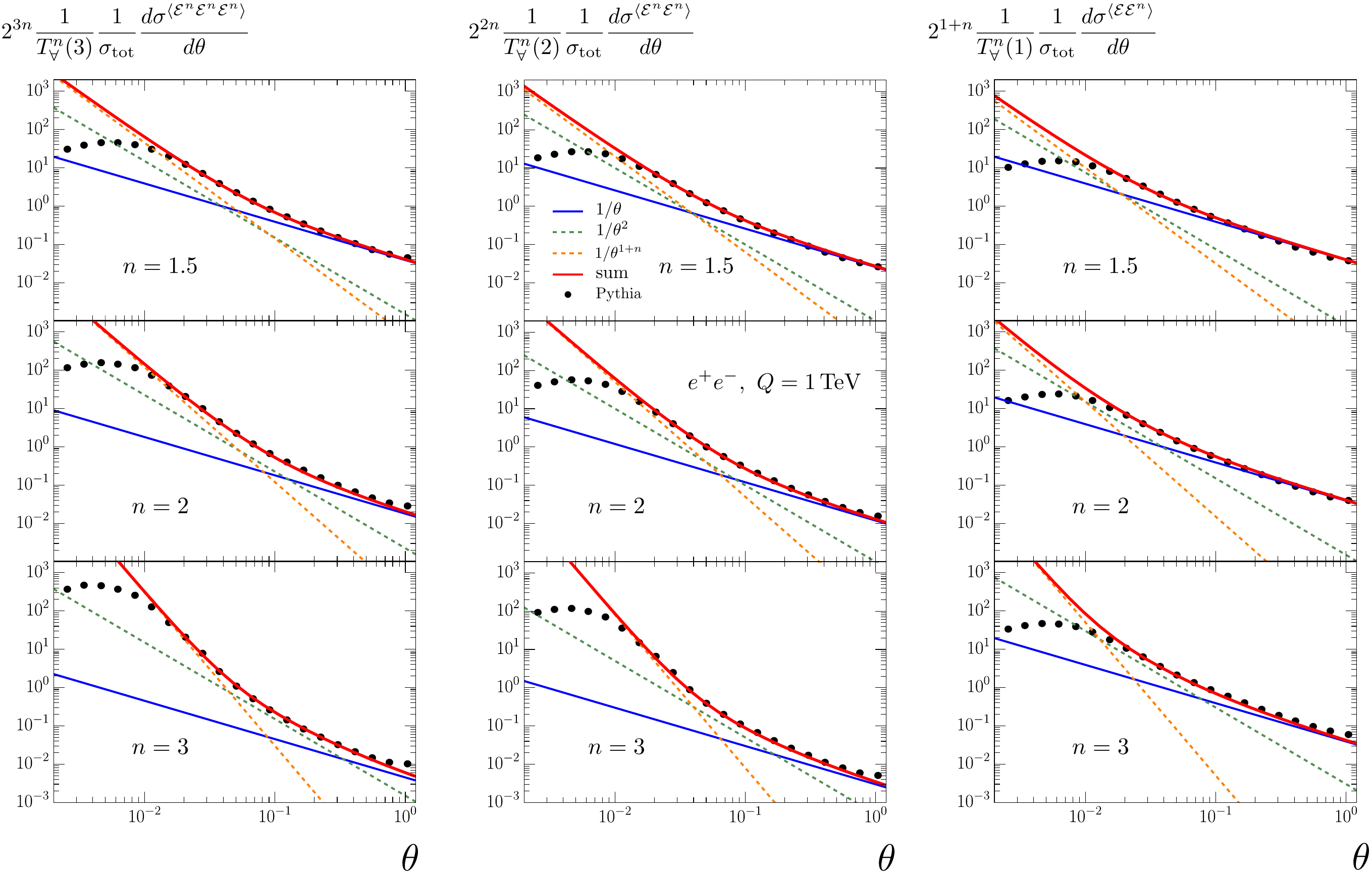}
\end{center}
\caption{Three-point correlators $\langle \mathcal{E}^n \mathcal{E}^n \mathcal{E}^n \rangle_{x_L}$ (first column), two-point correlators $\langle \mathcal{E}^n \mathcal{E}^n \rangle_{x_L}$ (second column), and mixed two-point correlators $\langle \mathcal{E} \mathcal{E}^n \rangle_{x_L}$ (third column) for $n=1.5,2,3$. In each panel, we compare the Pythia simulation with the scaling behaviors predicted by our framework: the classical perturbative scaling $1/\theta$ (blue), the $1/\theta^{2}$ scaling of the nonperturbative power correction of first order in $\Lambda_{\rm QCD}$ (green dashed), the $1/\theta^{1+n}$ scaling of the nonperturbative power correction of order $\Lambda_{\rm QCD}^n$ (orange dashed), and their sum (red), with appropriately tuned normalizations. When all detector weights are larger than one, the nonperturbative power corrections are large, and significantly modify the scaling behavior of the observable.}\label{fig:analytic}
\end{figure}

\begin{figure}[t!]
\begin{center}
	\includegraphics[width =6 in]{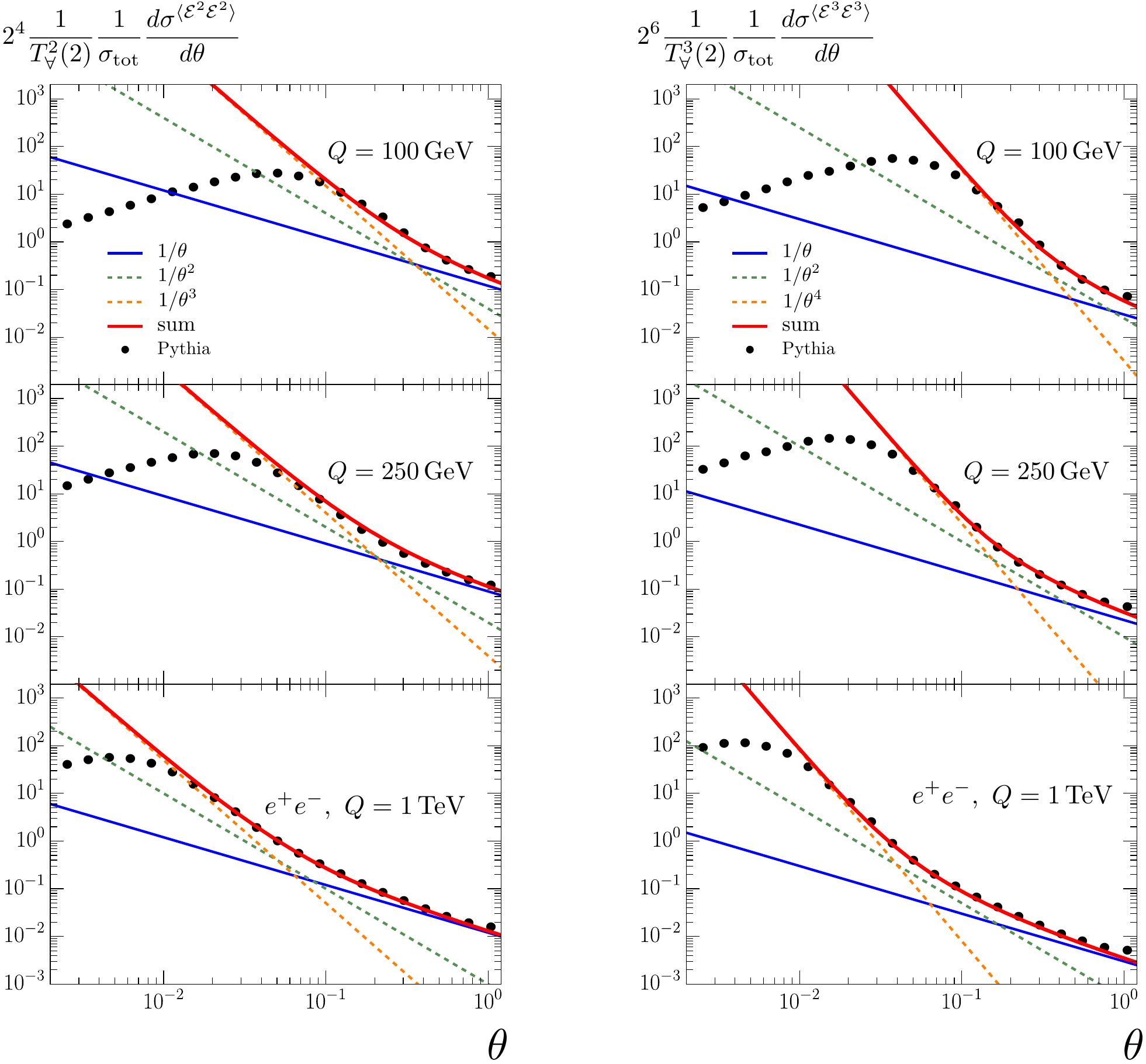}
\end{center}
\caption{Pythia simulations of $\langle\mathcal{E}^2\mathcal{E}^2\rangle_{x_L}$ and $\langle\mathcal{E}^3\mathcal{E}^3\rangle_{x_L}$ at $Q=100~\rm{GeV}$, $250~\rm{GeV}$, and $1~\rm{TeV}$, compared with the same scaling curves as in \fig{analytic}: the perturbative $1/\theta$ (blue), the nonperturbative $1/\theta^{2}$ (green dashed) and $1/\theta^{1+n}$ (orange dashed), and their sum (red). As $Q$ is increased, the size of the region described by the perturbative $1/\theta$ scaling (blue curve), where the power corrections are negligible, is enhanced.}\label{fig:Qdep}
\end{figure}

In the first two columns of~\fig{analytic}, we plot $\langle \mathcal{E}^n \mathcal{E}^n\mathcal{E}^n\rangle_{x_L}$ and $\langle \mathcal{E}^n \mathcal{E}^n\rangle_{x_L}$ for $n=1.5,2,3$, each in a separate panel.\footnote{In \figs{analytic}{Qdep}, the distributions are additionally normalized by the moments of the generalized track functions, $1/T^n_{\forall}(3)$, $1/T^n_{\forall}(2)$, and $1/T^n_{\forall}(1)$ for the three-point, two-point, and mixed correlators, respectively, to place all curves on a similar scale. These moments are the full-angle integrals of the corresponding unrescaled distributions: for instance, for the equal-weight two-point correlators, the event-level moment of~\eq{enemnorm} reduces to the second moment of the generalized track function, $T^{n\,n}_{\forall}(1,1)=T^{n}_{\forall}(2)$.} While we do not perform the full calculation, we test the scaling behaviors predicted by our framework by overlaying simple scaling curves, with normalizations appropriately tuned; the tuning plays the role of the extraction of the corresponding nonperturbative parameters, such as the power-correction matrix elements and the track function moments. The blue curve denotes the classical perturbative scaling $1/\theta$. The green dashed curve denotes the $1/\theta^{2}$ scaling of the nonperturbative power correction of first order in $\Lambda_{\rm QCD}$, which for the equal-weight correlators with $n>1$ is the $\alpha_s$-suppressed $\Delta$-type contribution identified in \Sec{sec:NP}. The orange dashed curve denotes the $1/\theta^{1+n}$ scaling of the universal $\Omega_n$ power correction of order $\Lambda_{\rm QCD}^n$, which, despite its stronger suppression in $\Lambda_{\rm QCD}$, is enhanced at small angles. Finally, the red curve denotes their sum. The nonperturbative power corrections significantly modify the scaling as $\theta$ decreases: the simulations depart from the perturbative $1/\theta$ curve well before the hadronization region, steepening first toward the $1/\theta^{2}$ scaling and then toward the $1/\theta^{1+n}$ scaling at yet smaller angles. The sum describes the simulations precisely down to the transition into the hadronization region, and the simulations qualitatively exhibit the distinct scaling behaviors in the different angular regions predicted by our framework.

In the third column of~\fig{analytic}, we consider the mixed correlators $\langle \mathcal{E} \mathcal{E}^n\rangle_{x_L}$. Since the smallest energy weight is now one, the leading power correction in $\Lambda_{\rm QCD}$ is the $\Omega_1$ term, corresponding to the configuration in which the weight-one detector is triggered by soft radiation, with the angular scaling $1/\theta^{2}$ (green dashed). In this case, the $1/\theta^{2}$ contribution dominates the power corrections throughout the perturbative region: relative to the $\Omega_1$ term, the $\Omega_n$ contribution (orange dashed) is suppressed by $\left(\Lambda_{\rm QCD}/(Q\sqrt{x_L})\right)^{n-1}$, and becomes relevant only close to the hadronization scale $Q\sqrt{x_L}\sim \Lambda_{\rm QCD}$.

While we do not pursue it here, it would also be interesting to study ratios of correlators that are expected to have similar anomalous scaling. For example, we saw in \Sec{sec:LROPE} that the anomalous scaling of $\langle\mathcal{E}^m\mathcal{E}^m\rangle_{x_L}$ is given by
\begin{align}
\label{eq:EmEmScale}
\gamma_S(2m+1-4\gamma_T(m+1))\,,
\end{align}
while that of $\langle\mathcal{E}^m\mathcal{E}^m\mathcal{E}^m\rangle_{x_L}$ is given by
\begin{align}
\label{eq:EmEmEmScale}
\gamma_S(3m+1-6\gamma_T(m+1))\quad \text{ and }\quad \gamma_S(3m+1-2\gamma_T(2m+1)-2\gamma_T(m+1))\,.
\end{align}
Substituting $m=3$ into~\eq{EmEmScale} and $m=2$ into~\eq{EmEmEmScale}, we observe that the scalings of $\langle\mathcal{E}^3\mathcal{E}^3\rangle_{x_L}$ and $\langle\mathcal{E}^2\mathcal{E}^2\mathcal{E}^2\rangle_{x_L}$ share the same classical part, differing only through the anomalous-dimension contributions to the argument. While the nonperturbative power corrections are sizable at very small angles, it would be interesting to study the ratio of these correlators at larger angles (still within the collinear limit), where the power corrections are negligible and such a small difference in anomalous scaling could be probed. This region of negligible power corrections is enlarged by increasing $Q$, which we confirm numerically for $\langle\mathcal{E}^2\mathcal{E}^2\rangle_{x_L}$ and $\langle\mathcal{E}^3\mathcal{E}^3\rangle_{x_L}$ in~\fig{Qdep}.
\section{Conclusions}\label{sec:conc}

In this paper, we have developed a systematic framework for multi-point correlation functions of generalized detector operators, $\mathcal{E}_{R_i}^{m_i}(\vec{n}_i)$, with general energy weights $m_i$ and restrictions $R_i$ on the measured hadrons. These observables enlarge the usual space of energy correlators by resolving both how energy is shared among hadrons and which hadrons carry it. Unlike correlators of the average null energy operator, however, they are generically not infrared and collinear safe. Their description therefore requires a nonperturbative matching between the hadronic detectors measured in the infrared and the partonic detectors used in short-distance calculations.

We formulated this matching in terms of universal nonperturbative detector functions, using this term for coefficients that match hadronic detectors onto partonic ones. For the $\mathcal{E}_R^m$ operators studied here, a single-detector function is represented by an energy-weighted single-hadron fragmentation-function moment. Products of detectors have a richer structure: in addition to the separate matching of each detector, they contain contact terms whenever a subset of the hadronic detectors cannot be resolved by the ultraviolet theory. These contact terms are immaterial for a two-point distribution at finite angular separation, but contribute at finite $x_L$ to higher-point projections, where an unresolved cluster can remain separated from the other detectors. Their matching coefficients are represented by the appropriate RG-diagonalized multi-hadron fragmentation-function moments. This multi-detector matching also enabled us to derive the light-ray OPE of hadronic detectors at weak coupling, which we carried out explicitly for pure Yang-Mills theory with fixed coupling.

We also showed that one- and multi-dimensional generalized track functions efficiently package the combinations of fragmentation-function moments required by detector matching. Since their evolution kernels depend only on the underlying partonic splittings, this representation allows the NLO evolution of these combinations to be obtained directly from the known NLO track-function kernels. Generalized track functions also expose an important difference from ordinary track functions. At unit energy weight, momentum conservation gives rise to a shift symmetry, and the relevant RG-diagonalized combinations reduce to central moments. For $m\neq 1$, this symmetry is lost and for $m>1$, the generalized track functions evolve toward a delta function at the origin in the deep ultraviolet. Their nonzero moments therefore all decrease with the scale, increasingly rapidly for larger energy weights and higher moments, so the evolution can have moderate impact on the angular dependence of generalized correlators.

Using these detector functions as universal nonperturbative boundary data, we presented QCD collinear factorization theorems for general $x_L$-projected correlators in the perturbative regime $Q\sqrt{x_L}\gg\Lambda_{\rm QCD}$. The factorization incorporates quark-gluon mixing and the running coupling, with jet functions universal across hard processes. We computed the two-point jet functions at NLO for arbitrary detector weights, as well as the NLO jet functions for projected three- and four-point correlators with distinct weights. Together with the NLO evolution kernels, the jet functions provide the ingredients for NLL resummation. The explicit calculation also verifies that the infrared poles are absorbed by the renormalization of the generalized track functions. In the pure Yang-Mills, fixed-coupling limit, reciprocity reproduces the anomalous scalings derived from the light-ray OPE and makes clear that higher-point correlators receive distinct contributions from fully separated detectors and from the OPE of unresolved contact clusters with the remaining detectors.

Beyond the leading-power perturbative factorization, we identified the soft power corrections at lowest order in $\alpha_s$. They are governed by generalized Wilson-line matrix elements $\Omega^R_{ka}\sim\Lambda_{\rm QCD}^k$, in which the transverse energy flow is weighted to the $k$-th power and, when appropriate, restricted to the hadron class $R$. Although suppressed by powers of $\Lambda_{\rm QCD}/Q$, these terms come with an enhanced small-angle scaling. For equal-weight two-point correlators, the $\Omega_n$ channel drives the differential distribution toward $d\sigma/d\theta\sim 1/\theta^{1+n}$, rather than the classical perturbative $1/\theta$ behavior. For mixed correlators $\langle\mathcal{E}\mathcal{E}^n\rangle_{x_L}$, the smallest detector weight controls the leading correction, producing $1/\theta^2$ scaling. At higher orders in $\alpha_s$, corrections of order $\Lambda_{\rm QCD}$, represented here by the matrix elements $\Delta^R_{1i}$, can also arise for $n,m>1$. Determining their operator definitions and universality will be necessary for a complete precision treatment.

Our Pythia study provides a qualitative test of these predictions. For $\langle\mathcal{E}^n\mathcal{E}^n\rangle_{x_L}$, the simulations exhibit the perturbative $1/\theta$ behavior at larger collinear angles, steepen toward $1/\theta^{1+n}$ as the power corrections become important, and finally transition to linear scaling when the hadronization scale is reached. The same transition is visible in equal-weight two- and three-point correlators with $n=1.5,2,3$, while the mixed correlators $\langle\mathcal{E}\mathcal{E}^n\rangle_{x_L}$ follow the predicted $1/\theta^2$ scaling. Varying $Q$ further confirms that the angular region with negligible power corrections grows with the hard scale.

The correlator $\langle\mathcal{E}^2\mathcal{E}^2\rangle_{x_L}$ has already been measured in both $pp$ and Pb--Pb collisions and used to study the nuclear modification of jet substructure~\cite{CMS:2025ydi}. Our framework supplies a field-theoretic interpretation of its distinctive angular behavior and the ingredients for systematic calculations. Immediate directions include precision extraction and evolution of generalized track functions, NLL phenomenology with power corrections, and measurements involving charged or particle-identified hadrons, as well as ratios such as $\langle\mathcal{E}^3\mathcal{E}^3\rangle_{x_L}/\langle\mathcal{E}^2\mathcal{E}^2\mathcal{E}^2\rangle_{x_L}$ at high $Q$, which offer a sensitive probe of the small difference between their perturbative anomalous scalings.

More broadly, the results of this paper greatly extend how the space of collider observables can be organized. Collider experiments fundamentally measure correlations of detector operators, yet systematic calculations have long been confined to the infrared and collinear safe corner of this space, occupied by the correlators of the average null energy operator. Our results remove this restriction, bridging the hadronic detectors that experiments measure and the partonic light-ray operators that theory computes, promoting the full space of energy weights and hadron selections to systematically improvable observables, with the light-ray OPE organizing their angular dependence and the detector functions packaging their nonperturbative data. Each choice of weights and quantum numbers images the same scattering event through a different filter, resolving not only how energy flows, but how it is apportioned among hadrons and which hadrons carry it --- turning hadronization from a source of uncertainty into a target of study. We anticipate that this enlarged and organized space of observables will open new doors for the study of the Standard Model at colliders, from precision QCD to the quark-gluon plasma, and for the search for new physics.

\acknowledgments
We thank Carlota Andres, Philipp Aretz, Cyuan-Han Chang, Hao Chen, Jack Holguin, Yue-Zhou Li, Iain Stewart, and Wouter Waalewijn for useful discussions. We thank Philipp Aretz and Cyuan-Han Chang for useful comments on the manuscript. K.L. is supported by the U.S. Department of Energy under contract DE-AC02-06CH11357. I.M. and M.G. are supported by the DOE Early Career Award DE-SC0025581. I.M. is additionally supported by the Sloan Foundation and the Simons Collaboration on Confinement and QCD Strings.
\appendix

\section{One-loop Jet Function for the Three and Four-point Projected Correlators with Distinct Detector Weights}
\label{Appendix}

In this appendix, we provide the one-loop jet function for the three and four-point projected correlators with distinct detector weights.
\subsection{Three-Point Projected Correlators}
The one-loop jet function for the three-point projected correlator $\langle \cE_R^n \cE_R^m \cE_R^r \rangle_{x_L}$ is given for gluon jets as
\begin{align}
2^{n+m+r} &J_{g(1,0)}^{\langle \cE_R^n \cE_R^m \cE_R^r \rangle} = \Bigg\{ C_A T^n_{gR}(1) T^{m\,r}_{gR}(1,1) \Bigg[\frac{2 \Gamma (n) \Gamma (m+r)}{\Gamma(m+n+r)} \left(\psi ^{(0)}(n)+ \psi ^{(0)}(m+r) \right. \\
&- \left. 2 \psi ^{(0)}(n+m+r)\right) +\frac{\Gamma (n) \Gamma (m+r+1)}{\Gamma (n+m+r+2)} \Big((m+r-2 n+1) \left(\psi ^{(0)}(n) + \psi ^{(0)}(m+r+1) \right. \nn \\
&- \left. 2 \psi^{(0)}(n+m+r+2)\right)-1\Big) -\frac{\Gamma (n) \Gamma (m+r+3)}{\Gamma (n+m+r+4)} \Big((m+r+2 n+3) \left(\psi ^{(0)}(n)+\psi ^{(0)}(m+r+3) \right. \nn \\
&- \left. 2 \psi^{(0)}(n+m+r+4)\right)+3\Big)-\frac{\Gamma(n+1) \Gamma(m+r)}{\Gamma(n+m+r+2)} \Big((2 (m+r)-n-1) \left(\psi ^{(0)}(n+1) \right. \nn \\
&+ \left. \psi ^{(0)}(m+r)-2 \psi^{(0)}(n+m+r+2)\right)+1\Big)- \frac{\Gamma(m+r) \Gamma(n+3)}{\Gamma(n+m+r+4)} \Big((2(m+r)+n+3) \nn \\
&\times \left(\psi ^{(0)}(m+r)+\psi ^{(0)}(n+3)-2 \psi^{(0)}(m+n+r+4)\right)+3\Big) \Bigg]+\Bigg[\left(n^2+m^2+r^2 \right. \nn \\
&+ \left. 3(n+m+r)+2 m r+4\right) \left(\psi ^{(0)}(n+1)+ \psi^{(0)}(m+r+1) - 2 \psi ^{(0)}(n+m+r+4)\right) \nn \\
&+ 3(n+m+r)+n(m+r)+7\Bigg] \frac{8n_f\, T_F\, n(m+r) \Gamma(n) \Gamma(m+r)}{\Gamma(n+m+r+4)} T^n_{qR}(1) T^{m\,r}_{qR}(1,1) \Bigg\} \nn \\
&+ (n \leftrightarrow r) + (n \leftrightarrow m)\,,\nn
\end{align}
and for quark jets as
\begin{align}
2^{n+m+r} &J_{q(1,0)}^{\langle \cE_R^n \cE_R^m \cE_R^r \rangle} = \Bigg\{\Bigg[2 \left((n+m+r)^2+n^2+3(2n+m+r)+4\right)\left(\psi^{(0)}(n+1) \right.  \\
&+ \left. \psi ^{(0)}(m+r) - 2 \psi ^{(0)}(n+m+r+3)\right)+ (m+r)(m+r+9)+12 n+18\Bigg] \nn \\
&\times \frac{2C_F\, \Gamma(n+1) \Gamma(m+r)}{\Gamma(n+m+r+3)} T^n_{qR}(1) T^{m\,r}_{gR}(1,1) + \Bigg[2\left((n+m+r)^2+(m+r)^2 \right. \nn \\
&+ \left. 3(n+2m+2r)+4\right)\left(\psi^{(0)}(n)+\psi ^{(0)}(m+r+1) - 2 \psi ^{(0)}(n+m+r+3)\right) \nn \\
&+12 (m+r)+n(n+9)+18\Bigg] \frac{2C_F\, \Gamma(n) \Gamma(1+m+r)}{\Gamma(n+m+r+3)} T^n_{gR}(1) T^{m\,r}_{qR}(1,1) \Bigg\} \nn \\
&+ (n \leftrightarrow r) + (n \leftrightarrow m)\,.\nn
\end{align}

\subsection{Four-Point Projected Correlators}

The one-loop jet function for the four-point projected correlator $\langle \cE_R^n \cE_R^m \cE_R^r \cE_R^q \rangle_{x_L}$ is given for gluon jets as
\begin{align}
&2^{n+m+r+q}J_{g(1,0)}^{\langle \cE_R^n \cE_R^m \cE_R^r \cE_R^q \rangle} = \Bigg\{\Bigg[2 \Gamma (n) \Gamma (m+r+q) \Gamma (n+m+r+q+2) \Gamma (n+m+r+q+4) \nn\\
&\times \left(\psi ^{(0)}(n)+\psi^{(0)}(m+r+q)-2 \psi ^{(0)}(n+m+r+q)\right)  \\
&-\Gamma (n+1) \Gamma (m+r+q) \Gamma (n+m+r+q)\Gamma (n+m+r+q+4) \nn \\
&\times\left(1 +(2 m-n+2 q+2 r-1) (\psi ^{(0)}(n+1)+\psi ^{(0)}(m+r+q)-2 \psi^{(0)}(n+m+r+q+2))\right) \nn \\
&-\Gamma (n+3) \Gamma (m+r+q) \Gamma (n+m+r+q) \Gamma (n+m+r+q+2) \nn \\
&\times\left(3+(2 m+n+2 q+2 r+3) (\psi ^{(0)}(n+3)+\psi ^{(0)}(m+r+q)-2 \psi^{(0)}(n+m+r+q+4))\right) \nn \\
&-\Gamma (n) \Gamma (m+r+q+1) \Gamma (n+m+r+q) \Gamma (n+m+r+q+4) \nn \\
&\times\left(1-(m-2 n+q+r+1) (\psi ^{(0)}(n)+\psi ^{(0)}(m+r+q+1)-2 \psi^{(0)}(n+m+r+q+2))\right) \nn \\
&-\Gamma (n) \Gamma (m+r+q+3) \Gamma (n+m+r+q) \Gamma (n+m+r+q+2) \nn \\
&\times\left(3+(m+2 n+q+r+3) (\psi ^{(0)}(n)+\psi ^{(0)}(m+r+q+3)-2 \psi^{(0)}(n+m+r+q+4))\right)\Bigg] \nn \\
&\times \frac{4\, C_A}{\Gamma(n+m+r+q) \Gamma(n+m+r+q+2) \Gamma(n+m+r+q+4)} T_{gR}^n(1) T_{gR}^{m\,r\,q}(1,1,1) \nn \\
&+ \Bigg[\left(4+3n+n^2+3(m+q+r)+(m+q+r)^2\right) \left(\psi^{(0)}(n+1)+\psi ^{(0)}(m+q+r+1)\right. \nn \\
&- \left.2 \psi ^{(0)}(n+m+r+q+4)\right)+7+3n+3(m+r+q)+n(m+r+q)\Bigg] \nn \\
&\times \frac{8\, n_f\, T_F\, n (m+q+r) \Gamma(n) \Gamma(m+r+q) }{\Gamma(n+m+r+q+4)} T_{qR}^n(1) T_{qR}^{m\,r\,q}(1,1,1)\Bigg\} + (n \leftrightarrow m) + (n \leftrightarrow r) + (n \leftrightarrow q) \nn \\
&+ \Bigg\{\Bigg[\Gamma (n+q) \Gamma (m+r) \Gamma (n+m+r+q+2) \Gamma (n+m+r+q+4) \left(\psi ^{(0)}(n+q) +  \psi ^{(0)}(m+r) \right. \nn \\
&- \left. 2 \psi^{(0)}(n+m+r+q)\right)-\Gamma (m+r)\Gamma (n+q+1) \Gamma (n+m+r+q) \Gamma (n+m+r+q+4) \nn \\
&\times \left(1+(2 m-n-q+2 r-1) \left(\psi ^{(0)}(m+r)+\psi ^{(0)}(n+q+1)-2
\psi ^{(0)}(n+m+r+q+2)\right)\right) \nn \\
&- \Gamma (m+r) \Gamma (n+q+3) \Gamma (n+m+r+q) \Gamma (n+m+r+q+2) \nn \\
&\times \left(3+(2 m+n+q+2 r+3) \left(\psi ^{(0)}(m+r)+\psi ^{(0)}(n+q+3)-2\psi ^{(0)}(n+m+r+q+4)\right)\right) \nn \\
&+\Gamma (n+q)\Gamma (m+r)\Gamma (n+m+r+q+2) \Gamma (n+m+r+q+4) \left(\psi ^{(0)}(n+q)+\psi ^{(0)}(m+r) \right. \nn \\
&- \left. 2 \psi^{(0)}(n+m+r+q)\right) - \Gamma (n+q) \Gamma (m+r+1)  \Gamma (n+m+r+q) \Gamma (n+m+r+q+4) \nn \\
&\times \left(1+(2 n+2 q-m-r-1) \left(\psi ^{(0)}(n+q)+\psi ^{(0)}(m+r+1)-2\psi ^{(0)}(n+m+r+q+2)\right)\right) \nn \\
&-\Gamma (n+q) \Gamma (m+r+3) \Gamma (n+m+r+q) \Gamma (n+m+r+q+2) \nn \\
&\times \left(3+(m+2 n+2 q+r+3) \left(\psi ^{(0)}(n+q)+\psi ^{(0)}(m+r+3)-2\psi ^{(0)}(n+m+r+q+4)\right)\right)\Bigg] \nn \\
&\times \frac{4 \, C_A}{\Gamma(n+m+r+q)\Gamma(n+m+r+q+2)\Gamma(n+m+r+q+4)} T_{gR}^{n\,q}(1,1) T_{gR}^{m\,r}(1,1)\nn \\
&+ \Bigg[\left((m+r)^2+3 (m+r)+(n+q)^2+3 (n+q)+4\right) \left(\psi ^{(0)}(n+q+1)+\psi^{(0)}(m+r+1) \right. \nn \\
&- \left. 2 \psi ^{(0)}(n+m+r+q+4)\right)+(m+r) (n+q)+3 (m+r)+3 (n+q)+7\Bigg] \nn \\
&\times \frac{8\, n_f\, T_F\, (n+q)(m+r) \Gamma(n+q) \Gamma(m+r)}{\Gamma(n+m+r+q+4)} T_{qR}^{n\,q}(1,1) T_{qR}^{m\,r}(1,1)\Bigg\} + (n \leftrightarrow m) + (m \leftrightarrow q)\,,\nn
\end{align}
and for quark jets as
\begin{align}
2^{n+m+r+q}&J_{q,(1,0)}^{\langle \cE_R^n \cE_R^m \cE_R^r \cE_R^q \rangle} = \Bigg\{\Bigg[\left((m+q+r) (2 (m+q+r)+4 n+6)+4 n^2+12 n+8\right)  \\
&\times (\psi^{(0)}(m+q+r)+\psi ^{(0)}(n+1)-2 \psi ^{(0)}(m+n+q+r+3)) +(m+q+r)^2 \nn \\
&+9 (m+q+r)+12 n+18\Bigg] \frac{2 C_F \Gamma(n+1) \Gamma(m+r+q)}{\Gamma(n+m+r+q+3)} T_{qR}^n(1) T_{gR}^{m\,r\,q}(1,1,1)\nn \\
&+ \Bigg[\left((m+q+r) (4 (m+q+r)+4 n+12)+2 n^2+6 n+8\right) \nn \\
& \times (\psi^{(0)}(m+q+r+1)+\psi ^{(0)}(n)-2 \psi ^{(0)}(m+n+q+r+3))+12 (m+q+r) \nn \\
&+n^2+9 n+18\Bigg] \frac{2 C_F \Gamma(n) \Gamma(m+r+q+1)}{\Gamma(n+m+r+q+3)} T_{gR}^n(1) T_{qR}^{m\,r\,q}(1,1,1)\Bigg\}\nn \\
&+ (n \leftrightarrow m) +(n \leftrightarrow r) + (n \leftrightarrow q)\nn \\
&+ \Bigg\{\Bigg[\left(8+6(n+q)+2(n+q)^2+4(m+r)(n+m+r+q+3)\right) \nn \\
&\times \left(\psi^{(0)}(n+q)+\psi ^{(0)}(m+r+1)-2 \psi ^{(0)}(n+m+r+q+3)\right) + (n+q)^2+9(n+q) \nn \\
&+12(m+r)+18\Bigg]\frac{2 C_F \Gamma(n+q) \Gamma(m+r+1)}{\Gamma(n+m+r+q+3)} T_{gR}^{n\,q}(1,1) T_{qR}^{m\,r}(1,1) \nn \\
&+ \Bigg[\left(8+6(m+r)+2(m+r)^2+4(n+q)(n+m+r+q+3)\right) \nn \\
&\times \left(\psi^{(0)}(m+r)+\psi ^{(0)}(n+q+1)-2 \psi ^{(0)}(n+m+r+q+3)\right) + (m+r)^2+9(m+r) \nn \\
&+12(n+q)+18\Bigg]\frac{2 C_F \Gamma(m+r) \Gamma(n+q+1)}{\Gamma(n+m+r+q+3)} T_{qR}^{n\,q}(1,1) T_{gR}^{m\,r}(1,1)\Bigg\}\nn\\
&+ \left(n \leftrightarrow m\right) + \left(m \leftrightarrow q\right)\,.\nn
\end{align}

\bibliography{main.bib}{}
\bibliographystyle{JHEP}

\end{document}